\documentclass[12pt]{article}

\usepackage{amssymb,epsfig}

\newcommand{\deriv}{\stackrel{\leftrightarrow}{D}}
\newcommand{\derleft}{\stackrel{\leftarrow}{D}}
\newcommand{\derright}{\stackrel{\rightarrow}{D}}

\def\bea{\begin{eqnarray}}
\def\eea{\end{eqnarray}}

\def\lsim{ {\ \lower-1.2pt\vbox{\hbox{\rlap{$<$}\lower5pt\vbox{\hbox{$\sim$}
}}}\ } }
\def\gsim{ {\ \lower-1.2pt\vbox{\hbox{\rlap{$>$}\lower5pt\vbox{\hbox{$\sim$}
}}}\ } }

\begin{document}
\font\el=cmbx10 scaled \magstep2{\obeylines\hfill February, 2007}

 \begin{center}
    {\Large \bf
    Light-Cone Distribution Amplitudes of Axial-vector Mesons}
 \end{center}

\bigskip
\centerline{\sc Kwei-Chou Yang} \centerline{Department of Physics,
Chung Yuan Christian University} \centerline{Chung-Li, Taiwan 320,
Republic of China}
\bigskip
\bigskip
\bigskip
\bigskip
\centerline{\bf Abstract} \vspace{0.2cm} \noindent We have presented
a detailed study of twist-2 and twist-3 light-cone distribution
amplitudes of $1^3P_1$ and $1^1P_1$ axial-vector mesons, based on
QCD conformal partial wave expansion. Applying equations of motion,
the twist-three two-parton light-cone distribution amplitudes can be
expressed in terms of leading-twist and twist-three three-parton
light-cone distribution amplitudes. The relevant G-parity invariant
and violating parameters, containing the corrections due to the
SU(3) breaking effects, are evaluated from the QCD sum rule method.
The results for axial-vector decay constants of $1^3P_1$ states are
presented. The values of tensor decay constants and Gegenbauer
moments of the leading twist distribution amplitudes for $1^1P_1$
states are updated. Using Gell-Mann-Okubo mass formula, the mixing
angle for the $f_8$ and $f_1$ of $1^3P_1$ states is
$\theta_{^3P_1}\sim 38^\circ$, and that for $h_8$ and $h_1$ of
$1^1P_1$ states is $\theta_{^1P_1}\sim 10^\circ$. The detailed
properties for physical states $f_1(1285), f_1(1420), h_1(1170)$,
and $h_1(1380)$ are given. Assuming the mixing angle between
$K_{1A}$ and $K_{1B}$ to be $\theta_K=45^\circ$ or $-45^\circ$, we
also give the detailed study for $K_1(1270)$ and $K_1(1400)$. Using
the conformal partial wave expansion, we obtain the models for
light-cone distribution amplitudes, containing contributions up to
conformal spin 9/2. It is interesting to note that some distribution
amplitudes have significant asymmetric behaviors, which should be
phenomenologically attractive.

\pagebreak

\section{Introduction}

For an energetic light hadron moving nearly on the light-cone in the
reference frame, its distribution amplitudes can be described by a set
of light-cone distribution amplitudes (LCDAs)
\cite{BF2,Braun:2003rp}. The LCDAs are governed by the special
collinear subgroup $SL(2,\mathbb{R})$ of the conformal
group\cite{BF2,Braun:2003rp}. There are four generators for the
$SL(2,\mathbb{R})$ group. Three of generators describe the
generalized rotations in the $SL(2,\mathbb{R})$ space, where the
rotational invariance is characterized by the so-called conformal
spin $j$, in analogy to the orbital quantum number in quantum
mechanics of having spherically symmetric potential. The remaining
generator counting the {\it collinear twist} commutes with the rest
generators in the $SL(2,\mathbb{R})$ group; thus, in the
mathematical spirit, the concept of ``collinear twist'' is
equivalent to the ``energy" in quantum mechanics, and, in other
words, the $SL(2,\mathbb{R})$ group are satisfied by the same Lie
algebra as the $O(3)$ group.

The conformal partial wave expansion of a light-cone distribution
amplitude is fully analogous to the partial wave expansion of a wave
function in quantum mechanics. Each conformal partial wave is
labeled by the specific conformal spin $j$. On the other hand, for a
hadron moving nearly on the light-cone, the transverse separation of
the partons corresponding to the two-dimensional transverse plane
$(0, x_1, x_2, 0)$ is characterized by the $SL(2,\mathbb{C})$ group.
Moreover, the collinear $SL(2,\mathbb{R})$ and transverse
$SL(2,\mathbb{C})$ groups are not independent because they share the
same dilation generator. Integrating out the transverse degrees of
freedom, which is governed by renormalization group (RG) equation,
yields the scale-dependent behavior for each conformal partial wave
amplitude of a hadronic LCDA. The conformal invariance in QCD
exhibits that for leading twist LCDAs there is no mixing among
conformal partial wave amplitudes with different conformal spins to
leading logarithmic accuracy, while, for higher twist LCDAs, because
there may exist multiple operators with the same conformal spin, the
corresponding conformal partial amplitudes with the same conformal
spin can mix as the scale is changed. Roughly speaking, to leading
logarithmic accuracy, the anomalous dimensions rise logarithmically
with conformal spin. This implies that contributions of conformal
operators with high conformal spins are suppressed at large scale;
for instance, the well-known leading twist LCDA of the pion reads
$\phi_\pi(u,\mu\to \infty) \to 6u(1-u)$, i.e., the asymptotic
amplitude with the lowest conformal spin 2, where $u$ is the
momentum fraction carried by the quark (or anti-quark) in the pion.
It is interesting to note that in the BFKL approach
\cite{Fadin:1975cb,Kuraev:1976ge,Kuraev:1977fs,Balitsky:1978ic}, the
QCD scattering amplitude in the large energy limit is related to the
$SL(2,\mathbb{C})$ group (see, e.g., the discussion in
Ref.~\cite{Braun:2003rp}).

Understanding the hadronic LCDAs is very important since in the QCD
description of various exclusive processes the amplitudes can be
represented as the hadronic LCDAs convoluted with the interacting
hard kernel. The studies for the hadronic LCDAs were first done by
Chernyak and Zhinitsky \cite{Chernyak:1983ej}. They used the QCD sum
rule approach to calculate the moments of LCDAs of $\pi, K, \rho,
K^*$ and some baryons. Ten years later, a conformal description for
multi-parton LCDAs was presented in more details in Ref.~\cite{BF2}.
Most recently, using conformal expansion together with equations of
motion (EOMs), the different twist LCDAs for the light pseudoscalar
and vector mesons have been systematically studied in
Refs.~\cite{Ball99,Ball:1998sk,Ball:1998ff}.

In the present paper, we devote to examining twist-2 and twist-3
LCDAs of light axial-vector mesons with quantum numbers
$1^3P_1,1^1P_1$ and their mixtures. The motivation of this study is
as follows.
LCDAs of axial-vector mesons should be important for (exclusive)
phenomenologies involving axial-vector mesons. $B^0 \to K_1(1270)
\gamma$ has recently been measured by Belle \cite{Yang:2004as},
whereas the real physical states $K_1(1270)$ and $K_1(1400)$ are the
mixtures of ideal $1^3P_1$ ($K_{1A}$) and $1^1P_1$ ($K_{1B}$)
states. The first charmless hadronic $B$ decay involving a $1^3P_1$
meson that has been observed is $B^0 \to a_1^\pm (1260) \pi^\mp$
\cite{Abe:2005rf,Aubert:2006dd}, which is relevant to measurements
of the unitarity triangle $\alpha\equiv \phi_2$ of the
Cabibbo-Kobayashi-Maskawa (CKM). {\it BaBar} has recently extracted
the effective value $\alpha_{\rm eff}$ from the measurements of
$CP$-violating asymmetries in the decay $B^0 \to a_1^\pm (1260)
\pi^\mp$ \cite{Aubert:2006gb}, where the bound on the difference
$\Delta \alpha=\alpha -\alpha_{\rm eff}$ can be constrained by using
the broken SU(3) flavor symmetry \cite{Gronau:2005kw}. On the other
hand, ${\cal B}(B\to V~A, A~A)$ can be used to test QCD annihilation
topology in $B$ decays and probe the new-physics signals, where
$V\equiv$ vector meson and $A\equiv$ axial-vector meson
\cite{Yang:2005tv,kcymoments}.

So far, there is no literature for the calculations of LCDAs of
axial-vector mesons except my previous work about leading-twist
LCDAs of $1^1P_1$ states \cite{Yang:2005gk}. In
Ref.~\cite{Yang:2005gk}, we have calculated the Gegenbauer moments
(with conformal spins 0, 1, and 2) of the leading twist LCDAs for
the $b_1(1235), h_1(1380)$ and $K_{1B}$ as well as their tensor
decay constants. In this paper, we will derive all parameters,
relevant to the two- and three-parton LCDAs of twist-2 and -3, with
conformal spin up to $9/2$. We take into account SU(3) breaking
effects in the sum rule calculations for G-parity invariant and
violating parameters, where the latter parameters should vanish in
the SU(3) limit. The decay constants for axial-vector mesons can be
obtained by using the sum rules computed in Ref.~\cite{GRVW}, where
the authors focused on the studies of the masses and used a small
value of $\alpha_s$. Considering the current value of $\alpha_s$ and
the renormalization group improvement for sum rules, we also examine
the masses of the complete $1^3P_1$ set, which may offer information
for the quality of the sum rules. Our results for parameters
relevant to LCDAs of axial-vector mesons are essentially original.
To determine the relative signs for parameters, one of the
interpolating currents in the two-point correlation function is
adopted to be the local axial-vector current (or the local tensor
current) for $1^3P_1$ states (or $1^1P_1$ states). However, because
the resultant $f_{3,^3P_1}^{\perp}$ sum rule does not show reliable
quality, we thus resort to the diagonal correlation function to
calculate the sum rule for this parameter. Fortunately, the related
calculations can be found in Ref.~\cite{Ball:2006wn}, where the sum
rule is instead used to analyze the couplings for pseudoscalar
mesons. The detailed results are given in
Subsec.~\ref{sec:properties-3pdas-1-3}. Except the formula
$f_{3,^3P_1}^{\perp}$ can be adopted directly from
Ref.~\cite{Ball:2006wn}, we do not find any explicit results can be
applied directly, but only two equations can be compared with the
results in the massless quark limit in the literature. See
discussions before and after Eq.~(\ref{eq:ope-1}) and after
Eq.~(\ref{eq:1P1_SRf3t_sigma}).

We also update the numerical results for the decay constants and
leading twist LCDAs of the $1^3P_1$ states, due to the following
reasons. First, here we consider the uncertainties of the
condensates and strange quark mass, which were not fully included in
Ref.~\cite{Yang:2005gk}. Meanwhile, we re-examine the stability of
the sum rules with requirement that the contributions from excited
states and from the highest dimension term in OPE expansion can be
well under control within the working Borel window. Second, in
Ref.~\cite{Yang:2005gk} we assumed that the property of $h_1(1170)$
is the same as that of $b_1(1235)$, and the quark content of
$h_1(1370)$ is $\bar s s$. Instead, we study the pure $1^1P_1$
states here. We estimate the singlet-octet mixing angle by means of
Gell-Mann-Okubo mass formula (see the detailed discussions in
subsection~\ref{subsec:properties}). Consequently, the physical
properties for the real states $h_1(1170)$ and $h_1(1380)$ are
obtained. Third, the G-parity violating Gegenbauer moments were not
evaluated correctly in Ref.~\cite{Yang:2005gk} because the
corrections arising from the $1^3P_1$ states had been ignored. In
the sum rule calculations, for G-parity violating parameters, the
contributions originating from $1^1P_1$ and $1^3P_1$ are always of
the same order.

In the quark model, a $1^3P_1$ ($1^1P_1$) meson is represented as a
constituent quark-antiquark pair with total spin $S=1$ (0) and
angular momentum $L=1$ (1). Nevertheless, a real hadron in QCD
language should be described in terms of a set of Fock states for
which each state has the same quantum number as the hadron, and the
leading-twist LCDAs are thus interpreted as amplitudes of finding
the meson in states with a minimum number of partons. Interestingly,
due to the G-parity, the leading-twist LCDA $\Phi_\perp$
($\Phi_\parallel$) of a $1^3P_1$ ($1^1P_1$) meson defined by the
nonlocal tensor current (nonlocal axial-vector current) is
antisymmetric under the exchange of $quark$ and $anti$-$quark$
momentum fractions in the SU(3) limit, whereas the $\Phi_\parallel$
($\Phi_\perp$) defined by the nonlocal axial-vector current
(nonlocal tensor current) is symmetric. The large magnitude of the
first Gegenbauer moment of $\Phi_\parallel$ ($\Phi_\perp$) could
have a large impact on the longitudinal fraction of
factorization-suppressed $B$ decays involving a $1^1P_1$ (or
$1^3P_1$) meson evaluated in the QCD factorization framework
\cite{Yang:2005tv,kcymoments}. Furthermore, $\Phi_\perp$ is relevant
not only for exploring the tensor-type new-physics effects in $B$
decays~\cite{Yang:2005tv} but also for $B \to K_1 \gamma$
studies~\cite{kcy:BK1Gamma}.

This paper is organized as follows. A brief introduction for
conformal partial wave expansions of LCDAs is provided in
Sec.~\ref{sec:SL2}, where we first introduce the SL(2, $\mathbb{R}$)
group, and then discuss the asymptotic LCDAs, two-parton and
three-parton distribution amplitudes. In earlier days, Appell's
polynomials were used to form the conformal basis in expanding
three-parton distribution amplitudes. However, the Appell's
polynomials are not mutual orthogonal (see, e.g.,
Refs.~\cite{Braun:2003rp,Braun:1999te}). Following
Ref.~\cite{Braun:2003rp,Braun:1999te,Braun:2004bu}, we thus use the
orthogonal basis, which will be introduced in
Sec.~\ref{subsec:3lcda}, to expand the three-parton LCDAs of
twist-3. The detailed collinear twist properties of LCDAs are
collected in Appendix~\ref{appsec:spin}.

Sec.~\ref{sec:def-lcda} presents the notations, that we use in the
paper, and definitions of LCDAs of axial-vector mesons. Moreover, we
apply the EOMs to obtain the relations among the twist-two and
twist-three light-cone distribution amplitudes, so that we can use a
minimal number of independent nonperturbative parameters to describe
the distribution amplitudes. EOMs in the QCD perturbative theory
respect all symmetries given in the classical level; in other words,
conformal operators related by EOMs should have the same conformal
spin. Appendix~\ref{appsec:EOM} contains some relevant formulas of
EOMs for the present work.

Using the QCD sum rule technique \cite{SVZ},
Secs.~\ref{sec:properties-ltdas} and \ref{sec:properties-3pdas} are
devoted to the determination of relevant parameters of leading-twist
and twist-3 three-parton LCDAs, respectively. In
Sec.~\ref{sec:properties-ltdas}, the physical properties of the
axial-vector mesons such as the quark contents and decay constants
are discussed. The relevant inputs in the calculation are collected
in Appendix~\ref{appsec:inputs}. Sec.~\ref{sec:lcda} contains
explicit models for the twist-2 and twist-3 LCDAs of the
axial-vector mesons. Finally, we summarize in
Sec.~\ref{sec:summary}.

\section{Conformal Partial Wave Expansions of LCDAs}\label{sec:SL2}

The properties of fields living on the light-cone satisfy the
$SL(2,\mathbb{R})$ group, which is a collinear subgroup of the full
conformal group. The $SL(2,\mathbb{R})$ group is governed by four
generators $\bf{P_+, M_{-+}, D}$ and $\bf{K}_-$. Here and below we
introduce two light-like vectors $n^\mu$ and $\bar n^\mu$ which
satisfy $n_\mu n^\mu=\bar n_\mu \bar n^\mu=0$ and $n_\mu \bar
n^\mu=1$, so that for a general vector $A$ we define $A_-\equiv
A_\mu \bar n^\mu$ and $A_+\equiv A_\mu n^\mu$. In analogy to the
space rotation in quantum mechanics, the three linear combinations
of the $SL(2,\mathbb{R})$ generators can describe hyperbolic
rotations, and have the following commutation relations:
\begin{eqnarray}\label{eq:SL2com}
 [{\bf L}_0,{\bf L}_\mp] =\mp  {\bf L}_\mp\,, &~~&
 [{\bf L}_-,{\bf L}_+] = -2 {\bf L}_0 \,,
\end{eqnarray} where
 \begin{eqnarray}\label{eq:DEFalgbra}
  {\bf L}_+ = - i {\bf P}_\mu n^\mu\,, \ \ \
  {\bf L}_- =  ({i}/{2}) {\bf K}_\mu \bar{n}^\mu\,,  \ \ \
 {\bf L}_0 = -({i}/{2}) ({\bf D} - {\bf M}_{\mu \nu} \bar{n}^\mu n^\nu)\,,
 \end{eqnarray}
with ${\bf P_\mu, K_\mu, D}$ and ${\bf M_{\mu\nu}}$ being the
translation, special conformal transformation, dilation, and Lorentz
generators, respectively \cite{book}. Following the standard quantum
mechanical technique, the above generators acting on a field
$\Phi(x_-)$, which lives along the light-cone $\bar n^\mu$ (in other
words, the state depends only on $x^\mu=x_- n^\mu$), yield
 \begin{eqnarray}
 {}[{\bf L}_+,\Phi(x_-)] &=& - \partial_+\Phi(x_-) \equiv L_+\Phi(x_-)\,,
 \label{eq:SL2relation1}\\
 {}[{\bf L}_-,\Phi(x_-)] &=&  \left(x_-^2\partial_+ +2 j x_- \right)\Phi(x_-)
          \equiv L_-\Phi(x_-)\,,
 \label{eq:SL2relation2}\\
 {}[{\bf L}_0,\Phi(x_-)] &=&
 -\left(x_-\partial_++ j\right)\Phi(x_-) \equiv L_0\Phi(x_-)\,,
 \label{eq:SL2relation3}
 \end{eqnarray}
where
$j=(\ell+s)/2$ with $\ell$ and $s$ being the field's scaling
dimension and spin projection on the light-cone, respectively. $j$
is called the conformal spin since
\begin{equation}
 [{\bf L}^2,\Phi(x_-)] = j(j-1)\Phi(x_-)\equiv L^2\Phi(x_-),
\label{eq:L2}
\end{equation}
where \begin{equation}
  {\bf L}^2 = {\bf L}_0^2+{\bf L}_1^2+{\bf L}_2^2 =
{\bf L}_0^2 -{\bf L}_0 +{\bf L}_+{\bf L}_-\,.\label{L2-expansion1}
\end{equation}
The remaining generator of the $SL(2,\mathbb{R}$) is symbolized as
${\bf E} = i ({\bf D} + {\bf M}_{- +})$ and satisfies
\begin{eqnarray}
 [{\bf E},\Phi(x_-)] &=& (\ell -s)\Phi(x_-), \label{eq:E1}\\
 {} [{\bf E}, {\bf L}^2 ]&=& 0 ,\ \ \
 {} [{\bf E}, {\bf L}_0] = 0 , \label{eq:E3}
\end{eqnarray}
where $\ell-s=t$ is called the {\it collinear twist}\footnote{For
simplicity, it will be just denoted as ``twist" in this paper.
Conventionally, we call $\ell^{\rm can}-s$ as the twist, instead of
using $\ell-s$, where $\ell^{\rm can}$ is the canonical dimension.}
of the field $\Phi(x_-)$. See Appendix~\ref{appsec:spin} for further
discussions. Therefore the role of ${\bf E}$ is analogous to the
Hamiltonian in quantum mechanics, and the twist corresponds to the
eigenenergy of the Hamiltonian. In short, a LCDA defined by
(non-local composite) field operators of a given {\it collinear
twist} can be expanded by using the basis, which is made of the
eigenstates of $L^2$ and $L_0$, as one does the partial wave
expansion in quantum mechanics. Such expansions for LCDAs are
so-called conformal partial wave expansions. In QCD, in the large
scale limit and to leading logarithmic accuracy, we have $\ell =
\ell^{\rm can} + \gamma$, where $\ell^{\rm can}$ and $\gamma$ are
the canonical and anomalous dimensions of the field, respectively.

A generic multi-parton distribution amplitude
is defined through the matrix element of a multi-local operator
 \begin{eqnarray}\label{eq:generic-da}
 \langle 0| \Phi_m(x_{m-})\cdots  \Phi_1(x_{1-})|p\rangle &\sim&
  \int [du]\,   {\rm e}^{-ip(x_{1} u_{1} +\cdots + x_m u_m)}
   \phi(u_1,\cdots, u_m)\,,
\end{eqnarray} where $
   \int [du] = \int_0^1\!\! du_1\cdots \!\int_0^1\!\! du_m\,
   \delta(\sum u_i-1)$,
$u_i=p_{i+}/p_+$ are momentum fractions carried by partons ``$i$'',
and the quantum fields $\Phi_i(x_{i-})$ live on the light-cone.
$\phi(u_i)$ can be expanded in the Hilbert space by using the basis
defined by $L^2$ and $L_0$.  In the expansion, the term  with the
lowest conformal spin and non-zero coefficient is the so-called the
asymptotic distribution amplitude, which is usually dominant at the
large scale for the general case \cite{BF2,Braun:2003rp},
\begin{equation}\label{eq:asDA}
   \phi_{\rm as}(u_i)
 = \frac{\Gamma(2j_1+\cdots+2j_m)}{\Gamma(2j_1)\cdots \Gamma(2j_m)}
     u_1^{2j_1-1}u_2^{2j_2-1}\cdots u_m^{2j_m-1}\,,
\end{equation} where the normalization is chosen such that $\int
[du_i]\,\phi_{\rm as}(u_i)=1$.

\subsection{Two-parton distribution amplitudes}

The study for conformal expansion of two-parton distribution
amplitudes has been given by Ohrndorf \cite{Ohrndorf} (see, e.g.,
the discussion in Ref.~\cite{Braun:2003rp}). Consider
\begin{eqnarray}\label{eq:da-2p}
\langle 0| \Phi_2^{j_2}(x_2) \Phi_1^{j_1}(x_1)|p\rangle &=&
  \int [du]\,
   {\rm e}^{-ip(x_1 u_1  + x_2 u_2)}
   \phi(u_1,u_2;\mu)\,.
 \end{eqnarray}
Using the conformal basis, the two-parton distribution amplitude
with a given {\it collinear twist} can be written as
\begin{eqnarray}\label{eq:conformal-expansion-2p}
\phi(u_1, u_2;\mu) = \phi_{as} \sum_{j-j_1-j_2=0}^\infty
\phi_{j-j_1-j_2}(\mu)\,\mathbb{P}_{j-j_1-j_2}^{j_1,j_2}(u_1, u_2)
\,.
 \end{eqnarray}
where \begin{eqnarray}\label{eq:Jacobi}
  {\mathbb P}_N^{j_1,j_2}(u_1,u_2) = \sum_{n_1+n_2=N}
 \frac{u_1^{n_1} (-u_2)^{n_2}
 \left(
      \begin{array}{c}
       N \\ n_2
      \end{array}
 \right)}{\Gamma(n_1+2j_1)\Gamma(n_2+2j_2)} = (u_1+u_2)^N
 P^{2j_1-1,2j_2-1}_N
 \left(\frac{u_1-u_2}{u_1+u_2}\right),
 \end{eqnarray}
with $P^{a,b}_N(x)$ being the so-called Jacobi polynomials.
 Here the conformal spins of $\Phi_1^{j_1}$ and $\Phi_2^{j_2}$ are
$j_1$ and $j_2$, respectively.

As for the case of the twist-2 (leading-twist) distribution
amplitude of an axial-vector meson, we have the conformal spins
$j_1=j_2=1$ for both quark and anti-quark fields. The leading-twist
LCDA thus reads
\begin{eqnarray}\label{eq:general-da-lt}
\phi(u_1,u_2;\mu)=6u_1 u_2 \sum_{l=0}^\infty a_l(\mu)
C^{3/2}_l(u_1-u_2) ,
\end{eqnarray}
where $P^{1,1}_{j-2} (u_1-u_2)\sim C^{3/2}_l(u_1-u_2)$ with
$j-2=l$.

\subsection{Three-parton distribution amplitudes}\label{subsec:3lcda}

We follow the method suggested by Braun {\it et al.}
\cite{Braun:1999te}, who first applied it to baryon distribution
amplitudes, to construct the three-parton distribution amplitudes of
an axial-vector meson. The traditional choice is  to expand
three-parton distribution amplitudes in terms of Appell's
polynomials. However, this basis is inconvenient for calculations
since Appell's polynomials are not mutually orthogonal. An
orthonormal ``conformal basis'' of functions
$J_{Nn}^{(12)3}(u_i)=(-1)^N Y_{Jj}^{(12)3}(u_i)/2$ can be defined by
requiring that the total three-parton conformal spin
$J=j_1+j_2+j_3+N$ of an eigenstate is fixed, and, moreover, the
definite value of the conformal spin of the two-parton channel
$(12)$ is given as $j=j_1+j_2+n$ with $n=0,\cdots,N$. The results of
${Y}_{Jj}^{(12)3}$ can be written as
 \begin{eqnarray}\label{eq:YJi123}
 Y_{Jj}^{(12)3}(u_i) &=&
(1-u_3)^{j-j_1-j_2}\,
  P_{J-j-j_3}^{(2j_3-1,2j-1)} (1-2u_3)
\,P_{j-j_1-j_2}^{(2j_1-1,2j_2-1)}
\left(\frac{u_1-u_2}{1-u_3}\right)\,.
 \end{eqnarray}
Note that this basis of $Y_{Jj}^{(12)3}(u_i)$ can be related to the
basis of $Y_{Jj}^{1(23)}(u_i)$ (or $Y_{Jj}^{2(31)}(u_i)$) through
the Racah $6j$-symbols of $SL(2,\mathbb{R})$ group
\cite{Braun:1999te,Braun:2003rp}.

For axial-vector mesons, the three-parton LCDAs  correspond to the
higher Fock states consist of the quark, anti-quark, and gluon. We
refer to $j_1=1, j_2=1$ and $j_3=3/2$ as the conformal spins of the
quark, anti-quark and gluon, respectively, in an axial-vector meson.
Using Eq.~(\ref{eq:YJi123}),  we redefine the orthonormal basis
functions,
\begin{equation}\label{eq:3p-t3-basis}
J_{Nn}^{(12)3}(u_i)=\frac{(-1)^N}{2} Y_{Jj}^{(12)3}(u_i) =
\frac{(-1)^N}{2}
 (1-u_3)^{n}\,  P_{N-n}^{(2,3+2n)}(1-2u_3)
 \,P_{n}^{(1,1)}\left(\frac{u_1-u_2}{1-u_3}\right),
\end{equation}
where $J$ is the total three-parton conformal spin,
$j$ is the total conformal spin of the quark and anti-quark,
$N=J-7/2$ and $n=j-2$ with $N, n=0,1,\cdots$, and
$N\geq n$.
Thus, the meson's three-parton distribution amplitudes of
twist-three can be represented as
\begin{equation}\label{eq:3p-da-t3}
 \phi^{\bar{q}q g}(u_i)= 360  u_1 u_2 u_3^2\sum_{N=0}^\infty
 \sum_{n=0}^N
\, \omega_{N-n,n} J_{Nn}^{(12)3}(u_i)\,. \end{equation}
If one takes into account the conformal spins of the distribution
amplitudes up to order of $J=9/2$, the twist-3 LCDA
reads\footnote{The contribution of conformal spin $11/2$ to the
meson's three-parton distribution amplitudes of twist-three is
\begin{eqnarray}
 && 360  u_1 u_2 u_3^2
 [\omega_{2,0} J_{20}^{(12)3}(u_i) +
 \omega_{1,1} J_{21}^{(12)3}(u_i)
 + \omega_{0,2} J_{22}^{(12)3}(u_i)]\nonumber\\
 &=& 360  u_1 u_2 u_3^2
 \Bigg[(18 u_3^2 -16 u_3 +3)\omega_{2,0}+ 3(1-3u_3)(u_1-u_2)\omega_{1,1}
 +
  \frac{3}{2}[(1-u_3)^2- 5u_1 u_2]\omega_{0,2} \Bigg]\,.
\end{eqnarray}
Comparing the above result with that given by Braun and
Filyanov~\cite{BF2}, we thus find
 \begin{eqnarray}
 \omega_{2,0} &=& \frac{2}{5} \omega_{2,0}^{\rm BF}
  + \frac{1}{5} \omega_{1,1}^{\rm BF},\\
 \omega_{0,2} &=& \frac{8}{15} \omega_{2,0}^{\rm BF}
  - \frac{2}{5} \omega_{1,1}^{\rm BF}.
 \end{eqnarray}
 Note that for $\omega_{1,1}$ there is no corresponding term given in
 Ref.~\cite{BF2}.}
 \begin{eqnarray}\label{eq:3p-da-t3-approx}
 \phi^{\bar{q}q g}(u_i)&\cong& 360  u_1 u_2 u_3^2
 [\omega_{0,0} J_{00}^{(12)3}(u_i) +
 \omega_{1,0} J_{10}^{(12)3}(u_i)
 + \omega_{0,1} J_{11}^{(12)3}(u_i)]\nonumber\\
 &=& 360  u_1 u_2 u_3^2
 \Bigg[\frac{1}{2}\omega_{0,0}+ \frac{1}{2}(7u_3-3)\omega_{1,0} -
 (u_1-u_2)\omega_{0,1} \Bigg]\,.
 \end{eqnarray}

\section{Definitions of light-cone distribution amplitudes}\label{sec:def-lcda}

In what follows, we define $z=y-x$ with $z^2=0$, and introduce the
light-like vector $p_\mu=P_\mu-m_A^2 z_\mu/(2 P\cdot z)$ with the
axial-vector meson's momentum ${P}^2=m_A^2$. The polarization vector
$\epsilon_\mu^{(\lambda)}$ of the axial-vector meson can be
decomposed into {\it longitudinal} and {\it transverse projections}
as~\cite{Ball:1998sk,Ball:1998ff}
\begin{eqnarray}\label{eq:polprojectiors}
 && \epsilon^{(\lambda)}_{\parallel\, \mu} \equiv
     \frac{\epsilon^{(\lambda)} \cdot z}{p\cdot z} \left(
      p_\mu-\frac{m_A^2}{2 p\cdot z} \,z_\mu\right), \qquad
 \epsilon^{(\lambda)}_{\perp\, \mu}
        = \epsilon^{(\lambda)}_\mu -\epsilon^{(\lambda)}_{\parallel\, \mu}\,,
\end{eqnarray}
respectively. In QCD description of hard processes involving axial
vector mesons, one encounter bilocal operators sandwiched between
the vacuum and the meson,
 \begin{eqnarray}\label{eq:general_op}
  \langle A(P,\lambda)|\bar q_1(y) \Gamma [y,x] q_2(x)|0\rangle ,
\end{eqnarray}
where $\Gamma$ is a generic notation for the Dirac matrix
structure and the path-ordered gauge factor is
\begin{equation}\label{eq:guagefactor}
[y,x]={\rm P}\exp\left[ig_s\!\!\int_0^1\!\! dt\,(x-y)_\mu
  A^\mu(t x+(1-t)y)\right].
\end{equation}
This factor is equal to unity in the light-cone gauge which is
equivalent to  the fixed-point gauge $(x-y)^\mu A_\mu(x-y)=0$, which
is also called the Fock-Schwinger gauge, as the quark-antiquark pair
is at the light-like separation. For simplicity, here and below we
do not show the gauge factor. For $f_1$ and $h_1$, the operators in
Eq.~(\ref{eq:general_op}) correspond to $(\bar u(y)\Gamma u(x) +
\bar d(y)\Gamma d(x) + \bar s(y)\Gamma s(x))/\sqrt{3} $, while for
$f_8$ and $h_8$, the relevant forms of operators are $(\bar
u(y)\Gamma u(x) + \bar d(y)\Gamma d(x) -2 \bar s(y)\Gamma
s(x))/\sqrt{6} $. In the present study, we adopt the conventions
$D_\alpha=\partial_\alpha +ig_s A^a_\alpha \lambda^a/2$,
$\widetilde{G}_{\alpha\beta}=(1/2)
\epsilon_{\alpha\beta\mu\nu}G^{\mu\nu}$, $\epsilon^{0123}=-1$.

In general, the LCDAs are scheme- and scale-dependent. One can
catalog the distribution amplitudes into two classes: (i)
chiral-even LCDAs for which the relevant current operators involve
chirality-conserving structures $\Gamma=\{\gamma_\mu,
\gamma_\mu\gamma_5 \}$, and (ii) chiral-odd LCDAs for which the
operators contain chirality-violating structures
$\Gamma=\{\sigma_{\mu\nu}(\gamma_5), 1\}$.

\subsection{Two-parton distribution amplitudes}\label{subsec:2da-def}
The chiral-even LCDAs are given by~\footnote{$g_\perp^{(v)}$ and
$g_\perp^{(a)}$ given in Ref.~\cite{Yang:2005gk} is respectively
redefined to be $g_\perp^{(a)}$ and $g_\perp^{(v)}$ in this paper.}
\begin{eqnarray}
  \langle A(P,\lambda)|\bar q_1(y) \gamma_\mu \gamma_5 q_2(x)|0\rangle
  && = if_A m_A \, \int_0^1
      du \,  e^{i (u \, p y +
    \bar u p x)}
   \left\{p_\mu \,
    \frac{\epsilon^{*(\lambda)} z}{p z} \, \Phi_\parallel(u)
         +\epsilon_{\perp\mu}^{*(\lambda)} \,
    g_\perp^{(a)}(u) \right. \nonumber\\
    &&\ \ \left. - \frac{1}{2}z_{\mu}
\frac{\epsilon^{*(\lambda)} z }{(p  z)^{2}} m_{A}^{2} g_{3}(u) \right\},
                                 \label{eq:evendef1} \\
  \langle A(P,\lambda)|\bar q_1(y) \gamma_\mu
  q_2(x)|0\rangle
  && = - i f_A m_A
\,\epsilon_{\mu\nu\rho\sigma} \,
      \epsilon^{*\nu}_{(\lambda)} p^{\rho} z^\sigma \, \int_0^1 du \,  e^{i (u \, p y +
    \bar u\, p  x)} \,
       \frac{g_\perp^{(v)}(u)}{4}, \label{eq:evendef2}
\end{eqnarray}
where $u$ and $\bar u\equiv 1-u$ are the momentum fractions of $q_1$
and $\bar q_2$ in the axial-vector meson, respectively. The
chiral-odd LCDAs are defined by~\footnote{$h_\parallel^{(s)}$ given
in Ref.~\cite{Yang:2005gk} is redefined to be $h_\parallel^{(p)}$ in
this paper.}
\begin{eqnarray}
  &&\langle A(P,\lambda)|\bar q_1(y) \sigma_{\mu\nu}\gamma_5 q_2(x)
            |0\rangle
  =  f_A^{\perp} \,\int_0^1 du \, e^{i (u \, p y +
    \bar u\, p x)} \,
\Bigg\{(\epsilon^{*(\lambda)}_{\perp\mu} p_{\nu} -
  \epsilon_{\perp\nu}^{*(\lambda)}  p_{\mu})
  \Phi_\perp(u)\nonumber\\
&& \hspace*{+5cm}
  + \,\frac{m_A^2\,\epsilon^{*(\lambda)} z}{(p z)^2} \,
   (p_\mu z_\nu -
    p_\nu  z_\mu) \, h_\parallel^{(t)}(u)\nonumber\\
 && \hspace*{+5cm} + \frac{1}{2}
(\epsilon^{*(\lambda)}_{\perp \mu} z_\nu
-\epsilon^{*(\lambda)}_{\perp \nu} z_\mu) \frac{m_{A}^{2}}{p\cdot z}
 h_{3}(u)
\Bigg\},\label{eq:odddef1}\\
&&\langle A(P,\lambda)|\bar q_1(y) \gamma_5 q_2(x) |0\rangle
  =  f_A^\perp
 m_{A}^2 (\epsilon^{*(\lambda)} z)\,\int_0^1 du \, e^{i (u \, p y +
    \bar u\, p x)}  \, \frac{h_\parallel^{(p)}(u)}{2}\,. \label{eq:odddef2}
\end{eqnarray}
Here $\Phi_\parallel, \Phi_\perp$ are of twist-2, $g_\perp^{(a)},
g_\perp^{(v)}, h_\parallel^{(t)}, h_\parallel^{(p)}$ of twist-3, and
$g_3, h_3$ of twist-4.
 In SU(3) limit, due to G-parity,
$\Phi_\parallel, g_\perp^{(a)}$, $g_\perp^{(v)}$, and $g_3$ are
symmetric [antisymmetric] under the replacement $u\to 1-u$ for the
$1^3P_1$ [$1^1P_1$] states, whereas $\Phi_\perp, h_\parallel^{(t)}$,
$h_\parallel^{(p)}$, and $h_3$ are antisymmetric [symmetric]. In
other words, in the SU(3) limit it follows that
\begin{equation}
 \int_0^1 du \Phi_\perp(u)=\int_0^1 du h_\parallel^{(t)}(u)
 =\int_0^1 du h_\parallel^{(p)}(u)=\int_0^1 du h_3(u)=0
 \end{equation}
for $1^3P_1$ states, but becomes
\begin{equation}
 \int_0^1 du \Phi_\parallel(u)=\int_0^1 du g_\perp^{(a)}(u)
 =\int_0^1 dug_\perp^{(v)}(u)=\int_0^1 du g_3(u)=0
 \end{equation}
for $1^1P_1$ states. The above integrals are not zero if
$m_{q_1}\neq m_{q_2}$, and the detailed results are summarized in
Secs.~\ref{subsec:2plcda-t2} and \ref{subsec:2plcda-t3}. We will not
further discuss the twist-4 LCDAs, $g_3$ and $h_3$, below. For
convenience, we therefore normalize the distribution amplitudes of
the $1^3P_1$ [$1^1P_1$] states to be subject to
 \begin{equation}
 \int_0^1 du\Phi_\parallel(u)=1 \ \ \
  \Bigg[\int_0^1 du\Phi_\perp(u)=1 \Bigg],
 \end{equation}
and take $f_{^3\! P_1}^\perp=f_{^3\! P_1}$\, [$f_{^1\! P_1}=f_{^1\!
P_1}^\perp(\mu=1~{\rm GeV})$] in the study, such that we have
 \begin{eqnarray}
 &&\langle 1^3P_1(P,\lambda)|
  \bar q_1(0) \sigma_{\mu\nu}\gamma_5 q_2(0)
   |0\rangle
  =  f_{^3P_1}^{\perp} a_0^{\perp,^3P_1} \,
(\epsilon^{*(\lambda)}_{\mu} P_{\nu} - \epsilon_{\nu}^{*(\lambda)}
P_{\mu}),
 \\
 &&  \langle 1^1P_1(P,\lambda)|\bar q_1(0) \gamma_\mu \gamma_5 q_2(0)|0\rangle
   = if_{^1P_1} a_0^{\parallel,^1P_1} \, m_{^1P_1} \,  \epsilon^{*(\lambda)}_\mu
   \,,
 \end{eqnarray}
where $a_0^{\perp,^3P_1}$ and  $a_0^{\parallel,^1P_1}$ are the
Gegenbauer zeroth moments, defined in Eqs.~(\ref{eq:generaldaperp})
and (\ref{eq:generaldaparallel}), and vanish in the SU(3) limit.
Using the results given in Eqs.~(\ref{eq:asDA}), (\ref{eq:da-2p})
and (\ref{eq:Jacobi}), the two-parton distribution amplitudes can be
expanded in a series of partial waves with different conformal
spins. However, in analogy to that discussed in
Refs.~\cite{BF2,Ball:1998sk}, by means of QCD EOMs, $g_\perp^{(a)}$,
$g_\perp^{(v)}, h_\parallel^{(t)}$, and $h_\parallel^{(p)}$ can be
expressed in terms of twist-two two-parton and twist-three
three-parton distribution amplitudes. The detailed results are given
in Secs.~\ref{subsec:da-even} and \ref{subsec:da-odd}. It should be
stressed that EOMs only relate terms with the same conformal spin so
that the relations between LCDAs can be satisfied order by order in
the conformal expansion~\cite{Braun:2003rp}.

\subsection{Three-parton distribution amplitudes of twist-three}\label{subsec:3da-def}

The three-parton chiral-even (${\cal A,V}$) and chiral-odd (${\cal T}$) distribution amplitudes of twist-3 are
defined by
\begin{eqnarray}
&& \langle A(P,\lambda)|\bar q_1(-x) \gamma_\alpha\gamma_5 g_s
G_{\mu\nu}(vx)
         q_2(x)|0\rangle =
 - p_\alpha[p_\mu \epsilon^{*(\lambda)}_{\perp\nu}
  -p_\nu\epsilon^{*(\lambda)}_{\perp\mu}]
      f_{3A}^A{\cal A}(v,-px)+\ldots \,,~~~~
      \label{eq:3plcda-t3-1}\\ [0.3em]
&& \langle A(P,\lambda)|\bar q_1(-x) \gamma_\alpha
         g_s\widetilde G_{\mu\nu}(vx) q_2(x)|0\rangle =
 i  p_\alpha[p_\mu \epsilon^{*(\lambda)}_{\perp\nu}-p_\nu
\epsilon^{*(\lambda)}_{\perp\mu}]
      f_{3A}^V{\cal V}(v,-px)+\ldots
      \,,\label{eq:3plcda-t3-2}\\[0.3em]
&&\langle A(P,\lambda)|\bar q_1(-x) \sigma_{\alpha\beta}\gamma_5
g_s G_{\mu\nu}(vx)  q_2(x)|0\rangle \nonumber\\
&&\ \ \ \ \ = -i \frac{\epsilon^{*(\lambda)} x }{2 (p x)}
    [ p_\alpha p_\mu g^\perp_{\beta\nu}
     -p_\beta p_\mu g^\perp_{\alpha\nu}
     -p_\alpha p_\nu g^\perp_{\beta\mu}
     +p_\beta p_\nu g^\perp_{\alpha\mu} ]
     f_{3A}^\perp m_{A} {\cal T}(v,-px)+\ldots,\label{eq:3plcda-t3-3}
\end{eqnarray}
where the ellipses stand for terms of twist higher than three, the
following shorthand notations are used:
\begin{equation}
   {\cal A}(v,-px) \equiv  \int {\cal D}\underline{\alpha}
   \,e^{ipx(\alpha_{q_2}-\alpha_{q_1}+v\alpha_g)}{\cal A}
   (\alpha_{q_1}\alpha_{q_2},\alpha_g), \label{eq:short}
\end{equation}
etc., and the integration measure is defined as
\begin{equation}
 \int {\cal D}\underline{\alpha} \equiv \int_0^1 d\alpha_{q_1}
  \int_0^1 d\alpha_{q_2}\int_0^1 d\alpha_g \,\delta(1-\sum \alpha_i).
\label{eq:measure}
\end{equation}
Here  $\alpha_{q_1}$, $\alpha_{q_2}$, and $\alpha_g$ are the
respective momentum fractions carried by $q_1$, $\bar q_2$ quarks
and gluon in the axial-vector meson. Due to G-parity, for a $1^3P_1$
[$1^1P_1$] state ${\cal A}$ is antisymmetric [symmetric] under the
interchange $\alpha_{q_2} \leftrightarrow \alpha_{q_1}$ in the SU(3)
limit, while ${\cal V}$ and ${\cal T}$ are symmetric [antisymmetric]
(cf. the case of the $\rho$ meson in Ref.~\cite{Ball:1998sk}).
Taking into account the contributions up to terms of conformal spin
$9/2$ and considering the corrections which arise from the quark
masses, the distribution amplitudes can be approximately written as
 \begin{eqnarray}
  {\cal A}&=&5040 (\alpha_{q_1}-\alpha_{q_2})\alpha_{q_1}\alpha_{q_2}\alpha_{g}^2
  +360\alpha_{q_1}\alpha_{q_2}\alpha_{g}^2
  \Big[ \lambda^A_{^3P_1}+ \sigma^A_{^3P_1}\frac{1}{2}(7\alpha_g-3)\Big],
 \label{eq:lcda-3p1-t3-1}\\
  {\cal V}&=&360\alpha_{q_1}\alpha_{q_2}\alpha_{g}^2
  \Big[ 1+ \omega^V_{^3P_1}\frac{1}{2}(7\alpha_g-3)\Big]
 +5040 (\alpha_{q_1}-\alpha_{q_2})\alpha_{q_1}\alpha_{q_2}\alpha_{g}^2
 \sigma^V_{^3P_1},\label{eq:lcda-3p1-t3-2}\\
  {\cal T}&=&360\alpha_{q_1}\alpha_{q_2}\alpha_{g}^2
\Big[ 1+ \omega^\perp_{^3P_1}\frac{1}{2}(7\alpha_g-3)\Big]+5040
(\alpha_{q_1}-\alpha_{q_2})\alpha_{q_1}\alpha_{q_2}\alpha_{g}^2
 \sigma^\perp_{^3P_1},
  \label{eq:lcda-3p1-t3-3}
 \end{eqnarray}
for the $1^3P_1$ states, and
 \begin{eqnarray}
  {\cal A}&=&360 \alpha_{q_1}\alpha_{q_2}\alpha_{g}^2
  \Big[ 1+ \omega^A_{^1P_1}\frac{1}{2}(7\alpha_g-3)\Big]
  +5040 (\alpha_{q_1}-\alpha_{q_2})\alpha_{q_1}\alpha_{q_2}\alpha_{g}^2
 \sigma^A_{^1P_1}, \label{eq:lcda-1p1-t3-1}\\
  {\cal V}&=&5040 (\alpha_{q_1}-\alpha_{q_2})\alpha_{q_1}\alpha_{q_2}\alpha_{g}^2
  +360\alpha_{q_1}\alpha_{q_2}\alpha_{g}^2
  \Big[ \lambda^V_{^1P_1}+ \sigma^V_{^1P_1}\frac{1}{2}(7\alpha_g-3)\Big],
  \label{eq:lcda-1p1-t3-2}\\
  {\cal T}&=&5040 (\alpha_{q_1}-\alpha_{q_2})\alpha_{q_1}\alpha_{q_2}\alpha_{g}^2
  +360\alpha_{q_1}\alpha_{q_2}\alpha_{g}^2
  \Big[ \lambda^\perp_{^1P_1}+ \sigma^\perp_{^1P_1}\frac{1}{2}(7\alpha_g-3)\Big],
  \label{eq:lcda-1p1-t3-3}
 \end{eqnarray}
for the $1^1P_1$ states, where $\lambda$'s correspond to conformal
spin 7/2, while $\omega$'s and $\sigma$'s are parameters with
conformal spin 9/2. As the SU(3)-symmetry (and G-parity) is
restored, we have $\lambda$'s=$\sigma$'s=0. The normalization
constants $f_{3A}^A, f_{3A}^V$ and $ f_{3A}^\perp$ are thus defined
in such a way that
\begin{eqnarray}
 \int\! {\cal D}\underline{\alpha}\, (\alpha_{q_1}-\alpha_{q_2})\,{\cal A}
    (\alpha_{q_1},\alpha_{q_2},\alpha_g) &=&1,~~~
 \int\! {\cal D}\underline{\alpha}\, \,{\cal A}
    (\alpha_{q_1},\alpha_{q_2},\alpha_g) = \lambda^A_{^3P_1},
\nonumber\\
 \int\! {\cal D}\underline{\alpha}\, {\cal V}
    (\alpha_{q_1},\alpha_{q_2},\alpha_g) &=&1, ~~~
 \int\! {\cal D}\underline{\alpha}\, (\alpha_{q_1}-\alpha_{q_2}){\cal V}
    (\alpha_{q_1},\alpha_{q_2},\alpha_g) = \sigma^V_{^3P_1},\nonumber\\
 \int\! {\cal D}\underline{\alpha}\, {\cal T}
    (\alpha_{q_1},\alpha_{q_2},\alpha_g) &=&1, ~~~
 \int\! {\cal D}\underline{\alpha}\, (\alpha_{q_1}-\alpha_{q_2}){\cal T}
    (\alpha_{q_1}, \alpha_{q_2},\alpha_g) = \sigma^\perp_{^3P_1},
    \label{eq:normalize3-3p1}
\end{eqnarray}
for the $1^3P_1$ states, and
 \begin{eqnarray}
 \int\! {\cal D}\underline{\alpha}\, {\cal A}
    (\alpha_{q_1},\alpha_{q_2},\alpha_g) &=&1, ~~~
 \int\! {\cal D}\underline{\alpha}\, (\alpha_{q_1}-\alpha_{q_2}){\cal A}
    (\alpha_{q_1},\alpha_{q_2},\alpha_g) = \sigma^A_{^1P_1},
\nonumber\\
 \int\! {\cal D}\underline{\alpha}\,(\alpha_{q_1}-\alpha_{q_2})\, {\cal V}
    (\alpha_{q_1},\alpha_{q_2},\alpha_g) &=&1,~~~
 \int\! {\cal D}\underline{\alpha}\, \,{\cal V}
    (\alpha_{q_1},\alpha_{q_2},\alpha_g) = \lambda^V_{^1P_1}, \nonumber\\
 \int\! {\cal D}\underline{\alpha}\,(\alpha_{q_1}-\alpha_{q_2})\, {\cal T}
    (\alpha_{q_1},\alpha_{q_2},\alpha_g) &=&1,~~~
 \int\! {\cal D}\underline{\alpha}\, \,{\cal T}
    (\alpha_{q_1},\alpha_{q_2},\alpha_g) = \lambda^\perp_{^1P_1},
    \label{eq:normalize3-1p1}
\end{eqnarray}
for the $1^1P_1$ states.

\subsection{Relations among chiral-even LCDAs}\label{subsec:da-even}

$\Phi_\parallel(u,\mu)$ can be expanded in a series of Gegenbauer
polynomials~\cite{BF2,Braun:2003rp}:
\begin{eqnarray}\label{eq:generaldaparallel}
\Phi_\parallel^A(u,\mu)=6u(1-u)\Bigg[ a_0^{\parallel,A}+
\sum_{l=1}^\infty a_l^{\parallel,A}(\mu) C^{3/2}_l(2u-1) \Bigg],
\end{eqnarray}
where $\mu$ is the normalization scale and the multiplicatively
renormalizable coefficients (or called Gegenbauer moments) are:
\begin{eqnarray}\label{eq:aparallel}
a_l^{\parallel,A}(\mu) = \frac{2(2l+3)}{3(l+1)(l+2)} \int_0^1 dx\,
C^{3/2}_l (2x-1) \Phi_\parallel^A(x,\mu).
\end{eqnarray}
In the limit of $m_{q_1}=m_{q_2}$, only terms with even (odd) $l$
survive due to G-parity invariance for the $1^3P_1$ ($1^1P_1$)
mesons. In the expansion of $\Phi_{\parallel}^A(u,\mu)$ in
Eq.~(\ref{eq:generaldaparallel}), the conformal invariance of the
light-cone QCD exhibits that partial waves with different conformal
spin cannot mix under renormalization to leading-order accuracy. As
a consequence, the Gegenbauer moments $a_l^{\parallel}$ renormalize
multiplicatively:
  \begin{equation}
    a_l^{\parallel,A}(\mu) = a_l^{\parallel,A}(\mu_0)
  \left(\frac{\alpha_s(\mu_0)}{\alpha_s(\mu)}\right)^{-\gamma_{(l)}^\parallel/{b}},
  \label{eq:RGparallel}
   \end{equation}
where $b=(11 N_c -2n_f)/3$ and the one-loop anomalous dimensions
are \cite{GW}
  \begin{eqnarray}
  \gamma_{(l)}^\parallel  = C_F
  \left(1-\frac{2}{(l+1)(l+2)}+4 \sum_{j=2}^{l+1} \frac{1}{j}\right),
  \label{eq:1loopandim}
  \end{eqnarray}
with $C_F=(N_c^2-1)/(2N_c )$.

Applying the QCD equations of motion, as the case for the vector
mesons in Refs.~\cite{Ball:1998sk,Ball:1998ff}, one can obtain some
useful nonlocal operator identities (see Appendix \ref{appsec:EOM})
such that the two-parton distribution amplitudes $g_\perp^{(a)}$ and
$g_\perp^{(v)}$ can be represented in terms of
$\Phi_{\parallel,\perp}$ and three-parton distribution amplitudes.
Setting $y=-x$ and adapting the formulas derived in
Ref.~\cite{Ball:1998sk} for vector mesons to the present case, we
find\footnote{Here and below we use the notations close to that
given in Ref.~\cite{Ball:1998sk}.}
\begin{eqnarray}
\lefteqn{ \int_0^1du\,e^{-i\xi px}g_\perp^{(a)}(u)\
 =\ \int^1_0 dt\int_0^1du\,e^{-it\xi px}\Phi_\parallel(u)
 }
 \makebox[1cm]{\ }\nonumber\\
& &{} -\zeta_{3,A}^V(px)^2\int_0^1\,t^2dt\,\int_{-1}^1dv\,{\cal
V}(v,-tpx)-\zeta_{3,A}^A(px)^2\int_0^1 dt\,t^2\int_{-1}^1
dv\,v\,{\cal A}(v,-tpx)
 \nonumber\\
& &{} -{1\over 2}(px)^2 \int_0^1dt\,t^2\,\int_0^1du\,e^{-it\xi
px}g_\perp^{(v)}(u)- i\widetilde{\delta}_+ (px)\int_0^1dt\,t
\int_0^1du\,e^{-it\xi px}\Phi_\perp(u), \label{eq:eom-even1}
\end{eqnarray}
 and
\begin{eqnarray}
 \lefteqn{{1\over 2} \int_0^1du\,e^{-i\xi px}g_\perp^{(v)}(u)\ = \
 \int_0^1dt\,t\int_0^1du\,e^{-it\xi px}g_\perp^{(a)}(u)}\makebox[2cm]{\ }
\nonumber\\
 & &{} - i\zeta_{3,A}^A(px)\int_0^1dt\,t^2\int_{-1}^1dv{\cal A}(v,-tpx)
 -i\zeta_{3,A}^V(px)\int_0^1dt\,t^2\int_{-1}^1dv\,v{\cal V}(v,-tpx)\nonumber\\
 & &{} + \widetilde{\delta}_- \int_0^1dt\,t\int_0^1du\,e^{-it\xi px}
 \Phi_\perp(u), \label{eq:eom-even2}
 \end{eqnarray}
where $\xi\equiv 2u-1$, and we introduced the abbreviation
\begin{equation}
\widetilde{\delta}_\pm  ={f_A^{\perp}\over f_A}{m_{q_2} \pm m_{q_1}
\over m_A},\qquad \zeta_{3,A}^{V,A} = \frac{f^{V,A}_{3A}}{f_A m_A}.
\label{eq:parameters3}
\end{equation}

Solving Eqs.~(\ref{eq:eom-even1}) and (\ref{eq:eom-even2}), we
obtain the solutions \cite{Ball:1998sk} for $g_\perp^{(v)}(u)$:
\begin{eqnarray}
 g^{(v)}_\perp(u) & = &
 \bar u \int\limits_0^u\!\! dv\, \frac{1}{\bar v}\,\Psi(v) + u
\int\limits_u^1\!\! dv\, \frac{1}{v}\,\Psi(v) .
\label{eq:even-sol1}
 \end{eqnarray}
and for $g_\perp^{(a)}(u)$:
\begin{eqnarray}
g_\perp^{(a)}(u) & = &
 \frac{1}{4} \left[ \int\limits_0^u\!\! dv\,
 \frac{1}{\bar v}\,\Psi(v) + \int\limits_u^1\!\! dv\,
 \frac{1}{v}\,\Psi(v)\right]
 - \widetilde{\delta}_- \Phi_\perp(u)
 \nonumber\\
&&{} + \zeta_{3,A}^{V}\int\limits_0^u \!\! d\alpha_{q_1}\!\!
 \int\limits_0^{\bar {u}}\!\! d\alpha_{q_2}\,
 \frac{1}{1-\alpha_{q_1}-\alpha_{q_2}}\,
 \left(\frac{d}{d\alpha_{q_1}} + \frac{d}{d\alpha_{q_2}} \right)
 {\cal V}(\underline{\alpha})\nonumber\\
&&{}+\zeta_{3,A}^{A}\,\frac{d}{du}\int\limits_0^u \!\!
 d\alpha_{q_1}\!\! \int\limits_0^{\bar u}\!\! d\alpha_{q_2}\,
\frac{{\cal A}(\underline{\alpha})} {1-\alpha_{q_1}-\alpha_{q_2}},
\label{eq:even-sol2}
\end{eqnarray}
where
 \begin{eqnarray}
 \Psi(u)  &=&  2\Phi_\parallel(u) - \widetilde{\delta}_- \xi\Phi'_\perp(u)
 - \widetilde{\delta}_+ \Phi'_\perp(u)\nonumber\\
& &
 {}+2\zeta_{3,A}^A\,\frac{d}{du}\,\int\limits_0^u\!\!d\alpha_{q_1}
\int\limits_0^{\bar u}\!\! d\alpha_{q_2}
\,\frac{1}{1-\alpha_{q_1}-\alpha_{q_2}}\left(\alpha_{q_1}\,\frac{d}{d\alpha_{q_1}}
+ \alpha_{q_2}\,\frac{d}{d\alpha_{q_2}}\right)
{\cal A}(\underline{\alpha})\nonumber\\
&&
 {} +
 2\zeta_{3,A}^V\,\frac{d}{du}\,\int\limits_0^u\!\!d\alpha_{q_1}
\int\limits_0^{\bar u}\!\! d\alpha_{q_2}
\,\frac{1}{1-\alpha_{q_1}-\alpha_{q_2}}\left(\alpha_{q_1}\,\frac{d}{d\alpha_{q_1}}
- \alpha_{q_2}\,\frac{d}{d\alpha_{q_2}}\right) {\cal V}(\underline{\alpha}). \label{eq:Psi}
\end{eqnarray}
Neglecting the three-parton distribution amplitudes containing
gluons and terms proportional to light quark masses,
$g_\perp^{(a)}$ and $g_\perp^{(v)}$ are thus related to the
twist-2 ones by Wandzura-Wilczek--type relations:
\begin{eqnarray}
 g_\perp^{(a)WW}(u)&\simeq&{1\over 2}\left[ \int_0^udv\,
 {1\over \bar{v}}\Phi_\parallel(v) + \int_u^1dv\,{1\over v}\Phi_\parallel(v) \right],
\label{eq:even-WW1}\\
 g_\perp^{(v)WW}(u)&\simeq&2\bar{u} \int_0^udv\,
 {1\over \bar{v}}\Phi_\parallel(v) + 2u\int_u^1dv\,{1\over v}\Phi_\parallel(v).
 \label{eq:even-WW2}
 \end{eqnarray}

\subsection{Relations among chiral-odd LCDAs}\label{subsec:da-odd}

The leading-twist LCDAs $\Phi_\perp^A(u,\mu)$ can be expanded
as~\cite{BF2,Braun:2003rp}
\begin{eqnarray}\label{eq:generaldaperp}
\Phi_\perp^A(u,\mu)=6u(1-u) \Bigg[ a_0^\perp +
   \sum_{l=1}^\infty a_l^{\perp,A}(\mu) C^{3/2}_l(2u-1) \Bigg],
\end{eqnarray}
where the multiplicatively renormalizable Gegenbauer moments, in
analogy to Eq.~(\ref{eq:aparallel}), read
\begin{eqnarray}\label{eq:aperp}
a_l^{\perp, A}(\mu) = \frac{2(2l+3)}{3(l+1)(l+2)} \int_0^1 dx\,
C^{3/2}_l (2x-1) \Phi_\perp^A(x,\mu)\,,
\end{eqnarray}
which satisfy
\begin{eqnarray}
  \bigg(f^\perp_A a_l^{\perp, A}\bigg)(\mu)
  &=& \bigg(f^\perp_A a_l^{\perp, A}\bigg)(\mu_0)
  \left(\frac{\alpha_s(\mu_0)}{\alpha_s(\mu)}\right)^{-\gamma_{(l)}^\perp/b}\,,
  \label{eq:RGperp}
   \end{eqnarray}
with the one-loop anomalous dimensions being~\cite{GW}
\begin{eqnarray}
  \gamma_{(l)}^\perp  = C_F
  \left(1+4 \sum_{j=2}^{l+1} \frac{1}{j}\right).
  \label{eq:gamma_perp}
  \end{eqnarray}
In the limit of $m_{q_1}=m_{q_2}$, $a_l^{\perp, A}$ with even (odd) $l$
vanish due to G-parity invariance for the $1^3P_1$ ($1^1P_1$)
mesons.

In analogy to Eqs.~(\ref{eq:eom-even1}) and (\ref{eq:eom-even1}),
using the formulas developed in Ref.~\cite{Ball:1998sk} for vector
mesons to the present case, we obtain the integral equations:
\begin{eqnarray}
\lefteqn{i px \int_{0}^{1}\!du\, e^{-i \xi px} \xi\, h_\parallel^{(t)}(u)
 - 2 \int_{0}^{1}\!du\, e^{-i \xi px}
 \left(h_\parallel^{(t)} (u) - \Phi_{\perp}(u) \right)}
 \nonumber \\
&& ~~~~ = \zeta_{3,A}^{\perp}(px)^{2} \!\int_{-1}^{1}\!dv\, v {\cal T}(v, -px)
 + (px)^{2} \!\int_{0}^{1}
 e^{-i\xi px} h_\parallel^{(p)}(u)
 + i \delta_{+} px \!\int_{0}^{1}\!du\, e^{-i \xi px}
\Phi_{\parallel}(u), \makebox[0.7cm]{\ } \label{eq:eom-odd1}
\end{eqnarray}
\begin{eqnarray}
\int_{0}^{1}\! du\, e^{-i \xi px} h_\parallel^{(p)} (u) &=& - i
\zeta_{3,A}^{\perp} px\! \int_{0}^{1}tdt \int_{-1}^{1}dv\, {\cal
T}(v, -tpx) + \int_{0}^{1}\!dt \int_{0}^{1}\!du \,e^{-i \xi t px}
h_\parallel^{(t)}  (u)
\nonumber \\
& & +\delta_{-} \int_{0}^{1}\!dt \int_{0}^{1}\!du\, e^{-i \xi tpx}
\Phi_{\parallel}(u), \label{eq:eom-odd2}
\end{eqnarray}
where we have introduced the notations
\begin{equation}
 \delta_{\pm} = \frac{f_{A}}{f_{A}^{\perp}}
 \frac{m_{q_2} \pm m_{q_1}}{m_{A}}, \qquad \zeta_{3,A}^{\perp}
 = \frac{f_{3,A}^{\perp}}{f_{A}^{\perp}m_A}\,. \label{eq:3pm}
\end{equation}

The solutions of the integral equations \cite{Ball:1998sk} given in
Eqs.~(\ref{eq:eom-odd1}) and (\ref{eq:eom-odd2}) can be obtained to
be
\begin{equation}
h_\parallel^{(p)}(u)  =  \bar u \int\limits_0^u\!\!
dv\, \frac{1}{\bar v}\,{\Theta}(v) + u \int\limits_u^1\!\! dv\,
\frac{1}{v}\,\Theta(v) \,,\label{eq:odd-sol1}
\end{equation}
and
\begin{eqnarray}
h_\parallel^{(t)}(u) & = & \frac{1}{2}\,\xi
\left(\int\limits_0^u\!\! dv\frac{1}{\bar v}\,\Theta(v) -
\int\limits_u^1\!\! dv\frac{1}{v}\,\Theta(v)\right) -
\delta_- \Phi_\parallel(u)
\nonumber\\
& & {} + \zeta_{3,A}^{\perp}\frac{d}{du}\,\int\limits_0^u\!\!d\alpha_{q_1}
\int\limits_0^{\bar u}\!\! d\alpha_{q_2}
\,\frac{1}{1-\alpha_{q_2}-\alpha_{q_1}}\,{\cal T}(\underline{\alpha}),
\label{eq:odd-sol2}
\end{eqnarray}
where
\begin{eqnarray}
\Theta(u) & = & 2\Phi_\perp(u) + \delta_- \left( \Phi_\parallel(u)
- \frac{1}{2}\,\xi\Phi'_\parallel(u)\right) -
\frac{1}{2}\,\delta_+ \Phi_\parallel'(u)\nonumber\\
& & {}+\zeta_{3,A}^{\perp}
\frac{d}{du}\,\int\limits_0^u\!\!d\alpha_{q_1}\int\limits_0^{\bar
u}\!\! d\alpha_{q_2}
\,\frac{1}{1-\alpha_{q_2}-\alpha_{q_1}}\left(\alpha_{q_1}\,\frac{d}{d\alpha_{q_1}}
+ \alpha_{q_2}\,\frac{d}{d\alpha_{q_2}}\, - 1\right) {\cal
T}(\underline{\alpha}). \label{eq:Phi}
\end{eqnarray}
The two-parton twist-3 distribution amplitudes are thus related to
the twist-2 ones approximately by Wandzura-Wilczek--type relations
\begin{eqnarray}
h_\parallel^{(t)WW}(u) &=& \xi \left( \int_{0}^{u} dv
\frac{\Phi_{\perp}(v)}{\bar v} - \int_{u}^{1} dv
\frac{\Phi_{\perp}(v)}{v} \right),
\label{eq:3hww} \\
h_\parallel^{(p)WW}(u) &=& 2 \left( \bar u \int_{0}^{u} dv
\frac{\Phi_{\perp}(v)}{\bar v} + u \int_{u}^{1} dv
\frac{\Phi_{\perp}(v)}{v} \right). \label{eq:3eww}
\end{eqnarray}

\subsection{Comparison of the vector mesons and axial-vector mesons}

In comparison with the vector mesons, we summarize in
Table~\ref{tab:symmetric} the (anti-)symmetric properties of LCDAs
of axial-vector mesons in the SU(3) limit under the interchange of
the momentum fractions of the quark and antiquark. On the other
hand, comparing results for axial-vector mesons given in
Eqs.~(\ref{eq:evendef1})-(\ref{eq:odddef2}), (\ref{eq:eom-even1}),
(\ref{eq:eom-even2}), (\ref{eq:eom-odd1}), (\ref{eq:eom-odd2}), and
(\ref{app:vector-id})-(\ref{app:pseudo-id}), and for vector mesons
given in Ref.~\cite{Ball:1998sk}, we can see an analogy between the
two cases. One can relate the mathematical forms of two-parton LCDAs
of twist-3 for vector mesons and axial-vector mesons in the following way:\\
\begin{centerline}
{
$\begin{array}{ccc}
{\rm vector~ mesons}: & & {\rm axial\hbox{-}vector~ mesons}:\\
\begin{array}{c} g_\perp^{(v)}\\ (1-\widetilde\delta_+)g_\perp^{(a)}
                  \\ h_\parallel^{(t)} \\
                 (1-\delta_+)h_\parallel^{(s)}\end{array} &
\begin{array}{c} \longleftrightarrow \\ \longleftrightarrow\\
                 \longleftrightarrow \\  \longleftrightarrow
\end{array} &
\begin{array}{cl} g_\perp^{(a)}
\\ g_\perp^{(v)} \\ h_\parallel^{(t)} \\ h_\parallel^{(p)}
\end{array}
\end{array}$
}
\end{centerline}
with the replacement\\
\begin{centerline}
{
$\begin{array}{ccc}
{\rm ~~~ ~~~~~~~~~~~~~~}~~ & & {\rm ~~~~~~~ ~~~~~~~ ~~~~~~~~~~~}~\\
\begin{array}{c} +\widetilde\delta_\pm\\ +\delta_\pm \\ {\cal A} \\
                 {\cal V} \\ {\cal T} \\ \Phi_{\parallel, \perp}\end{array} &
\begin{array}{c} \longleftrightarrow \\ \longleftrightarrow\\
                 \longleftrightarrow \\  \longleftrightarrow\\
                 \longleftrightarrow \\  \longleftrightarrow
\end{array} &
\begin{array}{c} -\widetilde\delta_\mp \\ -\delta_\mp \\ {\cal V}\\
                 {\cal A} \\{\cal T} \\ \Phi_{\parallel, \perp} \end{array}
\end{array}$
}
\end{centerline}
where the notations for LCDAs of vector mesons follow from
Refs.~\cite{Ball:1998sk,Ball:1998ff}. For vector mesons, additional
factors $(1-\widetilde\delta_+)$ and $(1-\delta_+)$ are considered
\cite{Ball:1998sk} in the definitions of $g_\perp^{(a)}$ and
$h_\parallel^{(s)}$, respectively, due to the normalizations
$\int_0^1 g_\perp^{(a)}(u) du = \int_0^1 h_\parallel^{(s)} du= 1$.
Nevertheless, we do not need to put such factors in the definitions
of axial-vector mesons; see results shown in
Eqs.~(\ref{eq:nomalization-3P1-1}), (\ref{eq:nomalization-3P1-2}),
(\ref{eq:nomalization-3P1-3}), (\ref{eq:nomalization-1P1-1}),
(\ref{eq:nomalization-1P1-2}), and (\ref{eq:nomalization-1P1-3}).
%
\begin{table}[thb!]
\caption[]{\label{tab:symmetric}The symmetric/antisymmetric
properties of LCDAs under the interchange of the momentum fractions
of the quark and antiquark in vector mesons and axial-vector mesons
in the SU(3) limit, where the definitions of LCDAs of vector mesons
follow from Refs.~\cite{Ball:1998sk,Ball:1998ff}. }
\renewcommand{\arraystretch}{1.5}
\addtolength{\arraycolsep}{4pt}
$$
\begin{array}{|c||c|c|c|}\hline
 {\rm LCDA} & {\rm Vector\ meson}& \hbox{\rm Axial-vector\ meson}~({1^3P_1})
 & \hbox{\rm Axial-vector\ meson}~({1^1P_1})
 \\ \hline
\begin{array}{c}\Phi_\parallel \\ \Phi_\perp \\ g_\perp^{(a)} \\ g_\perp^{(v)} \\
                h_\parallel^{(t)} \\ h_\parallel^{(s)}\, {\rm or}\, h_\parallel^{(p)}\\
                {\cal A} \\ {\cal V} \\ {\cal T} \end{array}&
\begin{array}{c} {\rm symmetric}\\ {\rm symmetric}\\  {\rm symmetric}\\
                  {\rm symmetric}\\ {\rm symmetric}\\ {\rm symmetric}\\
                 {\rm symmetric}\\ {\rm antisymmetric}\\ {\rm antisymmetric} \end{array}&
\begin{array}{c} {\rm symmetric}\\ {\rm antisymmetric}\\  {\rm symmetric}\\
                  {\rm symmetric}\\ {\rm antisymmetric}\\ {\rm antisymmetric}\\
                 {\rm antisymmetric}\\ {\rm symmetric}\\ {\rm symmetric} \end{array}&
\begin{array}{c} {\rm antisymmetric}\\ {\rm symmetric}\\  {\rm antisymmetric}\\
                  {\rm antisymmetric}\\ {\rm symmetric}\\ {\rm symmetric}\\
                 {\rm symmetric}\\ {\rm antisymmetric}\\ {\rm antisymmetric} \end{array}
\\ \hline
\end{array}
$$
\end{table}

\section{Determinations of leading-twist LCDAs}\label{sec:properties-ltdas}

\subsection{Physical properties for axial-vector mesons}\label{subsec:properties}

In the quark model, the light $1^1P_1$ states, referred to as
$b_1(1235)$, $h_1(1170)$, $h_1(1380)$, and $K_{1B}$, form the
$1^{+-}$ nonets, whereas the light $1^3P_1$ mesons, denoted as
$a_1(1260)$, $f_1(1285)$, $f_1(1420)$, and $K_{1A}$, form the
$1^{++}$ nonets. $h_1(1380)$ is not experimentally
well-established~\cite{PDG} and its quark content was suggested as
$\bar s s$ in the QCD sum rule calculation~\cite{GRVW}. It should be
noted that the real physical states $K_1(1270)$ and $K_1(1400)$ are
the mixtures of $1^3P_1$ ($K_{1A}$) and $1^1P_1$ ($K_{1B}$) states;
following the convention in Ref.~\cite{Suzuki:1993yc}, the relations
can be written as
 \begin{eqnarray}
 \label{eq:mixing}
 |K_1(1270)\rangle &=& |K_{1A}\rangle\sin\theta_K+ |K_{1B}\rangle\cos\theta_K, \nonumber \\
 |K_1(1400)\rangle &=& |K_{1A}\rangle\cos\theta_K -
 |K_{1B}\rangle\sin\theta_K.
 \end{eqnarray}
In Ref.~\cite{Suzuki:1993yc}, two possible solutions with two-fold
ambiguity  $|\theta_K|\approx 33^\circ$ and $57^\circ$ were
obtained. A similar constraint $35^\circ\lesssim |\theta_K| \lesssim
55^\circ$ was found in Ref. \cite{Burakovsky:1997ci}.  Therefore,
the favor values may lie in the range $|\theta_K| \simeq (45\pm
12)^\circ$. Just for simplicity, we will take $\theta_K =45^\circ,
-45^\circ$ as the reference points. Analogous to $\eta$ and
$\eta^\prime$, for $1^3P_1$ states, $f_1(1285)$ and $f_1(1420)$ are
mixed in terms of the pure octet $f_8$ and singlet $f_1$ due to
SU(3) breaking effects, and can be parameterized as
 \begin{eqnarray}
 |f_1(1285))\rangle = |f_1\rangle\cos\theta_{^3P_1}+|f_8\rangle\sin\theta_{^3P_1},
 \quad |f_1(1420)\rangle =
 -|f_1\rangle\sin\theta_{^3P_1} +|f_8\rangle\cos\theta_{^3P_1} \,.
 \end{eqnarray}
From the Gell-Mann-Okubo mass formula \cite{CloseBook,PDG}, it
follows that
\begin{equation}
\cos^2\theta_{^3P_1} =
 \frac{3m_{f_1(1285)}^2 -\bigg(4m_{K_{1A}}^2-m_{a_1}^2\bigg)}
{3\bigg(m_{f_1(1285)}^2-m_{f_1(1420)}^2\bigg)}\,,\label{eq:GMOkubo}
\end{equation}
where
 \begin{equation}
 m_{K_{1A}}^2 = m_{K_1(1400)}^2 \cos^2\theta_K + m_{K_1(1270)}^2
 \sin^2\theta_K \,.
 \end{equation}
Substituting into Eq.~(\ref{eq:GMOkubo}) with $\theta_K =(45\pm
12)^\circ$, we then obtain\footnote{\label{footnote:GMO1} Since
 \begin{equation}
\tan\theta_{^3P_1} =\frac{4m_{K_{1A}}^2-m_{a_1}^2-3m_{f_1(1420)}^2}
{3 m_{18}^2}>0\,,\label{eq:GMOkubo2}
\end{equation}
where $m_{18}^2=\langle f_1|{\cal H} |f_8 \rangle <0$ with ${\cal
H}$ being the Hamiltonian, we find that $\theta_{^3P_1} >0$. In the
present paper, we can extend the study in the traditional quark
model to a field-theoretical consideration that each Fock state of
$f_1$ [or $f_8$] is proportional to $(\bar{q}q)^m g^n(\bar u u +
\bar dd + \bar ss)/\sqrt{3}$ [or $(\bar{q}q)^m g^n (\bar uu + \bar
dd -2 \bar ss)/\sqrt{6}$], where there is a relative sign difference
between the $\bar{s}s$ contents of $f_1$ and $f_8$ in our
convention.} $\theta_{^3P_1}= 38^\circ {}^{+14^\circ}_{-16^\circ}$
which is consistent with the value of replacing $m^2$ by $m$
throughout Eq.~(\ref{eq:GMOkubo}). The previous phenomenological
analyses suggested $\theta_{^3P_1} \simeq 50^\circ$
\cite{Close:1997nm}. In the present paper, we will take
$\theta_{^3P_1}=38^\circ$ or $50^\circ$ as the reference input for
LCDA studies.

Similarly, for $1^1P_1$ states, $h_1(1170)$ and $h_1(1380)$ may be
mixed in terms of the pure octet $h_8$ ad singlet $h_1$,
 \begin{eqnarray}\label{eq:GMOkubo3}
 |h_1(1170))\rangle = |h_1\rangle\cos\theta_{^1P_1}+|h_8\rangle\sin\theta_{^1P_1},
 \quad
 |h_1(1380)\rangle = -|h_1\rangle\sin\theta_{^1P_1} +|h_8\rangle\cos\theta_{^1P_1} \,.
 \end{eqnarray}
Again from the Gell-Mann-Okubo mass formula, we obtain
\begin{equation}
\cos^2\theta_{^1P_1} =
 \frac{ 3m_{h_1(1170)}^2 - \bigg(4m_{K_{1B}}^2-m_{b_1}^2\bigg)}
{3\bigg(m_{h_1(1170)}^2-m_{h_1(1380)}^2\bigg)}\,,\label{eq:GMOkubo4}
\end{equation}
where
 \begin{equation}
 m_{K_{1B}}^2 =
 m_{K_1(1400)}^2 \sin^2\theta_K + m_{K_1(1270)}^2 \cos^2\theta_K \,.\label{eq:GMOkubo5}
 \end{equation}
We thus get $\theta_{^1P_1}\simeq 10^\circ
{}^{+15^\circ}_{-10^\circ}$ which coincides with the value of
replacing $m^2$ by $m$ throughout Eq.~(\ref{eq:GMOkubo4}). Note that
$|\theta_K| \geq 50^\circ$ is disfavored because of the constraint
$0\leq \cos^2\theta_{^1P_1}\leq 1$. Note also that the QCD sum
calculation suggested $\theta_{^1P_1}\simeq 45^\circ$, from which
the predictive content of $h_1(1380)$ is predominated by the $\bar s
s$ pair\footnote{\label{footnote:GMO2} With the same reason as
footnote~\ref{footnote:GMO1}, $\theta_{^1P_1}$ should be positive in
sign.}~\cite{GRVW}.  By comparing results with
$\theta_{^1P_1}=10^\circ$ and $45^\circ$ in the phenomenological
LCDA analysis, we will show the effects induced by different mixing
angles.

\subsection{Axial-vector couplings of $1^3P_1$
mesons}\label{subsec:decay-constant-1}

In this subsection, we
calculate renormalization-group (RG) improved QCD sum rules for the
axial-vector couplings of $1^3P_1$ mesons. Because the QCD sum rule
approach may not sufficiently determine the singlet-octet mixing
angle, we thus calculate the QCD sum rules for pure $1^3P_1$ states.

\subsubsection{QCD sum rules for axial-vector couplings}

To evaluate the axial-vector couplings $f_{^3P_1}$ for $1^3P_1$
mesons, we consider the two-point correlation function,
\begin{eqnarray}\label{eq:green-axial}
\Pi_{\mu\nu} (q^2)=i\int d^4x e^{iqx} \langle 0|{\rm T} (j_\mu(x)
j_\nu^{\dag} (0)|0\rangle =-\Pi_1(q^2) g_{\mu \nu} +\Pi_2(q^2)
q_\mu q_\nu \,,
\end{eqnarray}
where the interpolating current $j_\mu=\bar q_2\gamma_\mu\gamma_5
q_1$ satisfies
 \begin{eqnarray}
    \langle 0 |j_\mu(0) |1^3P_1(P,\lambda)\rangle
 =-if_{^3P_1} m_{^3P_1} \epsilon_\mu^{(\lambda)}.
  \label{eq:axial-decayconstant}
 \end{eqnarray}
In the massless quark limit, we have $\Pi_1=q^2\Pi_2$ due to
conservation of $j_\mu$. Here we focus on $\Pi_1$ since $\Pi_1$
receives contributions only from axial-vector ($^3P_1$) mesons,
whereas $\Pi_2$ contains effects from pseudoscalar mesons. The above
correlation function can be calculated from the hadron and
quark-gluon dynamical points of view, respectively. The lowest-lying
$1^3P_1$ meson contribution can be approximated via the dispersion
relation as
\begin{eqnarray}
\frac{m_{^3P_1}^2 f_{^3P_1}^2}{m_{^3P_1}^2-q^2}
 = \frac{1}{\pi}\int^{s_0^{^3P_1}}_0 ds
\frac{{\rm Im} \Pi_1^{\rm OPE}(s)}{s-q^2} \,, \label{eq:dispersion}
\end{eqnarray}
where $\Pi_1^{\rm OPE}$, the QCD operator-product-expansion (OPE)
result of $\Pi_1$ at the quark-gluon level, given in
Ref.~\cite{GRVW} up to dimension 6 and with ${\cal O}(\alpha_s)$
corrections, reads

\begin{eqnarray}
 \Pi_1^{\rm OPE}(q^2) &=&
 -\frac{1}{4\pi^2}q^2\ln\,\frac{-q^2}{\mu^2}
  \Bigg(1 +\frac{\alpha_s}{\pi}\Bigg)
  +\frac{1}{q^2} \Bigg(\frac{1}{12}\langle \frac{\alpha_s}{\pi}\,G^2\rangle
  -  m_2 \langle \bar q_1 q_1\rangle
  -  m_1 \langle \bar q_2 q_2\rangle\Bigg)\nonumber\\
  &&
  +  \frac{1}{q^4}\Bigg(\frac{2}{9}\pi\alpha_s
  \Big\langle (\bar q_1 \gamma_\mu \lambda^a q_1\,
  +\bar q_2 \gamma_\mu \lambda^a q_2)
  \sum_q \bar q \gamma^\mu \lambda^a q \Big\rangle
  + 2 \pi \alpha_s
  \langle \bar q_1 \lambda^a q_2\, \bar q_2 \lambda^a
  q_1\rangle\Bigg),\hspace{0.8cm}
  \label{eq:pipm}
\end{eqnarray}
and $s_0^{^3P_1}$ is the threshold of the higher resonant states,
such that the contributions of higher resonances are modeled  by
\begin{eqnarray}\label{eq:higher-res}
\frac{1}{\pi}\int_{s_0^{^3P_1}}^\infty ds \frac{{\rm Im} \Pi_1^{\rm
OPE}(s)}{s-q^2}\,.
\end{eqnarray}
We further apply the Borel (inverse-Laplace) transformation to both
sides of Eq.~(\ref{eq:dispersion})
 \begin{eqnarray}\label{eq:Borel}
 {\rm\bf B}[f(q^2)]=
\lim_{{\scriptstyle n\to \infty \atop\scriptstyle -q^2\to \infty}
\atop \scriptstyle -q^2/n^2 =M^2 {\rm fixed} } (-q^2)^{n+1}
\Bigg[{d\over dq^2}\Bigg]^n f(q^2),
 \end{eqnarray}
to improve the convergence of the OPE series and further suppress
the contributions from higher resonances. Moreover, we adopt the
vacuum saturation approximation for describing the four-quark
condensates in the present work, i.e.,
\begin{eqnarray}\label{eq:factorization}
\langle 0|\bar q \Gamma_i \lambda^a q \bar q \Gamma_i \lambda^a
q|0\rangle =-\frac{1}{16N_c^2}{\rm Tr}(\Gamma_i\Gamma_i) {\rm
Tr}(\lambda^a \lambda^a) \langle \bar qq\rangle^2 \,,
\end{eqnarray}
and neglect the possible corrections due to their anomalous
dimensions. Finally, we arrive at the $f_{^3P_1}$ sum rules, given
by
\begin{eqnarray}
  f_{^3P_1}^2 m_{^3P_1}^2e^{-m_{^3P_1}^2/M^2}
  &=&
  \frac{1}{4\pi^2}\int\limits_0^{s_0^{^3P_1}}\!\! s\, ds\,e^{-s/M^2}
  \left(1 + \frac{\alpha_s}{\pi}\right)
  -\frac{1}{12}\,\langle\frac{\alpha_s}{\pi}G^2\rangle
  + m_{q_2} \langle \bar q_1 q_1\rangle
  + m_{q_1} \langle \bar q_2 q_2\rangle \nonumber\\
  &&
  -\frac{1}{M^2} \Bigg[ \frac{32\pi\alpha_s}{81}\,
  (\langle\bar q_2 q_2\rangle^2 + \langle\bar q_1 q_1\rangle^2)
  + \frac{32\pi\alpha_s}{9}\,\langle\bar q_2 q_2\rangle
  \langle\bar q_1 q_1\rangle \Bigg]\,.\makebox[0.8cm]{}
  \label{eq:SR-3p1-decay-constant}
  \end{eqnarray}

\subsubsection{Results for $1^3P_1$ mesons}

\begin{figure}[thb!]
 \centerline{\epsfxsize=14.5cm \epsffile{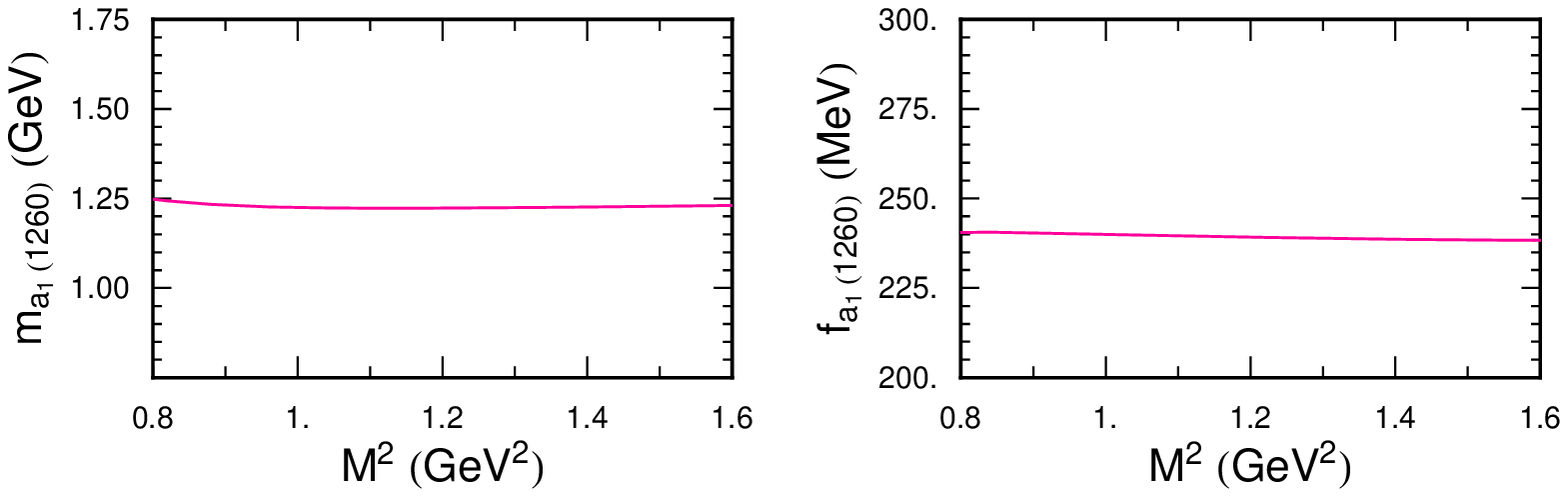}}
 \vskip0.4cm
 \centerline{\epsfxsize=14.5cm \epsffile{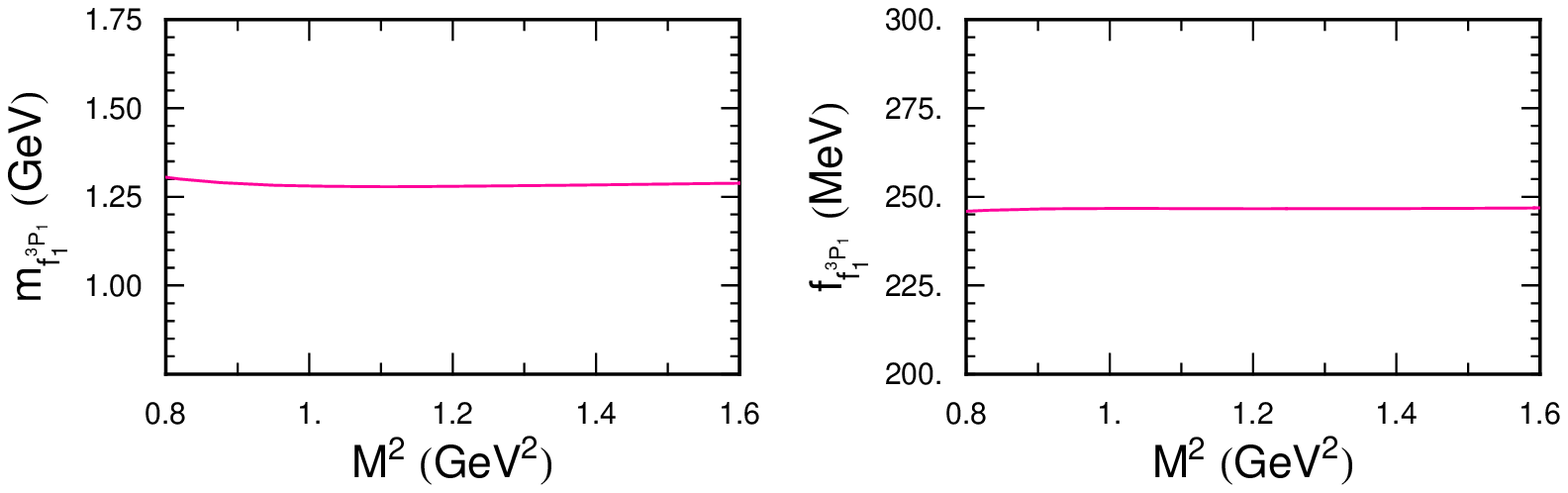}}
 \vskip0.4cm
 \centerline{\epsfxsize=14.5cm \epsffile{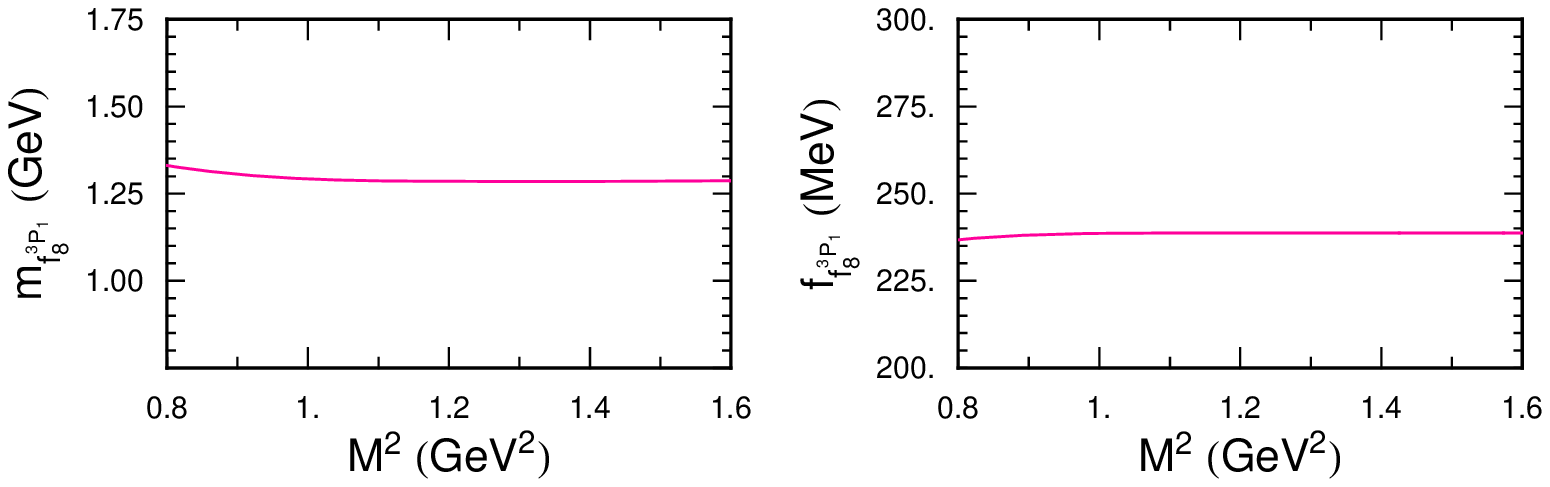}}
 \vskip0.4cm
 \centerline{\epsfxsize=14.5cm \epsffile{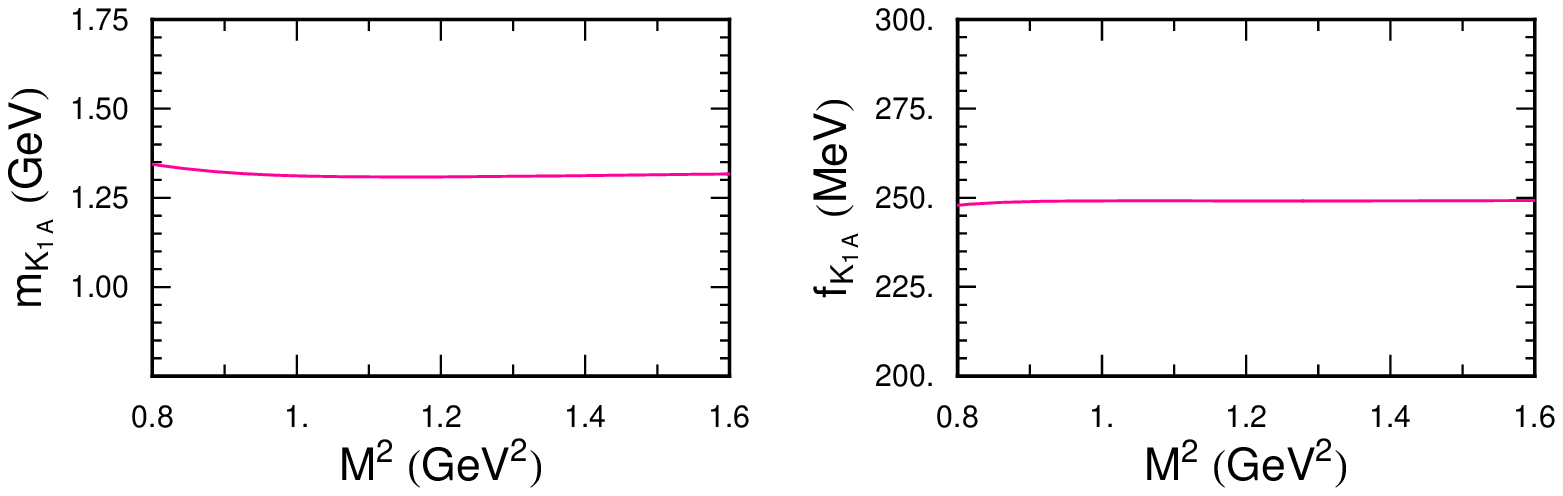}}
 \vskip0.4cm
 \centerline{\parbox{14cm}{\caption{\label{fig:fA-mass}
masses and decay constants of the $1^3P_1$ states as functions of
the Borel mass squared, where the central values of input parameters
given in Appendix \ref{appsec:inputs} have been used.}}}
\end{figure}

We start with the analysis of the $f_{^3P_1}$ sum rules. To examine
the quality of the sum rules, we also give the mass sum rule
results. The mass sum rule for the $1^3P_1$ lowest-lying resonance
can be obtained by taking the logarithm of both sides of
Eq.~(\ref{eq:SR-3p1-decay-constant}) and then applying the
differential operator $M^4 \partial /\partial M^2$ to them, where
$s_0^{^3P_1}$ is determined by the maximum stability of the sum
rules. Substituting the obtained $s_0^{^3P_1}$ and masses into
Eq.~(\ref{eq:SR-3p1-decay-constant}), one arrives at the sum rules
for the decay constants $f_{^3P_1}$. In the numerical analysis, we
use the parameters which are given in Appendix~\ref{appsec:inputs}
and choose the Borel window $0.8$~GeV$^2< M^2 < 1.3$~GeV$^2$, where
the contribution originating from higher resonances (and the
continuum) is less than 44\% and the highest OPE term at the quark
level is no more than 14\%. Note that in the sum rules, the
contributions from higher resonances are modeled by
\begin{eqnarray}\label{eq:model-higher-res}
\frac{1}{\pi}\int_{s_0}^\infty ds\, e^{-s/M^2}\, {\rm Im} \Pi_1^{\rm
OPE}(s) \,.
\end{eqnarray}
In Fig.~\ref{fig:fA-mass}, the masses and decay constants are
plotted as functions of the Borel mass squared $M^2$. The results
are summarized in Table \ref{tab:mass-decay-constant-3p1}.
Introducing the decay constants $f_{f_1(1285)}^q$ and
$f_{f_1(1420)}^q$ by
\begin{eqnarray}
\langle 0| \bar q\gamma_\mu \gamma_5 q | f_1(1285)(P,\lambda)\rangle
 &=& -i m_{f_1(1285)} f_{f_1(1285)}^q \epsilon_\mu^{(\lambda)}\,,\label{eq:decay-def1}\\
\langle 0| \bar q\gamma_\mu \gamma_5 q | f_1(1420)(P,\lambda)\rangle
 &=& -i m_{f_1(1420)} f_{f_1(1420)}^q \epsilon_\mu^{(\lambda)}\,, \label{eq:decay-def2}
\end{eqnarray}
we get
\begin{eqnarray}
f_{f_1(1285)}^u &=& \frac{f_{f_1}}{\sqrt{3}}\frac{m_{f_1}}{m_{f_1(1285)}}\cos\theta_{^3P_1}
 +  \frac{f_{f_8}}{\sqrt{6}}\frac{m_{f_8}}{m_{f_1(1285)}}\sin\theta_{^3P_1}
 = 173\pm 23 ~(167\pm 22)~{\rm MeV} \,,~~~~~ \label{eq:decay-f1-1}\\
f_{f_1(1285)}^s &=&
\frac{f_{f_1}}{\sqrt{3}}\frac{m_{f_1}}{m_{f_1(1285)}}\cos\theta_{^3P_1}
 -  \frac{2 f_{f_8}}{\sqrt{6}}\frac{m_{f_8}}{m_{f_1(1285)}}\sin\theta_{^3P_1}
 = -9\pm 13~(-59\pm 18)~{\rm MeV} \,,  \label{eq:decay-f1-2}\\
f_{f_1(1420)}^u &=&
-\frac{f_{f_1}}{\sqrt{3}}\frac{m_{f_1}}{m_{f_1(1420)}}\sin\theta_{^3P_1}
 +  \frac{f_{f_8}}{\sqrt{6}}\frac{m_{f_8}}{m_{f_1(1420)}}\cos\theta_{^3P_1}
 = -9 \pm 10~(-41\pm 11)~{\rm MeV} \,,  \label{eq:decay-f1-3}\\
f_{f_1(1420)}^s &=&
-\frac{f_{f_1}}{\sqrt{3}}\frac{m_{f_1}}{m_{f_1(1420)}}\sin\theta_{^3P_1}
 -  \frac{2 f_{f_8}}{\sqrt{6}}\frac{m_{f_8}}{m_{f_1(1420)}}\cos\theta_{^3P_1}  \nonumber\\
 &=& -217\pm 27~(-211\pm 26)~{\rm MeV}\,, \label{eq:decay-f1-4}
\end{eqnarray}
corresponding to $\theta_{^3P_1}=38^\circ (50^\circ)$. In
particular, $f_1(1285)$ and $f_1(1420)$ are predominated by their
$\bar uu$ and $\bar ss$ contents, respectively.

Using Eq.~(\ref{eq:mixing}), the decay constants for $K_{1(1270)}$
and $K_{1(1400)}$ read (with $\bar q=\bar u$ or $\bar d$)
 \begin{eqnarray}\label{eq:k1-1}
 \langle 0 |\bar q\gamma_\mu \gamma_5 s |K_{1}(1270)(P,\lambda)\rangle
 &=& -i \, f_{K_{1}(1270)}\, m_{K_{1}(1270)}\,\epsilon_\mu^{(\lambda)}
  \nonumber\\
 &=& -i (f_{K_{1A}} m_{K_{1A}} \sin{\theta_K}
    + f_{K_{1B}} m_{K_{1B}} a_0^{\parallel, K_{1B}}
   \cos{\theta_K})\,\epsilon_\mu^{(\lambda)},~~~
 \end{eqnarray}
and
 \begin{eqnarray}\label{eq:k1-2}
\langle 0 |\bar q\gamma_\mu \gamma_5 s|K_{1}(1400)(P,\lambda)\rangle
  &=& -i \, f_{K_{1}(1400)}\, m_{K_{1}(1400)}\,\epsilon_\mu^{(\lambda)}
  \nonumber\\
  &=& - i (f_{K_{1A}} m_{K_{1A}} \cos {\theta_K}
     - f_{K_{1B}} m_{K_{1B}} a_0^{\parallel, K_{1B}}
   \sin {\theta_K})\, \epsilon_\mu^{(\lambda)},~~~
 \end{eqnarray}
where $f_{K_{1A}}$ and $m_{K_{1A}}$ are given in Table
\ref{tab:mass-decay-constant-3p1}, and $a_0^{\parallel, K_{1B}}$
(with $f_{K_{1B}}=f_{K_{1B}}^\perp(1~{\rm GeV})$ by definition in
this paper) and $m_{K_{1B}}$ can be found in Tables
\ref{tab:mass-decay-constant-1p1} and \ref{tab:Gegenbauer},
respectively. (See the detailed discussion in
Sec.~\ref{subsec:tensor_1p1}.) Numerically, we thus obtain
\begin{eqnarray}
f_{K_{1}(1270)} &=& 197 \pm 15~ {\rm MeV} \,,\\
f_{K_{1}(1400)} &=& 151 \pm 12~ {\rm MeV} \,,
 \qquad {\rm if\ taking\ }\theta_K=45^\circ,
\end{eqnarray}
and
\begin{eqnarray}
f_{K_{1}(1270)} &=& -166  \pm 11~ {\rm MeV} \,,\\
f_{K_{1}(1400)} &=&~~\,  179 \pm 12~ {\rm MeV} \,,
 \qquad {\rm if\ taking\ }\theta_K=-45^\circ\,,
\end{eqnarray}
where the correlation of the errors between the masses and decay
constants is considered. It is interesting to note that, if one sets
$a_0^{\parallel, K_{1B}}=0$, then the central value of
$f_{K_{1}(1270)}$ becomes $\pm 182$ GeV corresponding to
$\theta_K=\pm 45^\circ$, whereas that of $f_{K_{1}(1400)}$ is 165
GeV independent of the sign of $\theta_K$; the local axial-vector
current can couple only to $K_{1A}$, but not to $K_{1B}$ in the
SU(3) limit.
%
\begin{table}[t!]
\caption[]{The sum rule results of masses, decay constants,
and corresponding excited thresholds $s_0^{^3P_1}$ for $1^3P_1$ mesons.}
\label{tab:mass-decay-constant-3p1}
\renewcommand{\arraystretch}{1.5}
\addtolength{\arraycolsep}{4pt}
$$
\begin{array}{|c|ccc|}\hline
 {\rm State} & {\rm Mass} [{\rm GeV}]& {\rm Decay\ cosntant}~f_{^3P_1} [{\rm MeV}]
 & s_0^{^3P_1} [{\rm GeV}^2]
 \\ \hline
\begin{array}{c} a_1(1260)   \\ f_1 (1^3P_1) \\ f_8 (1^3P_1) \\ K_{1A}\end{array}&
\begin{array}{c} 1.23\pm 0.06\\ 1.28\pm 0.06 \\ 1.29\pm 0.05 \\ 1.31\pm 0.06\end{array}&
\begin{array}{c} 238\pm 10   \\ 245\pm 13    \\ 239\pm 13    \\ 250\pm 13\end{array}&
\begin{array}{c} 2.55\pm 0.15\\ 2.80\pm 0.20 \\ 2.70\pm 0.20 \\ 2.90\pm 0.20\end{array}
\\ \hline
\end{array}
$$
\end{table}
%
%

\subsection{Tensor couplings of $1^1P_1$
mesons}\label{subsec:tensor_1p1}

The tensor couplings of $1^1P_1$ mesons are defined as
  \begin{equation}
  \langle 0 |\bar q_2 \sigma_{\mu\nu}q_1
  |1^1P_1(P,\lambda)\rangle  = i f_{^1P_1}^\perp\,\epsilon_{\mu\nu\alpha\beta}
  \epsilon_{(\lambda)}^\alpha P^\beta\,,
  \label{eq:tensor-1p1-1}
  \end{equation}
i.e.,
\begin{equation}
   \langle 0 |\bar q_2 \sigma^{\mu\nu}\gamma_5q_1 |1^1P_1(P,\lambda)\rangle
 =-f_{^1P_1}^\perp (\epsilon_{(\lambda)}^{\mu} P^\nu -
\epsilon_{(\lambda)}^{\nu} P^\mu)\,.
   \label{eq:tensor-1p1-2}
\end{equation}
In Ref.~\cite{Yang:2005gk}, we have calculated RG-improved QCD sum
rules of the tensor couplings for $1^1P_1$ mesons, where we assumed
that $h_1(1170)$ is the same as $b_1(1235)$, while $h_1(1380)$ is
made of $\bar s s$. Here we instead examine the pure $1^1P_1$ octet
states. The results are collected in
Table~\ref{tab:mass-decay-constant-1p1}. We have estimated the
singlet-octet mixing angle by means of Gell-Mann-Okubo mass formula.
Thus, the decay constants for the real states $h_1(1170)$ and
$h_1(1380)$ can be obtained. See the results listed in
Eqs.~(\ref{eq:decay-h1-1})$-$(\ref{eq:decay-h1-4}). We have also
updated the values for $b_1(1235)$ and and $K_{1B}$ states due to
the following two more reasons. First, we take into account the
uncertainties of the condensates and quark mass which were not fully
considered in Ref.~\cite{Yang:2005tv}. Second, we re-examine the
stability of the sum rules and choose to use the Borel window
0.8~GeV$^2<M^2<1.3$~GeV$^2$ in the analysis. During this window,
which covers the plateau region for the masses and decay constants
versus $M^2$, the contributions from excited states (including the
continuum) and from the highest dimension term in OPE expansion are
less than 43\% and 5\%, respectively, while for the previous choice
with the upper bound 1.5~GeV$^2$ the contributions of excited states
can reach 52\%. Although the region for the present Borel window is
a little lower than the previous study in Ref.~\cite{Yang:2005gk},
the present results should be more reliable. Note that it will
become questionable for the sum rule results if one further reduces
the lower bound of the Borel window. The reason is because in the
lower Borel mass region the effects owing to the uncertainty of the
highest dimensions term, for which we have assumed the vacuum
saturation approximation, and radiative corrections become much more
important and are out of control.

Since the QCD sum rule approach may not sufficiently determine the
singlet-octet mixing angle, we thus study the QCD sum rules for pure
$1^1P_1$ states; in other words, we neglect the corrections arising
from $\langle h_1|{\cal H}|h_8\rangle$ due to $\langle h_1|{\cal
H}|h_8\rangle \ll \langle h_1|{\cal H} | h_1\rangle , \langle
h_8|{\cal H} | h_8\rangle$, where ${\cal H}$ is the Hamiltonian, in
the studies of the mass and tensor coupling sum rules for $h_1, h_8$
and $K_{1B}$ states. Employing the formulas given in
Ref.~\cite{Yang:2005gk}, we obtain the results shown in
Table~\ref{tab:mass-decay-constant-1p1}. In complete analogy to the
discussion for $1^3P_1$ states, we introduce
  \begin{eqnarray}
\langle 0| \bar q\sigma_{\mu\nu} q | h_1(1170)(P,\lambda)\rangle
 &=& i f_{h_1(1170)}^{\perp,q}\,\epsilon_{\mu\nu\alpha\beta}
  \epsilon_{(\lambda)}^\alpha P^\beta\,, \label{eq:decay-def3}\\
\langle 0| \bar q\sigma_{\mu\nu} q | h_1(1380)(P,\lambda)\rangle
 &=& i f_{h_1(1380)}^{\perp,q}\,\epsilon_{\mu\nu\alpha\beta}
  \epsilon_{(\lambda)}^\alpha P^\beta\,, \label{eq:decay-def4}
\end{eqnarray}
and then obtain
\begin{eqnarray}
f_{h_1(1170)}^{\perp,u} &=& \frac{f_{h_1}}{\sqrt{3}}\cos\theta_{^1P_1}
 +  \frac{f_{h_8}}{\sqrt{6}}\sin\theta_{^1P_1}
 = 116\pm 8 ~(128\pm 7)~{\rm MeV} \,,~~~~~  \label{eq:decay-h1-1}\\
f_{h_1(1170)}^{\perp,s} &=&
\frac{f_{h_1}}{\sqrt{3}}\cos\theta_{^1P_1}
 -  \frac{2 f_{h_8}}{\sqrt{6}}\sin\theta_{^1P_1}
 = 75\pm 8~(-36\pm 10)~{\rm MeV} \,,  \label{eq:decay-h1-2}\\
f_{h_1(1380)}^{\perp,u} &=& -\frac{f_{h_1}}{\sqrt{3}}
\sin\theta_{^1P_1}
 +  \frac{f_{h_8}}{\sqrt{6}} \cos\theta_{^1P_1}
 = 58\pm 5~(-19\pm 7)~{\rm MeV} \,,  \label{eq:decay-h1-3}\\
f_{h_1(1380)}^{\perp,s} &=& -\frac{f_{h_1}}{\sqrt{3}}
\sin\theta_{^1P_1}
 -  \frac{2 f_{h_8}}{\sqrt{6}} \cos\theta_{^1P_1}
 = -171\pm 9~(-183\pm 10)~{\rm MeV}\,,  \label{eq:decay-h1-4}
\end{eqnarray}
corresponding to $\theta_{^1P_1}=10^\circ (45^\circ)$. As
$\theta_{^1P_1}=45^\circ$, $h_1(1380)$ is therefore dominated by the
$\bar ss$ content, and $h_1(1170)$ can be approximated by $(\bar uu
+\bar dd)/\sqrt{2}$. However, the $\bar ss$ content of $h_1(1170)$
and $\bar uu$ content of $h_1(1380)$ become significant for
$\theta_{^1P_1}=10^\circ$.

As for strange mesons, we have (with $\bar q\equiv \bar u, \bar d$)
 \begin{eqnarray}\label{eq:K1-1270}
 \langle 0 |\bar q\sigma_{\mu\nu}s |K_{1}(1270)(P,\lambda)\rangle
 &=& i f_{K_{1}(1270)}^\perp
 \,\epsilon_{\mu\nu\alpha\beta} \epsilon_{(\lambda)}^\alpha
 P^\beta\nonumber\\
 &=& i (f_{K_{1A}}^\perp a_0^{\perp, K_{1A}}\sin {\theta_K}
    + f_{K_{1B}}^\perp\cos {\theta_K})
 \,\epsilon_{\mu\nu\alpha\beta} \epsilon_{(\lambda)}^\alpha P^\beta
 \end{eqnarray}
and
 \begin{eqnarray}\label{eq:K1-1400}
 \langle 0 |\bar q\sigma_{\mu\nu}s |K_{1}(1400)(P,\lambda)\rangle
 &=& i f_{K_{1}(1400)}^\perp
 \,\epsilon_{\mu\nu\alpha\beta} \epsilon_{(\lambda)}^\alpha
 P^\beta\nonumber\\
 &=& i (f_{K_{1A}}^\perp a_0^{\perp, K_{1A}}\cos {\theta_K}
     - f_{K_{1B}}^\perp \sin {\theta_K})
 \,\epsilon_{\mu\nu\alpha\beta}\epsilon_{(\lambda)}^\alpha P^\beta
 \end{eqnarray}
where $f_{K_{1B}}^\perp$ and $a_0^{\perp, K_{1A}}$ are given in
Tables \ref{tab:mass-decay-constant-1p1} and \ref{tab:Gegenbauer},
respectively, and use of $f_{K_{1A}}^\perp=f_{K_{1A}}=(250\pm 20)$
MeV is made in the following by definition. Consequently, we obtain
(at the scale $\mu=1$~GeV)
\begin{eqnarray}
f_{K_{1}(1270)}^\perp &=&~~\, 145  \pm 15~ {\rm MeV} \,,\\
f_{K_{1}(1400)}^\perp &=& -124 \pm 15~ {\rm MeV} \,,
 \qquad {\rm if\ taking\ }\theta_K=45^\circ,
\end{eqnarray}
and
\begin{eqnarray}
f_{K_{1}(1270)}^\perp &=& 124  \pm 14~ {\rm MeV} \,,\\
f_{K_{1}(1400)}^\perp &=& 145 \pm 14~ {\rm MeV} \,,
 \qquad {\rm if\ taking \ }\theta_K=-45^\circ,
\end{eqnarray}
where $a_0^{\perp,K_{1A}}$ gives about 8\% corrections to the decay
constants. If setting $a_0^{\perp,K_{1A}}=0$, all the magnitudes of
the central values of the decay constants are equal to $\sim 134$
GeV.
\begin{table}[t!]
\caption[]{The sum rule results of masses, decay constants, and
corresponding excited thresholds $s_0^{^1P_1}$ for $1^1P_1$ mesons.
The values of $f^\perp_{^1P_1}$ are given at the scale $\mu=1$~GeV.}
\label{tab:mass-decay-constant-1p1}
\renewcommand{\arraystretch}{1.5}
\addtolength{\arraycolsep}{4pt}
$$
\begin{array}{|c|ccc|}\hline
 {\rm State} & {\rm Mass} [{\rm GeV}]& {\rm Decay\ cosntant}~f^\perp_{^1P_1} [{\rm MeV}]
 & s_0^{^1P_1} [{\rm GeV}^2]
 \\ \hline
\begin{array}{c} b_1(1235) \\ h_1 (1^1P_1) \\ h_8 (1^1P_1) \\ K_{1B}\end{array}&
\begin{array}{c} 1.21\pm 0.07\\ 1.23\pm 0.07\\ 1.37\pm 0.07\\ 1.34\pm 0.08\end{array}&
\begin{array}{c} 180\pm 8\\ 180\pm 12\\ 190\pm 10\\ 190\pm 10\end{array}&
\begin{array}{c} 2.6\pm 0.2\\ 2.6\pm 0.2\\ 3.2\pm 0.2\\ 3.1\pm 0.2\end{array}
\\ \hline
\end{array}
$$
\end{table}

\subsection{Gegenbauer moments of leading-twist LCDAs}\label{subsec:moments-t2}

The Gegenbauer moments can be calculated from the standard QCD sum
rule approach by adopting a relevant two-point correlation function
as the starting point. It is interesting to note that in SU(3)
symmetry limit the decay constant for a $1 ^1P_1$ state transiting
to the vacuum via the local axial-vector current vanishes due to
G-parity mismatch between the current and states. On the other hand,
conventionally, the decay constants $f_{^3P_1}$ for the local
axial-vector currents coupling to the $1^3P_1$ states are chosen to
be positive as in the present paper. Although the diagonal
correlation functions may have good qualities for sum rule results
of G-parity invariant parameters, it cannot determine the relative
sign of the Gegenbauer moments. Moreover, the diagonal correlation
functions cannot use to evaluate the G-parity violating parameters
due to the mixing between $^3P_1$ and $^1P_1$ states. To determine
not only the magnitudes but also the relative signs for Gegenbauer
moments relevant to the leading-twist LCDAs of $1^3P_1$ states, we
thus choose one of the interpolating currents in the two-point
correlation function to be the local axial-vector current $\bar
q_2\gamma_\mu\gamma_5 q_1$, i.e., we consider the non-diagonal
correlation functions here. Note that we define the sum rule to be
diagonal here only if two interpolating currents in the correlation
function are exactly the same, but non-diagonal otherwise (See
footnote~\ref{foot:dia} for further discussions). With the same
reason, in the following section we also adopt the local
axial-vector current as one of the interpolating currents in QCD sum
rule studies to determine the relevant parameters for twist-3
three-parton LCDAs of $1^3P_1$ states, whereas we use the local
pseudo-tensor current as one of the interpolating currents since the
resulting contributions arising from $^3P_1$ states vanish in SU(3)
limit.

 However,
as we consider the non-diagonal correlation functions, G-parity
breaking contributions of $^3P_1$ and $^1P_1$ states always mix. To
obtain the G-parity violating Gegenbauer moments for LCDAs of
$^3P_1$ and $^1P_1$ mesons, we will assume an additional reasonable
constraint on the parameters (see Eq.~(\ref{eq:constraint})).

\subsubsection{Gegenbauer moments of $\Phi_\parallel^{^3P_1}$ for
 $1^3P_1$ mesons}\label{subsubsec:moments-parallel-3p1}

The LCDAs $\Phi_\parallel^{^3P_1}(u,\mu)$ corresponding to the $1^3P_1$
states, which are denoted by the superscript $^3P_1$,
with the quark contents $\bar q_2$ and $q_1$, are defined as
 \begin{equation}
\langle 1^3P_1(P,\lambda)|\bar q_1(y)\not\! \bar{z}\gamma_5
q_2(x)|0\rangle = i f_{^3P_1} m_{^3P_1}
(\epsilon^{*(\lambda)}\bar{z})\int^1_0 dx e^{i(up\cdot y +\bar
up\cdot x)}\Phi_\parallel^{^3P_1}(u,\mu),
 \end{equation}
where $u$ (or $\bar u=1-u$) is the momentum fraction carried by the
quark $q_1$ (or antiquark $\bar q_2$) and $\mu$ is the
renormalization scale of the LCDAs, and we have considered here and
below that $\bar{z}\propto y-x, \bar{z}^2=0, (y-x)^2=0$, but
$\bar{z}\not\to 0$ even for $y\to x$.
$\Phi_\parallel^{^3P_1}(u,\mu)$ can be expanded in a series of
Gegenbauer polynomials as given in Eq.~(\ref{eq:generaldaparallel}).
To evaluate the Gegenbauer moments of $\Phi_\parallel^{^3P_1}$, we
take into account the following two-point correlation function
\begin{eqnarray}
 \Pi_{\mu}^{(l)} (q)
 = i\int d^4x e^{iqx} \langle 0| T(\Omega_{^3P_1}^{(l)}(x)
 \ O_{\mu}^\dagger(0) |0 \rangle = (\bar{z} q)^{l}
 [-I_{1}^{(l)} (q^2)\, \bar{z}_\mu + I_{2}^{(l)}(q^2) (\bar{z} q) q_\mu ] ,
\end{eqnarray}
where
\begin{eqnarray}
 O_{\mu}(0)= \bar q_2(0) \gamma_\mu \gamma_5 q_1(0)
\end{eqnarray}
 and the relevant
multiplicatively renormalizable operator, to leading logarithmic
(LO) accuracy, is
  \begin{eqnarray}
\Omega_{^3P_1}^{(l)}(x) &=& \sum\limits_{j=0}^l c_{l,j}
(i\bar{z}\partial)^{l-j}
  \bar{q}_2(x)\!\not\!\bar{z} \gamma_5\,(i\bar{z}\deriv)^j q_1(x)\,,
  \label{eq:optensor}
\end{eqnarray}
with $\deriv_\mu = \derright_\mu - \derleft_\mu =
(\stackrel{\rightarrow}{\partial} +ig_s A^a(x)\lambda^a/2)_\mu-
(\stackrel{\leftarrow}{\partial} -ig_s A^a(x)\lambda^a/2)_\mu$ and
$c_{l,k}$ being the coefficients of the Gegenbauer polynomials such
that $C_l^{3/2}(x) = \sum_{k=0}^{k=l} c_{l,k}x^k$.
$\Omega_{^3P_1}^{(l)}$ and $O_\mu$ satisfy the following relations:
\begin{eqnarray} \langle 0| \Omega_{^3P_1}^{(l)}(0)|1^3P_1(P,\lambda)\rangle
&=& -if_{^3P_1} m_{^3P_1}(\epsilon^{(\lambda)} \bar{z})(P\bar{z})^l
\frac{3(l+1)(l+2)}{2(2l+3)}\, a_l^{\parallel,^3P_1}(\mu)\,,
\\
\langle 0| O_{\mu}(0)|1^3P_1(P,\lambda)\rangle &=& - i f_{^3P_1} m_{^3P_1}
\epsilon^{(\lambda)}_\mu .
\end{eqnarray}
We are interested only in $I_1^{(l)}(q^2)$ since only states with
quantum numbers of $^3P_1$ contribute to $I_1^{(l)}(q^2)$, whereas
$I_2^{(l)}(q^2)$ receives contributions from states with quantum
numbers of pseudoscalar mesons and $^3P_1$. $I_2^{(l)}(q^2)$ was
already given in Ref.~\cite{Ball:2003sc}. In the massless quark
limit, one has $I_1^{(l)} =q^2 I_2^{(l)}$. Nevertheless, the above
relation is broken even for $m_{q_1}=m_{q_2}\neq 0$. The OPE result
of $I_1^{(l)}$, up to dimension 6 and with ${\cal O}(\alpha_s)$
corrections, is given by (for $l\geq 1$)
\begin{eqnarray}\label{eq:ope-1}
I_1^{(l)\rm OPE} &=& -\frac{\alpha_s}{2\pi^3}\, q^2
\ln\frac{-q^2}{\mu^2}\int_0^1\!\! du\, u\bar u\,
C_l^{3/2}(2u-1)\, \ln^2\,\frac{u}{\bar u}\nonumber\\
& & {}  -\frac{C^{3/2}_l(1)}{q^2}\, [m_{q_1}\langle \bar q_2
q_2\rangle + m_{q_2}\langle \bar q_1 q_1\rangle (-1)^l ]
  +\frac{C^{3/2}_l(1)}{24q^2}\, \langle \frac{\alpha_s}{\pi}
  G^2\rangle [1+(-1)^l]\nonumber\\
 && + \frac{2}{q^4}\, C^{5/2}_{l-1}(1)\, \theta(l-1)
 [m_{q_2}\langle \bar q_1 g_s\sigma G q_1\rangle (-1)^l
 + m_{q_1}\langle \bar q_2 g_s\sigma G q_2\rangle ] L^{-14/(3b)}\nonumber\\
 & & -\frac{16\pi\alpha_s}{9q^4} C^{3/2}_l(1)\,
 \langle \bar q_1 q_1\rangle \langle \bar q_2 q_2\rangle [1+(-1)^l]\nonumber\\
 & & - \frac{32\pi\alpha_s}{81q^4}\, [C^{3/2}_l(1)]^2
 [ \langle \bar q_1 q_1\rangle^2 (-1)^l + \langle \bar q_2
 q_2\rangle^2]\,,
\end{eqnarray}
where the terms containing quark mass corrections and the results
with odd $l$ are new. For even $l$ and in the massless quark limit,
the above result for $I_1^{(l) \rm OPE}$ is consistent with $q^2
I_2^{(l) \rm OPE}$ given in Ref.~\cite{Ball:2003sc}.

The quality of the sum rules obtained directly from $I_1^{(l) \rm
OPE}$ is not good. See the discussion below. Another way to obtain
the sum rules is to take into account the dispersion relation with
one subtraction. This method was introduced in
Refs.~\cite{Balitsky:1982dt,Braun:1985ah}. Consider $\tilde
I_1^{(l)}(q^2)= I_1^{(l)}(q^2)- I_1^{(l)\,,pert}(q^2)$. $\tilde
I_1^{(l)}(q^2)$ is finite in $-q^2\to \infty$, and therefore we can
write down
\begin{eqnarray}\label{eq:ope-2}
\tilde I_1^{(l)} (q^2) &=& \tilde
I_1^{(l)}(0)-\frac{-q^2}{\pi}\int_0^\infty\frac{ds}{s(s-q^2)}
 [\rho_{phys} (s) - {\rm Im}I_1^{(l)\,,pert}(s)] \,, \nonumber\\
 &=& \tilde
I_1^{(l)}(0)-\frac{-q^2}{\pi}\int_0^\infty\frac{ds}{s(s-q^2)}
 [\rho_{1^3P_1} (s) - \theta(s_0^\parallel-s){\rm Im}I_1^{(l)\,,pert}(s)] \,,
\end{eqnarray}
where $\rho_{phys}$ and $\rho_{1^3P_1}$ are the total physical
spectral density and the lowest-lying ($1^3P_1$) spectral
density\footnote{Here we also need to consider $1^1P_1$ state which
will contribute to the sum rules due to the G-parity violating
effect. See Eq.~(\ref{eq:g-violating}).}, respectively, and where we
have modeled the higher resonance states as
 \begin{eqnarray}
\rho_{phys}(s)=\rho_{1^3P_1} (s) +\theta(s-s_0^\parallel){\rm
Im}I_1^{(l)\,,pert}(s)\,.
 \end{eqnarray}
 On the other hand, one should note that $^1P_1$ states can still enter
the sum results by the G-parity violating effect, which is due to
$m_{q_1}-m_{q_2}\neq 0$ ,
\begin{eqnarray}\label{eq:g-violating}
 &&  \langle 1^1P_1(P,\lambda)|\bar q_1(0) \gamma_\mu \gamma_5 q_2(0)|0\rangle
   = if_{^1P_1} a_0^{\parallel,^1P_1} \, m_{^1P_1} \,  \epsilon^{*(\lambda)}_\mu
   \,.
 \end{eqnarray}
If one derives the Gegenbauer moment sum rules directly from
Eq.~(\ref{eq:ope-1}), then the results are
\begin{eqnarray}\label{eq:sr-1.1}
 &&
 \Bigg[a_l^{\parallel,^3P_1} m_{^3P_1}^2 f_{^3P_1}^2
e^{-m_{^3P_1}^2/M^2}
 + a_l^{\parallel,^1P_1} a_0^{\parallel,^1P_1}m_{^1P_1}^2 f_{^1P_1}^2
e^{-m_{^1P_1}^2/M^2} \Bigg] \frac{3(l+1)(l+2)}{2(2l+3)}
L^{-\gamma_{(l)}^\parallel/b}
 \nonumber\\
&& { } = \frac{\alpha_s}{2\pi^3}\, M^4
\Bigg[1-e^{-s_0^\parallel/M^2} \Big( 1+
\frac{s_0^{\parallel,^3P_1}}{M^2} \Big) \Bigg] \int_0^1\!\! du\,
u\bar u\,
C_l^{3/2}(2u-1)\,\,\ln^2\,\frac{u}{\bar u}\nonumber\\
& & {}~~~  + C^{3/2}_l(1)\, (m_{q_1}\langle \bar q_2 q_2\rangle +
m_{q_2}\langle \bar q_1 q_1\rangle )
  -\frac{C^{3/2}_l(1)}{12}\, \langle \frac{\alpha_s}{\pi}
  G^2\rangle\nonumber\\
 && ~~~ + \frac{2}{M^2}\, C^{5/2}_{l-1}(1)\, \theta(l-1)
 (m_{q_1}\langle \bar q_2 g_s\sigma G q_2\rangle
 + m_{q_2}\langle \bar q_1 g_s\sigma G q_1\rangle ) L^{-14/(3b)}\nonumber\\
 & & ~~~ -\frac{32\pi\alpha_s}{9M^2} C^{3/2}_l(1)\,
 \langle \bar q_1 q_1\rangle \langle \bar q_2 q_2\rangle
 - \frac{32\pi\alpha_s}{81M^2}\, [C^{3/2}_l(1)]^2
 ( \langle \bar q_1 q_1\rangle^2 + \langle \bar q_2 q_2\rangle^2)
\end{eqnarray}
for even $l\geq 2$ and
\begin{eqnarray}\label{eq:sr-1.2}
&&
 \Bigg[a_l^{\parallel,^3P_1} m_{^3P_1}^2 f_{^3P_1}^2
e^{-m_{^3P_1}^2/M^2}
 + a_l^{\parallel,^1P_1} a_0^{\parallel,^1P_1}m_{^1P_1}^2 f_{^1P_1}^2
e^{-m_{^1P_1}^2/M^2} \Bigg] \frac{3(l+1)(l+2)}{2(2l+3)}
L^{-\gamma_{(l)}^\parallel/b}
 \nonumber\\
&& { } =  C^{3/2}_l(1)\,
 (m_{q_1}\langle \bar q_2 q_2\rangle - m_{q_2}\langle \bar q_1
 q_1\rangle)\nonumber\\
 && ~~~ + \frac{2}{M^2}\, C^{5/2}_{l-1}(1)\, \theta(l-1)
 (m_{q_1}\langle \bar q_{2} g_s\sigma G q_2\rangle
  - m_{q_2}\langle \bar q_1 g_s\sigma G q_1\rangle) L^{-14/(3b)}
   \nonumber\\
 && ~~~  + \frac{32\pi\alpha_s}{81M^2}\, [C^{3/2}_l(1)]^2
 (\langle \bar q_1 q_1 \rangle^2 - \langle \bar q_2 q_2\rangle^2)
\end{eqnarray}
for odd $l$ (of the $K_{1A}$ meson), where $\theta(x)=1$ for $x\geq
0$ or 0 otherwise. However, if one chooses to use
Eq.~(\ref{eq:ope-2}) and divides the relation in
Eq.~(\ref{eq:ope-2}) by $-q^2$ before applying the Borel transform,
then the Gegenbauer moment sum rules are
\begin{eqnarray}\label{eq:sr-2.1}
 &&
 \Bigg[a_l^{\parallel,^3P_1} f_{^3P_1}^2
\bigg(e^{-m_{^3P_1}^2/M^2}-1 \bigg)
 + a_l^{\parallel,^1P_1} a_0^{\parallel,^1P_1}f_{^1P_1}^2
\bigg(e^{-m_{^1P_1}^2/M^2}-1 \bigg) \Bigg]
\frac{3(l+1)(l+2)}{2(2l+3)} L^{-\gamma_{(l)}^\parallel/b}
 \nonumber\\
&& { } = \frac{\alpha_s}{2\pi^3}\, M^2
\Bigg[1-e^{-s_0^\parallel/M^2} -\frac{s_0}{M^2} \Bigg] \int_0^1\!\!
du\, u\bar u\,
C_l^{3/2}(2u-1)\,\,\ln^2\,\frac{u}{\bar u}\nonumber\\
& & {}~~~  - \frac{C^{3/2}_l(1)}{M^2}\, (m_{q_1}\langle \bar q_2
q_2\rangle + m_{q_2}\langle \bar q_1 q_1\rangle )
  + \frac{C^{3/2}_l(1)}{12M^2}\, \langle \frac{\alpha_s}{\pi}
  G^2\rangle\nonumber\\
 && ~~~ - \frac{1}{M^4}\, C^{5/2}_{l-1}(1)\, \theta(l-1)
 (m_{q_1}\langle \bar q_2 g_s\sigma G q_2\rangle
 + m_{q_2}\langle \bar q_1 g_s\sigma G q_1\rangle ) L^{-14/(3b)}\nonumber\\
 & & ~~~ +\frac{16\pi\alpha_s}{9M^4} C^{3/2}_l(1)\,
 \langle \bar q_1 q_1\rangle \langle \bar q_2 q_2\rangle
 + \frac{16\pi\alpha_s}{81M^4}\, [C^{3/2}_l(1)]^2
 ( \langle \bar q_1 q_1\rangle^2 + \langle \bar q_2 q_2\rangle^2)
\end{eqnarray}
for even $l\geq 2$ and
\begin{eqnarray}\label{eq:sr-2.2}
&&
 \Bigg[a_l^{\parallel,^3P_1} f_{^3P_1}^2
\bigg(e^{-m_{^3P_1}^2/M^2}-1 \bigg)
 + a_l^{\parallel,^1P_1} a_0^{\parallel,^1P_1} f_{^1P_1}^2
\bigg(e^{-m_{^1P_1}^2/M^2}-1 \bigg) \Bigg]
\frac{3(l+1)(l+2)}{2(2l+3)} L^{-\gamma_{(l)}^\parallel/b}
 \nonumber\\
&& { } =  -\frac{C^{3/2}_l(1)}{M^2} \,
 (m_{q_1}\langle \bar q_2 q_2\rangle - m_{q_2}\langle \bar q_1
 q_1\rangle)\nonumber\\
 && ~~~ - \frac{1}{M^4}\, C^{5/2}_{l-1}(1)\, \theta(l-1)
 (m_{q_1}\langle \bar q_{2} g_s\sigma G q_2\rangle
  - m_{q_2}\langle \bar q_1 g_s\sigma G q_1\rangle) L^{-14/(3b)}
   \nonumber\\
 && ~~~  - \frac{16\pi\alpha_s}{81M^4}\, [C^{3/2}_l(1)]^2
 (\langle \bar q_1 q_1 \rangle^2 - \langle \bar q_2 q_2\rangle^2)\,,
\end{eqnarray}
for odd $l$ (of the $K_{1A}$ meson), where we have substituted
 $$\tilde I_1^{(l)}(0)= \Bigg[ a_l^{\parallel,^3P_1} f_{^3P_1}^2
+ a_l^{\parallel,^1P_1} a_0^{\parallel,^1P_1} \Bigg]
\frac{3(l+1)(l+2)}{2(2l+3)} -
\frac{\alpha_s}{2\pi^3}\,s_0^\parallel\int_0^1\!\! du\, u\bar u\,
C_l^{3/2}(2u-1)\, \ln^2\,\frac{u}{\bar u}\,,
 $$
which can be determined in the limit $M^2\to \infty$, into the above
two equations.

Before we discuss the reliability of the sum rules about Eqs.
(\ref{eq:sr-1.1}), (\ref{eq:sr-1.2}) and (\ref{eq:sr-2.1}),
(\ref{eq:sr-2.2}), one should note that, for even $l$, the
contributions due to $^1P_1$ states are $\sim {\cal O} (m_q^2)$ and
thus negligible. Nevertheless, for odd $l$, the contributions for
$1^3P_1$ and $1^1P_1$ states are of the same order of magnitude. The
quality of the sum rules in Eqs.~(\ref{eq:sr-1.1}) and
(\ref{eq:sr-1.2}) is not good due to the following two reasons. (i)
The OPE series converges very slowly. For instance, if taking a
close look at Eq.~(\ref{eq:sr-1.1}) with $l=2$, the terms of
dimension-4 and -6 in the OPE series are still comparable to the
perturbative contribution even for choosing a quite large Borel mass
$ M^2\sim 2$~GeV$^2$. (ii) The terms of dimension-4 and -6 have the
opposite sign compared with the perturbative contribution, such that
the contributions of higher resonance states that we modeled have
also the opposite sign compared with the lowest-lying state. As a
result, it is difficult to choose a reliable windows for sum rules
given in Eqs.~(\ref{eq:sr-1.1})  and (\ref{eq:sr-1.2}).

Nevertheless, the sum rules, given Eqs.~(\ref{eq:sr-2.1})  and
(\ref{eq:sr-2.2}), converge much more quickly. Meanwhile, in the OPE
series in Eq.~(\ref{eq:sr-2.1}) the contribution of the perturbative
term have the same sign as the the terms of dimension-4 and -6.
Consequently, we can find a suitable Borel window, where the
contributions originating from higher resonances and the highest OPE
terms are well under control. For the time being, we will hence
focus on Eqs.~(\ref{eq:sr-2.1})  and (\ref{eq:sr-2.2}). The
numerical results are given in
Sec.~\ref{subsec:gegenbauer-3p1-result}. Considering
Eq.~(\ref{eq:sr-2.2}) with $l=1$ for the $K_{1A}$ (and $K_{1B}$)
mesons, the result approximately reads
\begin{eqnarray}
&&
 a_1^{\parallel,K_{1A}}
 + a_1^{\parallel,K_{1B}} a_0^{\parallel,K_{1B}}
 \frac{f_{K_{1B}}^2}{ f_{K_{1A}}^2}
  \simeq \frac{5}{9}
  \frac{1}{f_{K_{1A}}^2 \bigg(e^{-\bar m^2/M^2}-1\bigg)}
L^{\gamma_{(1)}^\parallel/b} \Bigg\{ -\frac{3}{M^2}\,
 (m_s\langle \bar q_2 q_2\rangle - m_{q_2}\langle \bar s s\rangle)
 \nonumber\\
 && ~~
 - \frac{1}{M^4}\,
 (m_{s}\langle \bar q_{2} g_s\sigma G q_2\rangle
  - m_{q_2}\langle \bar s g_s\sigma G s\rangle) L^{-14/(3b)}
   - \frac{16\pi\alpha_s}{9M^4}\,
 (\langle \bar s s\rangle^2 - \langle \bar q_2 q_2\rangle^2)
 \Bigg\}, \label{eq:mix-1}
\end{eqnarray}
with $\bar{m}=(m_{K_{1A}}+m_{K_{1B}})/2$.

\subsubsection{Gegenbauer moments of $\Phi_\perp^{^3P_1}$ for $1^3P_1$ mesons}

To calculate the Gegenbauer moments of $\Phi_\perp^{^3P_1}$ for $1^3P_1$
states, we consider the following correlation function
\begin{eqnarray}
 \int d^4x e^{iqx}
\langle 0| {\rm T}(\Omega^{\perp(l)}_{\mu}(x)\ O_{\nu}^\dagger(0) |0
\rangle
 =
\Bigg[g_{\mu\nu} - \frac{1}{q\bar{z}}(q_\mu \bar{z}_\nu + q_\nu
\bar{z}_\mu)\Bigg] (q\bar{z})^{l+1}T_1(q^2) +\cdots,
  \label{eq:Gmomentperp-t2}
\end{eqnarray}
where
 \begin{eqnarray}
 O_\nu(0)=\bar{q_1}(0)\gamma_\mu \gamma_5 q_2(0),
 \end{eqnarray}
and, to leading logarithmic accuracy, the relevant
multiplicatively renormalizable operator is
  \begin{eqnarray}
\Omega^{\perp(l)}_{\mu}(x)
 &=& \sum\limits_{j=0}^l c_{n,j} (i\bar{z} \partial)^{l-j}
\bar{q}_2(x)\!\sigma_{\mu\alpha}\gamma_5 \bar{z}^\alpha \,(i\bar{z}
\deriv)^j q_1(x)\,, \label{eq:optensor1}
\end{eqnarray}
with $\bar{z}$ being the light-like vector as defined previously.
$\Omega^{\perp(l)}_{\mu}$ and $O_\nu$ satisfy the relation:
\begin{eqnarray}
 &&\sum_\lambda
 \langle 0| \Omega^{\perp(l)}_{\mu}(0)|1^3P_1(P,\lambda)\rangle
 \langle 1^3P_1(P,\lambda)| O_{\nu}(0)|0 \rangle\nonumber\\
 &&
 \ \ \  = i f_{^3P_1}^\perp f_{^3P_1} m_{^3P_1}
 \frac{3(l+1)(l+2)}{2(2l+3)}
 \Bigg[g_{\mu\nu}
 - \frac{1}{P\bar{z}}(P_\mu \bar{z}_\nu + P_\nu \bar{z}_\mu)\Bigg]
(P\bar{z})^{l+1}a_l^{\perp,^3P_1} + \cdots\,. \hspace{0.8cm}
\end{eqnarray}
Using the dispersion relation for $T_1$, $a_l^{\perp,^3P_1}$ can be
represented in the form
\begin{eqnarray}\label{eq:Gmoment_alperpduality}
\frac{3(l+1)(l+2)}{2(2l+3)}\frac{f_{^3P_1}^\perp f_{^3P_1} m_{^3P_1}
a_l^{\perp,^3P_1}}{m_{^3P_1}^2-q^2}= \frac{1}{\pi}\int^{s_0^{\perp,^3P_1}}_0 ds
\frac{{\rm Im} T_1^{\rm OPE}(s)}{s-q^2} \,,
\label{eq:duality_a1perp}
\end{eqnarray}
where
\begin{eqnarray}
  T_1^{\rm OPE}(0) &=& \frac{3 }{4\pi^2 }\ln \frac{-q^2}{\mu^2}
 \Bigg( \int^1_0 d\alpha [m_{q_2} \alpha + m_{q_1}
 (\alpha-1)] C_l^{3/2}(2\alpha -1) \Bigg)  \nonumber\\
 && - \frac{1}{q^2} C_l^{3/2}(1)(\langle \bar q_2 q_2\rangle + \langle \bar q_1 q_1\rangle
  (-1)^{l+1}) \nonumber\\
 && + \Bigg[ \frac{1}{3} C_l^{3/2}(1) + 2 C_{l-1}^{5/2}(1)\theta(l-1)
  \Bigg] \frac{\langle \bar q_2 g_s \sigma\cdot G q_2\rangle
   + \langle \bar q_1 g_s \sigma\cdot G q_1 \rangle (-1)^{l+1}}{q^4}  \nonumber\\
 && -\frac{2\pi^2}{3 q^6}
  \Big[20 C_{l-2}^{7/2}(1)\theta(l-2)
  + C_{l-1}^{5/2}(1)\theta(l-1))\Big]\langle \frac{\alpha_s}{\pi} G^2\rangle
 [\langle \bar q_2 q_2\rangle + \langle \bar q_1 q_1\rangle
  (-1)^{l+1}] .\nonumber\\
\end{eqnarray}
Since $^1P_1$ states can have small axial-vector coupling constants
due the G-parity violating (SU(3)-breaking) effect, the RG-improved
sum rules for Gegenbauer moments $a_l^{\perp,^3P_1}$ thus read
\begin{eqnarray}\label{eq:Gmoment_alperpSR}
&& \Bigg[ a_l^{\perp,^3P_1} m_{^3P_1} f_{^3P_1} f_{^3P_1}^\perp
e^{-m_{^3P_1}^2/M^2}
 + a_l^{\perp,^1P_1} a_0^{\parallel,^1P_1}
 m_{^1P_1} f_{^1P_1} f_{^1P_1}^\perp
e^{-m_{^1P_1}^2/M^2}
 \Bigg]
 \frac{3(l+1)(l+2)}{2(2l+3)} L^{-\gamma_{(l)}^\perp/b}\nonumber\\
&& ~~~
 =  - \Bigg\{ \frac{3 }{4\pi^2 }M^2
(1-e^{-s_0^{\perp,^3P_1}/M^2})
 \Bigg( \int^1_0 d\alpha C_l^{3/2}(2\alpha -1) [m_{q_2} \alpha + m_{q_1}
 (\alpha-1)] \Bigg)  L^{-4/b} \nonumber\\
 && ~~~~~~
 - C_l^{3/2}(1)(\langle \bar q_2 q_2\rangle + \langle \bar q_1 q_1\rangle
  (-1)^{l+1}) L^{4/b} \nonumber\\
 && ~~~~~~
 - \Bigg[ \frac{1}{3} C_l^{3/2}(1) + 2 C_{l-1}^{5/2}(1)\theta(l-1)
  \Bigg]
   \frac{\langle \bar q_2 g_s \sigma\cdot G q_2\rangle
   + \langle \bar q_1 g_s \sigma\cdot G q_1
 \rangle (-1)^{l+1}}{M^2}  L^{-2/(3b)} \nonumber\\
 && ~~~~~~
 -\frac{\pi^2}{M^4}\Bigg[\frac{20}{3} C_{l-2}^{7/2}(1)\theta(l-2)
  +\frac{1}{3} C_{l-1}^{5/2}(1)\theta(l-1))\Bigg]\nonumber\\
 &&\ \ \ \ \ \ \ \ \ \times \langle \frac{\alpha_s}{\pi} G^2\rangle
 (\langle \bar q_2 q_2\rangle + \langle \bar q_1 q_1\rangle
  (-1)^{l+1})   L^{4/b} \Bigg\},
 \end{eqnarray}
where for odd $l$ the corrections coming from the $^1P_1$ states are
of order $m_q^2$ and can be neglected, whereas for even $l$ the
$1^3P_1$ and $1^1P_1$ states give contributions of the same order.
With $l=0,2 $, we can therefore obtain the following approximation
for the $K_{1A}$ and $K_{1B}$ mesons:
\begin{eqnarray}
 a_0^{\perp,K_{1A}}
 &+&   a_0^{\parallel,K_{1B}}
 \frac{m_{K_{1B}} f_{K_{1B}}^\perp f_{K_{1B}}}
      {m_{K_{1A}} f_{K_{1A}}^\perp f_{K_{1A}}}
  \simeq -\frac{1}{m_{K_{1A}} f_{K_{1A}}^\perp
  f_{K_{1A}}}
 e^{\bar m^2/M^2}
L^{\gamma_{(0)}^\perp/b} \nonumber\\
&& \times  \Bigg\{ \frac{3 }{8\pi^2 }M^2 (1-e^{-s_0/M^2})
 ( m_{q_2}- m_{s})  L^{-4/b}
 - (\langle\bar q_2 q_2\rangle - \langle \bar ss\rangle) L^{4/b}
 \nonumber\\
 && ~~~
 -\frac{\langle \bar q_2 g_s \sigma\cdot G q_2\rangle
   - \langle \bar s g_s \sigma\cdot G s
 \rangle}{3M^2}  L^{-2/(3b)}
 + \frac{0}{M^4}\langle \frac{\alpha_s}{\pi} G^2\rangle
 (\langle \bar q_2 q_2\rangle -\langle \bar ss\rangle)   L^{4/b} \Bigg\},
 \qquad
 \label{eq:mix-2}
\end{eqnarray}
\begin{eqnarray}
& & a_2^{\perp,K_{1A}}
 + a_2^{\perp,K_{1B}} a_0^{\parallel,K_{1B}}
 \frac{m_{K_{1B}} f_{K_{1B}}^\perp f_{K_{1B}}}
      {m_{K_{1A}} f_{K_{1A}}^\perp f_{K_{1A}}}
  \simeq -\frac{7}{18}\frac{1}{m_{K_{1A}} f_{K_{1A}}^\perp
  f_{K_{1A}}}
 e^{\bar m^2/M^2}
L^{\gamma_{(2)}^\perp/b} \nonumber\\
&&\ \ \ \ \
 \times  \Bigg\{ \frac{3 }{8\pi^2 }M^2 (1-e^{-s_0/M^2})
 ( m_{q_2}- m_{s})  L^{-4/b}
 - 6(\langle\bar q_2 q_2\rangle - \langle \bar ss\rangle) L^{4/b}
 \nonumber\\
 && ~~~~~~~~
 -12\frac{\langle \bar q_2 g_s \sigma\cdot G q_2\rangle
   - \langle \bar s g_s \sigma\cdot G s
 \rangle}{M^2}  L^{-2/(3b)}
 -\frac{25}{3} \frac{\pi^2}{M^4}\langle \frac{\alpha_s}{\pi} G^2\rangle
 (\langle \bar q_2 q_2\rangle -\langle \bar ss\rangle)   L^{4/b} \Bigg\},
 \qquad
 \label{eq:mix-2-2}
\end{eqnarray}
 with $s_0 \approx s_0^{\perp,K_{1A}}$.

\subsubsection{Results}\label{subsec:gegenbauer-3p1-result}

In the numerical analysis, we shall adopt parameters which are
collected in Appendix~\ref{appsec:inputs} and
Tables~\ref{tab:mass-decay-constant-3p1} and
\ref{tab:mass-decay-constant-1p1}. It should be noted that for the
moment sum rules  the actual expansion parameter is $M^2/l$ in the
large $l$ limit \cite{Chernyak:1983ej}. As a result, for
$a_l^\parallel$ and $a_l^\perp$ with a larger $l$ and fixed $M^2$,
the OPE series are convergent more slowly or even divergent as
compared with the sum rules for the masses or decay constants.

Consider $a_l^{\parallel,^3P_1}$ first. We adopt the sum rules given
in Eqs. (\ref{eq:sr-2.1}) and (\ref{eq:sr-2.2}). As discussed after
Eq. (\ref{eq:sr-2.2}), these two sum rules can give much more
reliable results than those in Eqs. (\ref{eq:sr-1.1}) and
(\ref{eq:sr-1.2}). For $a_2^{\parallel,^3P_1}$, as expected, we find
the higher Borel window to be $2.0$~GeV$^2< M^2 < 3.0$~GeV$^2$,
where the contributions originating from higher resonance states lie
between 15\% and 32\%, and moreover the correction arising from the
highest dimension term in OPE series is between 17\% and 6\%. For
$a_1^{\parallel,K_{1A}} + a_1^{\parallel,K_{1B}}
a_0^{\parallel,K_{1B}} f_{K_{1B}}^2/ f_{K_{1A}}^2$, the correction
from highest dimension is quite small, but it is hard to estimate
the contributions from higher resonances. Fortunately, the result of
this sum rule is quite stable and we will choose to use the Borel
window $1.5$~GeV$^2< M^2 < 2.5$~GeV$^2$, which is in between the
cases of the sum rules for decay constants and
$a_2^{\parallel,^3P_1}$.

The sum rules for $a_1^{\perp,^3P_1}$ are the typical cases about
the non-diagonal sum rules, for which the main contributions in OPE
may come from the quark and quark-gluon condensates. It has been
argued in Refs.~\cite{Belyaev:1984ic,Balitsky:1985aq,Ball:2002ps}
that these sum rules may suffer from contributions of higher
resonances, so that the value about the lowest-lying meson may be
{\it overestimate}. On the other hand, note that usually the
radiative corrections are constructive at the 10\% level for each
OPE term. Since we have neglected the radiative corrections, it
means that the real value for the lowest-lying meson may be {\it
underestimate}. The above two corrections may partially cancel each
other. Equivalently, these two corrections can be lumped into the
uncertainties of the condensates and quark masses. In summary, the
non-diagonal sum rules may suffer from above two corrections and the
net effects on results may be less than 10\%. Essentially, the above
estimate is suitable for all twist-3 and twist-3 parameters studied
in the present paper.  However, since we do not do the qualitative
estimates about the parameters, we thus do not include these
possible errors in Tables \ref{tab:Gegenbauer},
\ref{tab:para-3p-t3}, and \ref{tab:para-3p-t1}.

The best stability of the $a_1^{\perp,^3P_1}$ sum rules is reached
within the Borel window 1.5~GeV$^2<M^2<2.5$~GeV$^2$, the same as the
case of $a_1^{\parallel,K_{1A}}$, where the correction from the
highest dimension term at the quark-gluon level is quite small. For
G-parity violating Gegenbauer moments, $a_0^{\perp,K_{1A}}+ c_2\cdot
a_0^{\parallel,K_{1B}} $ and $a_2^{\perp,K_{1A}}+ c_2 \cdot
a_2^{\perp,K_{1B}} a_0^{\parallel,K_{1B}}$ with
$$c_2=(m_{K_{1B}}/m_{K_{1A}}) (f_{K_{1B}}^\perp f_{K_{1B}}/
f_{K_{1A}}^\perp f_{K_{1A}}),$$ we find the suitable Borel windows
to be 1.3~GeV$^2< M^2 < 2.0$~GeV$^2$ and 2.0~GeV$^2< M^2 <
3.0$~GeV$^2$, respectively. Besides, the contributions originating
from higher resonances (about 15\% $\sim$ 32\% and 15\% $\sim$ 32\%
for the former and latter ones, respectively) and the highest OPE
terms (zero and 5\% $\sim$ 2\% for the former and latter ones,
respectively) are well under control.

\begin{figure}[t!]
\epsfxsize=14.5cm \centerline{\epsffile{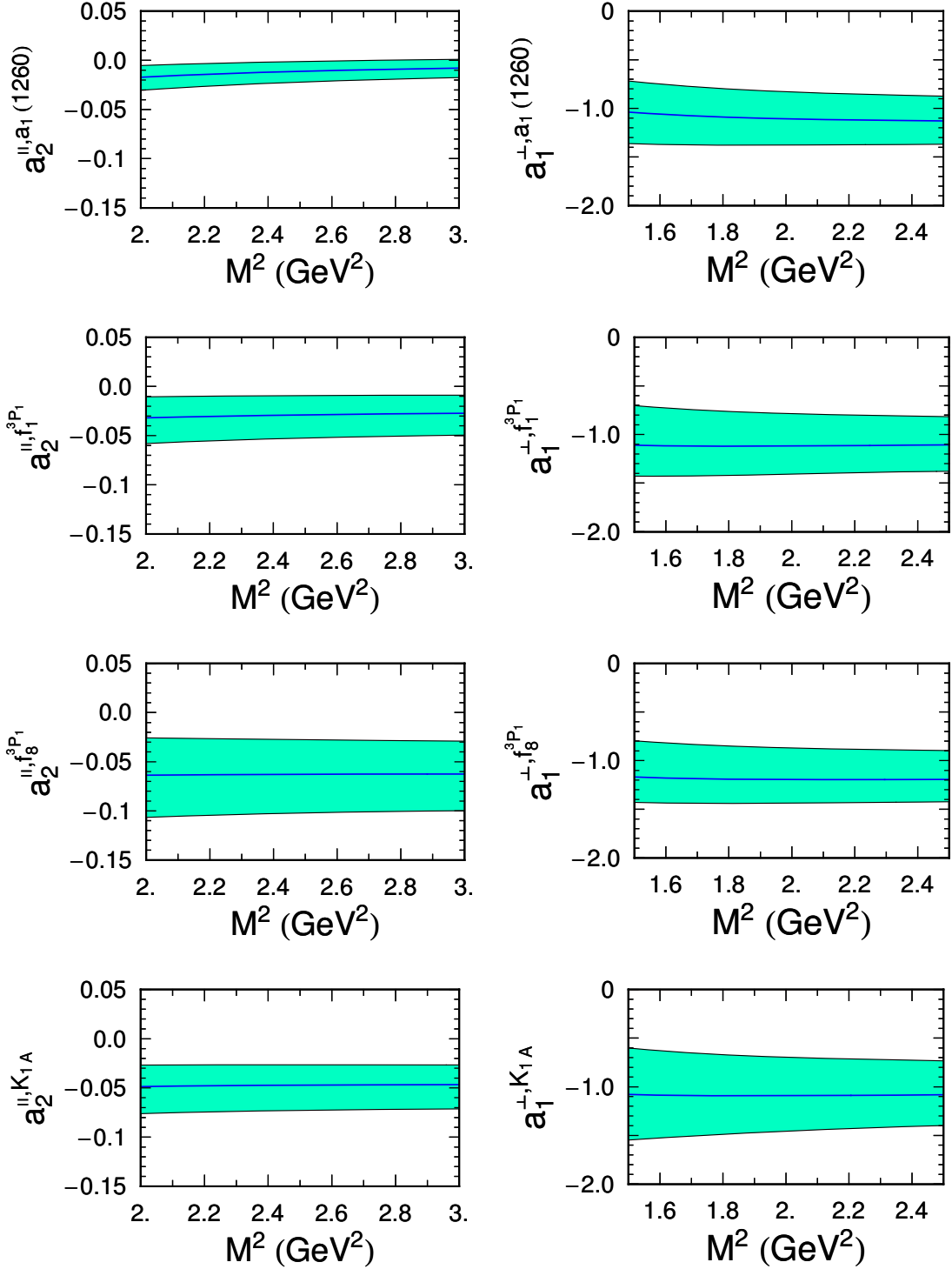}}
\centerline{\parbox{14cm}{\caption{\label{fig:a2-g-invariant}
G-parity invariant Gegenbauer moments of leading twist LCDAs for
$1^3P_1$ states, corresponding to the scale $\mu=1$~GeV, as
functions of the Borel mass squared. The solid curves and bands
correspond to the central values and uncertainties of the input
parameters, respectively.}}}
\end{figure}

\begin{figure}[t!]
\epsfxsize=14.5cm \centerline{\epsffile{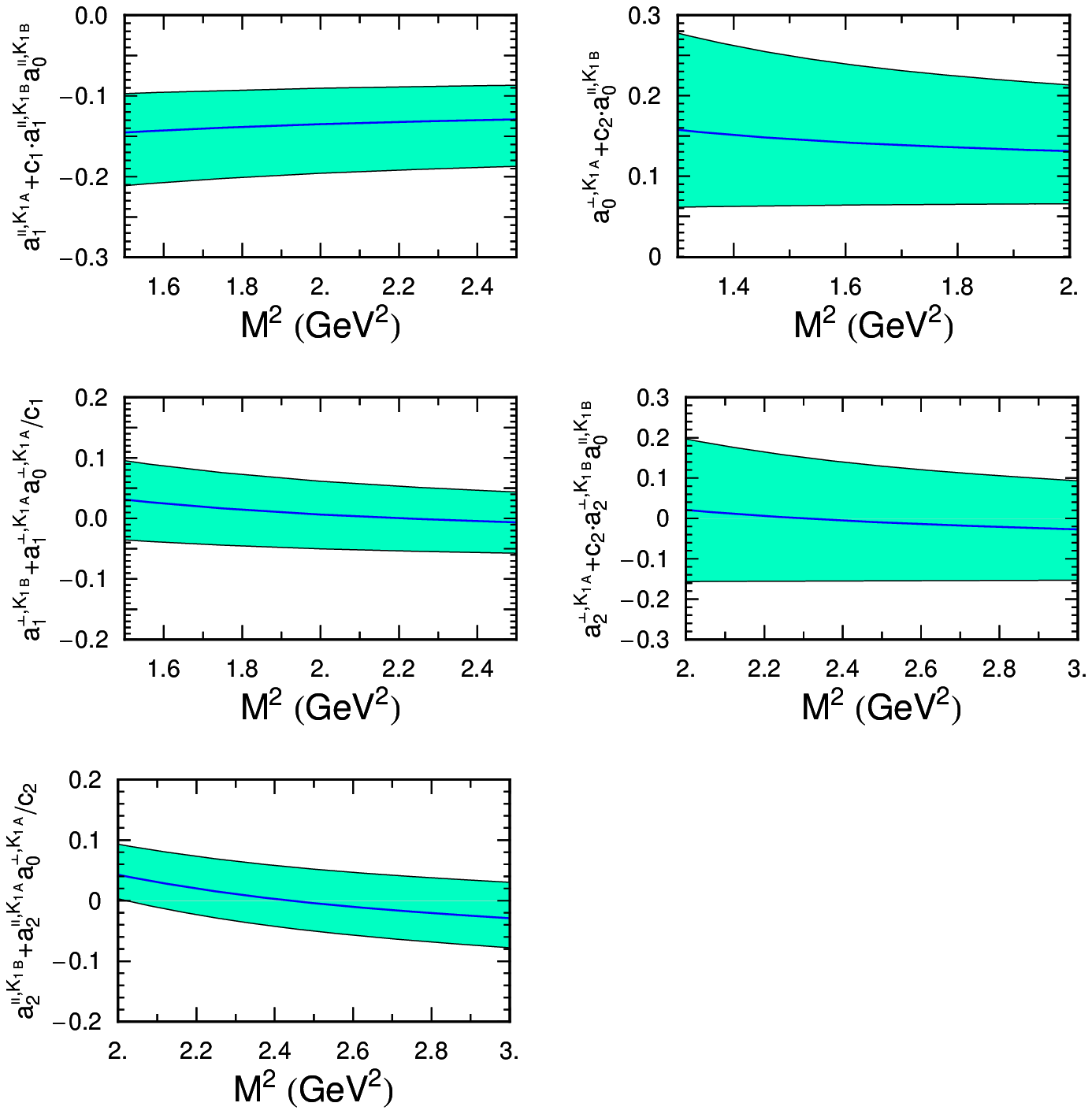}}
\centerline{\parbox{14cm}{\caption{\label{fig:a2-g-violating}
G-parity violating Gegenbauer moments of leading twist LCDAs,
corresponding to the scale $\mu=1$~GeV, as functions of the Borel
mass squared, where $c_1= f_{K_{1B}}^2/ f_{K_{1A}}^2$ and
$c_2=(m_{K_{1B}}/m_{K_{1A}}) (f_{K_{1B}}^\perp f_{K_{1B}}/
f_{K_{1A}}^\perp f_{K_{1A}})$. The solid curves and bands correspond
to the central values and uncertainties of the input parameters,
respectively.}}}
\end{figure}

For G-parity invariant parameters involving quark mass corrections,
because the results may be sensitive to the nonperturbative
parameters for which we have considered larger uncertainties of the
parameters here (see Appendix~\ref{appsec:inputs}), we therefore
re-examine the errors of $1^1P_1$ states as well. The formulas for
$1^1P_1$ states were given in Ref.~\cite{Yang:2005gk}. The results
for G-parity invariant Gegenbauer moments for $1^1P_1$ and $1^3P_1$
states are summarized in Table~\ref{tab:Gegenbauer}. Note that the
results for $h_1$ and $h_8$ (the $1^1P_1$ states) are new, where use
of $a_2^{\perp,\omega_1}\approx a_2^{\perp,\omega_8}\approx
a_2^{\perp,\phi}$ has been made in the numerical analysis and
$a_2^{\perp,\phi}=0.0\pm 0.1$ \cite{Ball:2004rg}. To exhibit the
quality of the sum rules for the $1^3P_1$ results, the Gegenbauer
moments versus the Borel mass squared are plotted in
Figs.~\ref{fig:a2-g-invariant} and \ref{fig:a2-g-violating}.

For G-parity violating parameters, the Gegenbauer moments for
$K_{1A}$ and $K_{1B}$ mix as given in Eqs. (\ref{eq:mix-1}),
(\ref{eq:mix-2}) and (\ref{eq:mix-2-2}). On the other hand, the
G-parity violating quantities were not considered correctly in
Ref.~\cite{Yang:2005gk} owing to the mentioned mixtures and should
read
\begin{eqnarray}
&&
 a_1^{\perp,K_{1B}}
 + a_1^{\perp,K_{1A}} a_0^{\perp,K_{1A}}
 \frac{(f_{K_{1A}}^{\perp})^2}{(f_{K_{1B}}^\perp)^2}
  \simeq \frac{5}{9}\frac{1}{(f_{K_{1A}}^\perp)^2}
 e^{\bar m^2/M^2}
L^{\gamma_{(1)}^\parallel/b} \Bigg\{ -\frac{3}{M^2}\,
 (m_s\langle \bar q_2 q_2\rangle - m_{q_2}\langle \bar s s\rangle)
 \nonumber\\
 && ~~
 - \frac{2}{M^4}\,
 (m_{s}\langle \bar q_{2} g_s\sigma G q_2\rangle
  - m_{q_2}\langle \bar s g_s\sigma G s\rangle) L^{-14/(3b)}
  -\frac{9}{5} {(f_{K^*}^\perp)^2} a_1^{\perp,K^*}
 e^{-m_{K^*}^2/M^2}
 \Bigg\}, \label{eq:mix-3}
\end{eqnarray}
and
\begin{eqnarray}
& & a_2^{\parallel,K_{1B}}
 + a_2^{\parallel,K_{1A}} a_0^{\perp,K_{1A}}
 \frac{m_{K_{1A}} f_{K_{1A}}^\perp f_{K_{1A}}}
      {m_{K_{1B}} f_{K_{1B}}^\perp f_{K_{1B}}}
  \simeq -\frac{7}{18}\frac{1}{m_{K_{1B}} f_{K_{1B}}^\perp f_{K_{1B}}}
 e^{\bar m^2/M^2}
L^{\gamma_{(2)}^\perp/b} \nonumber\\
&&\ \ \ \ \
 \times  \Bigg\{ \frac{3 }{8\pi^2 }M^2 (1-e^{-s_0/M^2})
 ( m_{q_2}- m_{s})  L^{-4/b}
 - 6(\langle\bar q_2 q_2\rangle - \langle \bar ss\rangle) L^{4/b}
 \nonumber\\
 && ~~~~~~~~
 -12\frac{\langle \bar q_2 g_s \sigma\cdot G q_2\rangle
   - \langle \bar s g_s \sigma\cdot G s
 \rangle}{M^2}  L^{-2/(3b)}
 -\frac{25}{3} \frac{\pi^2}{M^4}\langle \frac{\alpha_s}{\pi} G^2\rangle
 (\langle \bar q_2 q_2\rangle -\langle \bar ss\rangle)   L^{4/b}
 \Bigg\}\,.
 \qquad
 \label{eq:mix-3-2}
\end{eqnarray}
We use the updated values $a_1^{\perp,K^*}(1\ {\rm GeV})=0.04\pm
0.03$ \cite{Ball:2005vx} and $f_{K^*}^\perp =(0.185\pm 0.010)$~GeV
\cite{Ball:2005vx} in the numerical analysis. Eqs.~(\ref{eq:mix-1}),
(\ref{eq:mix-2}), (\ref{eq:mix-2-2}), (\ref{eq:mix-3}), and
(\ref{eq:mix-3-2}) do not offer sufficient relations to obtain
explicit solutions for G-parity violating parameters but can give
the following relations (at the scale $\mu=$ 1GeV):
\begin{eqnarray}
  a_1^{\parallel,K_{1A}}
 -(1.10 \pm0.30)\, a_0^{\parallel,K_{1B}}
  &=&  -0.15\pm 0.06\,, \nonumber\\
 a_0^{\perp,K_{1A}}
 + (0.59 \pm 0.15)\,  a_0^{\parallel,K_{1B}}
  &=& 0.17\pm 0.11\,,  \nonumber\\
a_1^{\perp,K_{1B}}
 - (1.87 \pm 0.56)\, a_0^{\perp,K_{1A}}
 &=& 0.02\pm 0.08\,, \nonumber\\
 a_2^{\perp,K_{1A}}
 - (0.01 \pm 0.16)\,  a_0^{\parallel,K_{1B}}
  &=& 0.02\pm 0.18\,,  \nonumber\\
  a_2^{\parallel,K_{1B}}
 - (0.08\pm 0.07)
  a_0^{\perp,K_{1A}}
 &=&  0.01 \pm 0.09\,,
\end{eqnarray}
where we have substituted the numerical values of mesons' masses and
decay constants given in Tables~\ref{tab:mass-decay-constant-3p1}
and \ref{tab:mass-decay-constant-1p1}. If the above G-parity
violating parameters are expected to be small as the results for $K$
and $K^*$ (for instance, see results in Ref.~\cite{Ball:2005vx}), it
is preferred that $a_0^{\perp,K_{1A}}$ and $a_0^{\parallel,K_{1B}}$
have the same positive sign. On the other hand, comparing
Eq.~(\ref{eq:Gmoment_alperpSR}) in this paper and Eq.~(3.23) in
Ref.~\cite{Yang:2005gk}, we obtain the good approximation between
G-parity invarint Gegenbauer moments (with odd $l$) of $1^3P_1$ and
$1^1P_1$ states:
\begin{eqnarray}
 a_l^{\perp,^3P_1}
 \simeq  a_l^{\parallel,^1P_1}
 \frac{m_{^1P_1} f_{^1P_1} f_{^1P_1}^\perp}
  { m_{^3P_1} f_{^3P_1} f_{^3P_1}^\perp},
\end{eqnarray}
which can be reconfirmed from the results in
Table~\ref{tab:Gegenbauer}. Therefore, we further assume that the
G-parity violating Gegenbauer moments satisfy a similar relation
with an enlarged uncertainty range:
\begin{eqnarray}\label{eq:constraint}
 \frac{a_0^{\perp,K_{1A}}}
  { a_0^{\parallel,K_{1B}}
 \frac{m_{K_{1B}} f_{K_{1B}} f_{K_{1B}}^\perp}
  { m_{K_{1A}} f_{K_{1A}} f_{K_{1A}}^\perp}}
 =1.0 \pm 0.3.
\end{eqnarray}
From the above estimate, it follows
\begin{eqnarray}
     a_0^{\perp,K_{1A}}
 &=&  0.08 \pm 0.09\,,~ \qquad
     a_0^{\parallel,K_{1B}}=0.14 \pm 0.15\,, \\
   a_1^{\parallel,K_{1A}}
 &=&  0.00\pm 0.26,  \qquad
   a_1^{\perp,K_{1B}} = 0.17 \pm 0.22\,,  \\
 a_2^{\perp,K_{1A}}
  &=& 0.02\pm 0.20,~ \qquad a_2^{\parallel,K_{1B}}
 = 0.02 \pm 0.10\,,
\end{eqnarray}
which are collected in Table~\ref{tab:Gegenbauer}.

 Finally, four remarks are in order. First, we will simply take
$f_{^3P_1}^\perp = f_{^3P_1}$, which is independent of the scale, in
the study since only the products of $f_{^3P_1}^\perp
a_l^{\perp,^3P_1}$ are relevant. Second, the sum rules obtained from
the nondiagonal correlation functions in
Eq.~(\ref{eq:Gmomentperp-t2}) can also determine the sign of
$f_{^3P_1}^\perp a_1^{\perp,^3P_1}$ relative to $f_{^3P_1}$. Third,
for the present case, the RG effects are relatively small compared
with the uncertainties of input parameters. Fourth, neglecting the
small isospin violation but considering the SU(3)-breaking
correction, $a_1^\parallel, a_3^\parallel$ $a_0^\perp$ and
$a_2^\perp$ are nonzero only for $K_{1A}$. Note that here we adopt
the convention that $q_1\equiv s$ for $K_{1A}$ and $K_{1B}$. For
$K_{1A}$ ($K_{1B}$) containing an $\bar s$ quark, we have the
following replacements $a_{1}^\parallel, a_{0,2}^\perp \to
-a_{1}^\parallel, -a_{0,2}^\perp$ ($a_{0,2}^\parallel, a_{1}^\perp
\to -a_{0,2}^\parallel, -a_{1}^\perp$) for G-parity violating
parameters.

%
\begin{table}[htb!]
\caption[]{Gegenbauer moments of $\Phi_\perp$ and $\Phi_\parallel$
for $1^3P_1$ and $1^1P_1$ mesons, where uses of $f_{^3P_1}^\perp =
f_{^3P_1}$ and $f_{^1P_1} = f_{^1P_1}^\perp(1~{\rm GeV}) $ have been
made. For $1^1P_1$ states, the results for $h_1$ and $h_8$ are new,
for $a_{0,2}^{\parallel, K_{1B}}$ and $a_1^{\perp, K_{1B}}$ are
corrected, and for the rest are updated.} \label{tab:Gegenbauer}
\renewcommand{\arraystretch}{1.8}
\addtolength{\arraycolsep}{0.4pt}
{\small
$$
\begin{array}{|c|c|c|c|c|c|c|}\hline
 \mu & a_2^{\parallel, a_1(1260)}& a_2^{\parallel,f_1^{^3P_1}}
 & a_2^{\parallel,f_8^{^3P_1}} & a_2^{\parallel, K_{1A}}
 & \multicolumn{2}{|c|}{a_1^{\parallel, K_{1A}}}
 \\ \hline
\begin{array}{c} {\rm 1~GeV}   \\ {\rm 2.2~GeV} \end{array}&
\begin{array}{c} -0.02\pm 0.02 \\ -0.01\pm 0.01  \end{array}&
\begin{array}{c}  -0.04\pm 0.03 \\ -0.03 \pm 0.02 \end{array}&
\begin{array}{c} -0.07\pm 0.04\\ -0.05 \pm  0.03 \end{array}&
\begin{array}{c} -0.05\pm 0.03\\  -0.04 \pm 0.02 \end{array}&
\multicolumn{2}{c|}{\begin{array}{c} {0.00\pm 0.26}\\ 0.00\pm
0.22\end{array}}
\\ \hline\hline
 \mu & a_1^{\perp, a_1(1260)}& a_1^{\perp,f_1^{^3P_1}}
 & a_1^{\perp,f_8^{^3P_1}} & a_1^{\perp, K_{1A}}
 & a_0^{\perp, K_{1A}}    & a_2^{\perp, K_{1A}}
 \\ \hline
\begin{array}{c} {\rm 1~GeV}   \\ {\rm 2.2~GeV} \end{array}&
\begin{array}{c} -1.04\pm 0.34 \\ -0.85\pm 0.28  \end{array}&
\begin{array}{c} -1.06\pm 0.36 \\ -0.86\pm 0.29   \end{array}&
\begin{array}{c} -1.11\pm 0.31 \\ -0.90\pm 0.25   \end{array}&
\begin{array}{c} -1.08\pm 0.48 \\ -0.88\pm 0.39   \end{array}&
\begin{array}{c}  0.08\pm 0.09\\   0.07\pm 0.08   \end{array}&
\begin{array}{c}  0.02\pm 0.20\\   0.01\pm 0.15   \end{array}
\\ \hline\hline
\mu & a_1^{\parallel, b_1(1235)}& a_1^{\parallel,h_1^{^1P_1}}
 & a_1^{\parallel,h_8^{^1P_1}} & a_1^{\parallel, K_{1B}}
 & a_0^{\parallel, K_{1B}}    & a_2^{\parallel, K_{1B}}
 \\ \hline
\begin{array}{c} {\rm 1~GeV}   \\   {\rm 2.2~GeV} \end{array}&
\begin{array}{c} -1.95\pm 0.35 \\  -1.61\pm 0.29  \end{array}&
\begin{array}{c} -2.00\pm 0.35 \\  -1.65\pm 0.29  \end{array}&
\begin{array}{c} -1.95\pm 0.35 \\  -1.61\pm 0.29  \end{array}&
\begin{array}{c} -1.95\pm 0.45 \\  -1.61\pm 0.37  \end{array}&
\begin{array}{c}  0.14\pm 0.15 \\   0.14\pm 0.15  \end{array}&
\begin{array}{c}  0.02\pm 0.10 \\   0.01\pm 0.07  \end{array}
\\ \hline\hline
 \mu & a_2^{\perp, b_1(1235)}& a_2^{\perp,h_1^{^1P_1}}
 & a_2^{\perp,h_8^{^1P_1}} & a_2^{\perp, K_{1B}}
 & \multicolumn{2}{|c|}{a_1^{\perp, K_{1B}}}
 \\ \hline
\begin{array}{c}  {\rm 1~GeV}  \\   {\rm 2.2~GeV} \end{array}&
\begin{array}{c}  0.03\pm 0.19 \\   0.02\pm 0.15  \end{array}&
\begin{array}{c}  0.18\pm 0.22 \\  0.14 \pm 0.17  \end{array}&
\begin{array}{c}   0.14\pm 0.22\\  0.11 \pm  0.17 \end{array}&
\begin{array}{c}  -0.02\pm 0.22\\ -0.02 \pm 0.17  \end{array}&
\multicolumn{2}{c|}{\begin{array}{c}0.17\pm 0.22\\
                                    0.14\pm 0.18\end{array}}
\\ \hline
\end{array}
$$}
\end{table}
%

%

\section{Determinations of three-parton LCDAs of twist-three}\label{sec:properties-3pdas}

In this section, using the QCD sum rule approach, we estimate the
relevant parameters involving SU(3)-breaking effects in
determinations of the three-parton LCDAs of twist-three. The quark
masses (SU(3)-breaking effects), can give contributions not only to
G-parity invariant conformal moments but also G-parity violating
ones of LCDAs. In the calculation, the OPE series in the QCD sum
rules are evaluated up to dimension-seven, but up to dimension-four
for terms proportional to the quark masses.

The parameters can be calculated from the standard QCD sum rule
approach by adopting a relevant two-point correlation function as
the starting point. The choices of suitable interpolating currents
may affect the qualities of the final estimates. Note again that in
SU(3) limit the decay constant for a $1 ^3P_1$ state transiting to
the vacuum via the local (pseudo-)tensor current vanishes due to the
G-parity mismatch between the current and states, whereas the
axial-vector decay constant vanishes for a $1 ^1P_1$ meson.
Motivated from the above properties and to determine not only the
magnitudes but also the relative signs for parameters relevant to
three-parton LCDAs of twist-3, we thus choose one of the
interpolating currents in the two-point correlation functions to be
the local axial-vector (or pseudo-tensor) current to calculate the
parameters for the $1 ^3P_1$ (or $1 ^1P_1$) state, i.e., we consider
the non-diagonal correlation functions here. We did not consider the
diagonal correlation functions in the beginning since it cannot
determine the sign~\footnote{\label{foot:dia} It is a little
different for the definition of the diagonal correlation function
here and in Ref.~\cite{Ball:2006wn}. In Ref.~\cite{Ball:2006wn} and
related studies, the authors choose the non-local light-ray
operators in calculating the correlation functions. According to
their definition, although some parameters' signs can be determined
in the diagonal sum rules which are actually non-diagonal in our
definition, the relative signs of $f_{3,A}^V,f_{3,A}^A$ and
$f_{3,A}^\perp$ still cannot be established.} of the parameters and
cannot be used to evaluate the G-parity violating parameters due to
the mixes of the $1 ^3P_1$ and $1 ^1P_1$ states.

Unlike results from diagonal correlation functions, where the
perturbative contribution may dominate in the OPE series, our
results show that the leading contributions are dominated by the
term with the quark or gluon condensate in the most cases, and we
may need to take into account a higher Borel window, so that the
contribution originating from the highest dimension term in the OPE
expansion can be well under control. However, a problem occurs in
the $f_{3, ^3P_1}^\perp$ study for which the quark condensate
contributions vanish in ${\cal O}(\alpha_s)$ after adding all the
diagrams, and therefore the OPE result become highly unreliable (see
the details in Sec. \ref{sec:properties-3pdas-1-3}). We therefore
resort to the diagonal sum rule for pursuing this parameter. However
the lowest lying pseudoscalar meson contributes to the diagonal sum
rule. Fortunately, although the pseudoscalar meson contribution is
involved in the diagonal sum rule, such effects can be subtracted by
using the non-diagonal sum rule for the pseudoscalar meson. These
two sum rules have been studied in Ref. \cite{Ball:2006wn} for
calculating the twist-3 parameter $f_{3K}$ relevant to the twist-3
three-parton LCDA of the kaon. Instead, we use these two sum rules
to extract the value of $f_{3, ^3P_1}^\perp$.

We study the parameters relevant to twist-3 three-parton LCDAs for
$1 ^3P_1$ and $1 ^1P_1$ states in Secs. \ref{sec:properties-3pdas-1}
and \ref{sec:properties-3pdas-2}, respectively. One should note
that, as calculating the sum rules for G-parity symmetric parameters
of twist-3 three-parton LCDAs of $1^3P_1$ ($1^1P_1$) mesons, the
G-parity violating parameters for $1^1P_1$ ($1^3P_1$) states,
relatively suppressed by ${\cal O}(m_q^2)$, contribute to them but
can be negligible. However, the sum rules for G-parity violating
parameters for $1^3P_1$ ($1^1P_1$) mesons receive the contributions
arising from G-parity symmetric parameters about $1^1P_1$ ($1^3P_1$)
states, which are of the same order of magnitude because the
axial-vector decay constants (pseudo-tensor decay constants) for
$1^1P_1$ ($1^3P_1$) states do not vanish due to the quark mass
corrections. The numerical results are given in
Sec.~\ref{sec:3pdas-3}. To LO approximation and including the quark
mass corrections, the complete RG evolutions of the parameters are
given in Appendix \ref{appsec:inputs}.

\subsection{Axial-vector mesons with quantum number
$1^3P_1$}\label{sec:properties-3pdas-1}

\subsubsection{$f^V_{3,^3\! P_1}$, $\omega_{^3\! P_1}^V$ and
 $\sigma_{^3\! P_1}^V$}\label{sec:properties-3pdas-1-1}

The coupling constants $f^V_{3,^3\! P_1}$, $\omega_{^3\! P_1}^V$ and
$\sigma_{^3\! P_1}^V$ can be obtained by considering the correlation
functions,
\begin{equation}
i \int d^4x\, e^{iqx}\,\langle 0 | T\Big\{J^{3,V}_{1,\, \mu}(x),\,
\bar q_1(0) \gamma_\nu\gamma_5 q_2(0) \Big\}|0\rangle =
 -T^V_{^3P_1}(q^2)\, (q\bar{z})^2 g_{\mu\nu}^\perp +\cdots,
\end{equation}
\begin{eqnarray}
i \int d^4x\, e^{iqx}\,\langle 0 | T\Big\{J^{3,V}_{2,\, \mu}(x),\,
\bar q_1(0) \gamma_\nu\gamma_5 q_2(0) \Big\}|0\rangle
 = -T^{V,\alpha_g}_{^3P_1}(q^2)\, (q\bar{z})^3 g_{\mu\nu}^\perp
+\cdots,
\end{eqnarray}
\begin{eqnarray}
i \int d^4x\, e^{iqx}\,\langle 0 | T\Big\{J^{3,V}_{3,\, \mu}(x),\,
\bar q_1(0) \gamma_\nu\gamma_5 q_2(0) \Big\}|0\rangle
 = -T^{V,\sigma}_{^3P_1}(q^2)\, (q\bar{z})^3 g_{\mu\nu}^\perp +\cdots,
\end{eqnarray}
where the currents are defined as
\begin{eqnarray}
 J^{3,V}_{1,\, \mu}(0)
 &=& \bar{z}^\alpha \bar{z}^\beta
   \bar q_2(0) \gamma_\alpha g_s \widetilde{G}_{\beta\mu}(0) q_1(0)\,,
   \label{eq:c1-1}\\
 J^{3,V}_{2,\, \mu}(0)
 &=& \bar{z}^\alpha \bar{z}^\beta
   \bar q_2(0) \gamma_\alpha g_s [iD\bar{z}\widetilde{G}_{\beta\mu}(0)]
   q_1(0)\,,\label{eq:c1-2}\\
 J^{3,V}_{3,\, \mu}(0)
 &=& \bar{z}^\alpha \bar{z}^\beta
   \bar q_2(0) \gamma_\alpha g_s i\bar{z}^\delta
 [\widetilde{G}_{\beta\mu}(0)\derright_\delta
 -\derleft_\delta \widetilde{G}_{\beta\mu}(0)] q_1(0)\,,
 \label{eq:c1-3}
\end{eqnarray}
which can couple to the $1^3P_1$ states as
 \begin{eqnarray}
 && \langle 0| J^{3,V}_{1,\, \mu}(0)| 1^3P_1(P,\lambda)\rangle
 = -i f^V_{3,^3\! P_1}(P\bar{z})^2 \epsilon^{(\lambda)}_{\perp,\mu}
 +{\cal O}(\bar{z}_\mu)\,,\\
 && \langle 0|J^{3,V}_{2,\, \mu}(0)|1^3P_1(P,\lambda)\rangle
 =-i f^V_{3,^3\! P_1} \langle \alpha_g^{V}\rangle
 (P\bar{z})^3  \epsilon^{(\lambda)}_{\perp,\mu}
 +{\cal O}(\bar{z}_\mu)\,,\\
 &&  \langle 0|J^{3,V}_{3,\, \mu}(0)|1^3P_1(P,\lambda)\rangle
 = -i f^V_{3,^3\! P_1} \sigma_{^3\! P_1}^V  (P\bar{z})^3 \epsilon^{(\lambda)}_{\perp,\mu}
 +{\cal O}(\bar{z}_\mu)\,,
 \end{eqnarray}
with ${\cal O}(\bar{z}_\mu)$ including the twist-4 correction and
the average gluon momentum fraction $\langle \alpha_g^V\rangle$
satisfying
\begin{equation}
 \langle \alpha_g^V\rangle
 = \frac{3}{7} +\frac{3}{28}\omega_{^3\! P_1}^V\,.
 \end{equation}
Here and below the ellipses denote terms irrelevant to the present
studies. It should be note that $T^{V}_{^3P_1}$, $T^{V,
\alpha_g}_{^3P_1}$, and $T^{V, \sigma}_{^3P_1}$ receive no
contributions from pseudoscalar states.

Assuming the quark-hadron duality, we can obtain the approximate
expressions
\begin{eqnarray}
\frac{1}{m_{^3\! P_1}^2-q^2}m_{^3\! P_1} f_{^3\! P_1}
 f_{3\, ^3\! P_1}^V
 =\frac{1}{\pi}\int_0^{s_{0}^{^3P_1}} ds
 \frac{{\rm Im}\, T^{V,\rm OPE}_{^3P_1}(s)}{s-q^2},
\end{eqnarray}
\begin{equation}
\frac{1}{m_{^3\! P_1}^2-q^2} m_{^3\! P_1} f_{^3\! P_1}
 f_{3, ^3\! P_1}^V \langle \alpha_g^V \rangle
 =\frac{1}{\pi}\int_0^{s_0^{^3\! P_1}}ds
 \frac{{\rm Im}\, T^{V,\alpha_g,\rm OPE}_{^3P_1}(s)}{s-q^2},
 \end{equation}
 and
 \begin{equation}
\frac{1}{m_{^3\! P_1}^2-q^2} m_{^3\! P_1} f_{^3\! P_1}
 f_{3, ^3\! P_1}^V \sigma_{^3\! P_1}^V
 =\frac{1}{\pi}\int_0^{s_0^{^3\! P_1}}ds
 \frac{{\rm Im}\, T^{V,\sigma,\rm OPE}_{^3P_1}(s)}{s-q^2},
 \end{equation}
where
 \begin{eqnarray}
 T^{V, {\rm OPE}}_{^3P_1}(q^2)&=&
 - \frac{\alpha_s}{144\pi^3}q^2\ln\frac{-q^2}{\mu^2}
 -\frac{1}{24 \pi q^2}\, \langle \alpha_s G^2\rangle
 + \frac{8\pi\alpha_s}{27 q^4}
 \bigg(\langle \bar q_1 q_1\rangle^2 + \langle\bar q_2 q_2\rangle^2
 -3 \langle \bar q_1 q_1\rangle \langle\bar q_2 q_2\rangle \bigg)
 \nonumber\\
 && \! \! \!\!\!\!\!\ +\frac{\alpha_s}{36\pi q^2}
  \Bigg[5 \bigg(m_1\langle \bar q_1 q_1\rangle+m_2\langle \bar q_2 q_2\rangle\bigg)
  +4\bigg(m_1\langle \bar q_2 q_2\rangle+m_2\langle \bar q_1 q_1\rangle\bigg)
  \Bigg(\ln\frac{-q^2}{\mu^2}-\frac{17}{6}\Bigg)\Bigg],\nonumber\\
 \end{eqnarray}
\begin{eqnarray}
 T^{V,\alpha_g, {\rm OPE}}_{^3P_1} &=&
 - \frac{\alpha_s}{120\pi^3}q^2\ln\frac{-q^2}{\mu^2}
 + (0+{\cal O}(\alpha_s)) \langle\alpha_s G^2\rangle
 - \frac{8\pi\alpha_s}{9 q^4} \langle \bar q_1 q_1\rangle
 \langle\bar q_2 q_2\rangle\nonumber\\
 &&
 +\frac{\alpha_s}{12\pi q^2}
  \Bigg[\frac{1}{2}
  \bigg(m_1\langle \bar q_1 q_1\rangle+m_2\langle \bar q_2 q_2\rangle\bigg)
 +\bigg(m_1\langle \bar q_2 q_2\rangle+m_2\langle \bar q_1 q_1\rangle\bigg)
  \Bigg(\ln\frac{-q^2}{\mu^2}-\frac{11}{3}\Bigg)\Bigg].\nonumber\\
\end{eqnarray}
and
\begin{eqnarray}
 T^{V, \sigma, {\rm OPE}}_{^3P_1}(q^2)&=&
 \frac{-\alpha_s}{36\pi q^2}
  \Bigg[\frac{7}{2} \bigg(m_1\langle \bar q_1 q_1\rangle-m_2\langle \bar q_2 q_2\rangle\bigg)
  -\bigg(m_1\langle \bar q_2 q_2\rangle-m_2\langle \bar q_1 q_1\rangle\bigg)
  \Bigg(\ln\frac{-q^2}{\mu^2}-\frac{11}{3}\Bigg)\Bigg]\nonumber\\
  && +0 \cdot \langle \alpha_s G^2\rangle
   + 0 \cdot \langle \bar q_1 q_1\rangle \langle \bar q_2 q_2\rangle
  +(0+{\cal O}(\alpha_s^2))
 \bigg(\langle \bar q_1 q_1\rangle^2, \langle \bar q_2
  q_2\rangle^2\bigg).
 \end{eqnarray}
Note that $T^{V}_{^3P_1}$, $T^{V, \alpha}_{^3P_1}$, and $T^{V,
\sigma}_{^3P_1}$ receive contributions from $^1P_1$ states because
$^1P_1$ states have small axial-vector coupling constants due to
$m_{q_1}-m_{q_2}\neq 0$. Consequently, after performing the Borel
transform, we obtain the sum rules:
\begin{eqnarray}
 && e^{-m_{^3\! P_1}^2/M^2} m_{^3\! P_1} f_{^3\! P_1} f^V_{3,^3\! P_1}
 + e^{-m_{^1\! P_1}^2/M^2} m_{^1\! P_1} f_{^1\! P_1} a_0^{\parallel,1^1P_1}
 f^V_{3,^1\! P_1}\lambda_{^1P_1}^V
 \nonumber\\
 & & \ =
\frac{\alpha_s}{144\pi^3} \int_0^{s_0^{^3\! P_1}}s\, e^{-s/M^2} ds
 +\frac{1}{24 \pi}\, \langle \alpha_s G^2\rangle
 + \frac{8\pi\alpha_s}{27 M^2}
 \bigg(\langle \bar q_1 q_1\rangle^2 + \langle\bar q_2 q_2\rangle^2
 -3 \langle \bar q_1 q_1\rangle \langle\bar q_2 q_2\rangle \bigg)
 \nonumber\\
 && \ \ \ \ \ -\frac{\alpha_s}{36\pi}
  \Bigg[5 \bigg(m_1\langle \bar q_1 q_1\rangle
    +m_2\langle \bar q_2 q_2\rangle\bigg)\nonumber\\
  &&\ \ \ \ \
  -4\bigg(m_1\langle \bar q_2 q_2\rangle+m_2\langle \bar q_1 q_1\rangle\bigg)
  \Bigg(\frac{17}{6}+\gamma_E +
 \ln\frac{\mu^2}{M^2}-{\rm Ei}\Bigg(-\frac{s_0}{M^2}\Bigg)\Bigg)
 \Bigg],
 \label{eq:3P1-SR-f3V}
\end{eqnarray}
\begin{eqnarray}
&& e^{-m_{^3\! P_1}^2/M^2} m_{^3\! P_1} f_{^3\! P_1}
 f^V_{3,^3\! P_1}
\Bigg(\frac{3}{7} +\frac{3}{28}\omega_{^3\! P_1}^V\Bigg)
 +e^{-m_{^1\! P_1}^2/M^2} m_{^1\! P_1} f_{^1\! P_1} a_0^{\parallel,1^1P_1}
 f^V_{3,^1\! P_1}
\Bigg(\frac{3}{7} \lambda_{^1\!P_1}^V+\frac{3}{28}\sigma_{^1\!
P_1}^V\Bigg)
\nonumber\\
 && ~~~= \frac{\alpha_s}{120\pi^3} \int_0^{s_0^{^3\! P_1}}s \,
e^{-s/M^2} ds - \frac{8\pi\alpha_s}{9 M^2} \langle \bar q_1
q_1\rangle
 \langle\bar q_2 q_2\rangle
+\frac{\alpha_s}{12\pi}
  \Bigg[-\frac{1}{2}
  \bigg(m_1\langle \bar q_1 q_1\rangle+m_2\langle \bar q_2 q_2\rangle\bigg)
 \nonumber\\
 && ~~~\ \ \
 +\bigg(m_1\langle \bar q_2 q_2\rangle+m_2\langle \bar q_1 q_1\rangle\bigg)
  \Bigg( \frac{11}{3} +\gamma_E +
 \ln\frac{\mu^2}{M^2}-{\rm Ei}\Bigg(-\frac{s_0}{M^2}\Bigg)\Bigg)
 \Bigg]\,,
 \label{eq:3P1_SRf3V_omega}
\end{eqnarray}
\begin{eqnarray}
 && e^{-m_{^3\! P_1}^2/M^2} m_{^3\! P_1} f_{^3\! P_1} f^V_{3,^3\! P_1}
 \, \sigma_{^3\! P_1}^V
 + e^{-m_{^1\! P_1}^2/M^2} m_{^1\! P_1} f_{^1\! P_1} a_0^{\parallel,1^1P_1}
 f^V_{3,^1\! P_1}\nonumber\\
 && ~~~~~~~~= \frac{\alpha_s}{36\pi}
  \Bigg[\frac{7}{2} \bigg(m_1\langle \bar q_1 q_1\rangle
    - m_2\langle \bar q_2 q_2\rangle\bigg)\nonumber\\
  &&~~~~~~~~~~~~
  + \bigg(m_1\langle \bar q_2 q_2\rangle - m_2\langle \bar q_1 q_1\rangle\bigg)
  \Bigg(\frac{11}{3}+\gamma_E +
 \ln\frac{\mu^2}{M^2}-{\rm Ei}\Bigg(-\frac{s_0}{M^2}\Bigg)\Bigg)
 \Bigg].\ \ \
 \label{eq:3P1-SR-f3V-sigma}
\end{eqnarray}
As for Eqs. (\ref{eq:3P1-SR-f3V}) and (\ref{eq:3P1_SRf3V_omega}),
the contributions, arising from G-parity breaking parameters
relevant to twist-3 three-parton LCDAs of the $1^1\!P_1$ state, are
relatively suppressed by ${\cal O}(m_q^2)$ and can be negligible.
Nevertheless, in Eq.~(\ref{eq:3P1-SR-f3V-sigma}), the contribution
originating from the $1^1\!P_1$ state is of the same order of
magnitude as the G-parity breaking parameter relevant to the
$1^3\!P_1$ state, and should be taken into account. Similar
situations occur in Eqs. (\ref{eq:3P1_SRf3A}),
(\ref{eq:3P1_SRf3A_lambda}), (\ref{eq:3P1_SRf3A_sigma}),
(\ref{eq:3P1-SR-f3T}), (\ref{eq:3P1-SR-f3T-omega}), and
(\ref{eq:3P1-SR-f3T-sigma}) in the following subsections
\ref{sec:properties-3pdas-1-2} and \ref{sec:properties-3pdas-1-3}.

\subsubsection{$f^A_{3, ^3\! P_1}$, $\lambda^A_{^3\! P_1}$, and $\sigma^A_{^3\!
P_1}$}\label{sec:properties-3pdas-1-2}

 The constants $f^A_{3, ^3\! P_1}$,
$\lambda^A_{^3\! P_1}$, and $\sigma^A_{^3\! P_1}$ for $1^3P_1$
mesons can be defined as the following matrix elements
\begin{eqnarray}
 \langle 0 |J^{3,A}_{3,\, \mu}(0)| A(P, \lambda)\rangle
 & = & -f_{3,^3\! P_1}^A\, (P\bar{z})^3
\epsilon^{(\lambda)}_{\perp,\mu} +{\cal O}(\bar{z}_\mu)\,,
\\
 \langle0 | J^{3,A}_{1,\, \mu}(0)| A(P, \lambda)\rangle
 &=&
 -f_{3,^3\!P_1}^A\, \lambda^A_{^3\! P_1} (P\bar{z})^2
  \epsilon^{(\lambda)}_{\perp,\mu}  +{\cal O}(\bar{z}_\mu)\,,
\\
 \langle 0 | J^{3,A}_{2,\, \mu}(0)| A(P, \lambda)\rangle
  &=& -f_{3,^3\! P_1}^A\, \Bigg( \frac{3}{7} \lambda^A_{^3\! P_1}
 + \frac{3}{28}\sigma^A_{^3\! P_1} \Bigg)
 (P\bar{z})^3 \epsilon^{(\lambda)}_{\perp,\mu}  +{\cal O}(\bar{z}_\mu)\,,
\end{eqnarray}
where ${\cal O}(\bar{z}_\mu)$ involves the twist-4 correction and
the interpolating currents are
\begin{eqnarray}
 & & J^{3,A}_{1,\, \mu}(0)
 = \bar{z}^\alpha \bar{z}^\beta
 \bar q_2(0)\gamma_\alpha\gamma_5 g_s  G_{\beta\mu}(0) q_1(0)\,,
 \label{eq:c2-1}\\
 & & J^{3,A}_{2,\, \mu}(0)
  = \bar{z}^\alpha\bar{z}^\beta  \bar q_2(0)\gamma_\alpha\gamma_5 g_s
 \Big[iD\bar{z} G_{\beta\mu}(0)\Big] q_1(0) \,,
  \label{eq:c2-2}\\
 & & J^{3,A}_{3,\, \mu}(0)
  = \bar{z}^\alpha\bar{z}^\beta  \bar q_2(0)\gamma_\alpha\gamma_5 g_s
 [G_{\beta\mu}(0)i\derright z- i\derleft z G_{\beta\mu}(0)]
 q_1(0)\,.  \label{eq:c2-3}
 \end{eqnarray}
Thus, to evaluate $f^A_{3, ^3\! P_1}$, $\lambda^A_{3, ^3\! P_1}$,
and $\sigma^A_{3, ^3\! P_1}$, we consider the non-diagonal
correlation functions,
\begin{eqnarray}
 &&  \int d^4x\, e^{iqx}\,
  \langle 0 | T\Big\{J^{3,A}_{3,\, \mu}(x),\,
  \bar q_1(0) \gamma_{\nu}\gamma_5 q_2(0) \Big\}|0\rangle
  = T^{A}_{^3P_1}(q^2)\, (q\bar{z})^3 g_{\mu\nu}^\perp
    +\cdots,\\
 && \int d^4x\, e^{iqx}\,
  \langle 0 | T\Big\{J^{3,A}_{1,\, \mu}(x),\,
  \bar q_1(0) \gamma_{\nu}\gamma_5
q_2(0) \Big\}|0\rangle  =
  T^{A,\lambda}_{^3P_1}(q^2)\, (q\bar{z})^2 g_{\mu\nu}^\perp +\cdots,
 \\
 && \int d^4x\, e^{iqx}\,
  \langle 0 | T\Big\{J^{3,A}_{2,\, \mu}(x),\,
  \bar q_1(0)\gamma_{\nu}\gamma_5 q_2(0) \Big\}|0\rangle
  = T^{A,\sigma}_{^3P_1}(q^2)\, (q\bar{z})^3 g_{\mu\nu}^\perp +\cdots ,
\end{eqnarray}
where we get
\begin{eqnarray}
  T^{A,\rm OPE}_{^3P_1}(q^2)&=&
 - \frac{\alpha_s}{1440\pi^3}q^2\ln\frac{-q^2}{\mu^2}
 -\frac{1}{72 \pi q^2} \langle \alpha_s G^2\rangle
 +\frac{\alpha_s}{36\pi q^2}
 \Bigg[\frac{9}{2}\bigg(m_1\langle \bar q_1 q_1\rangle
  +m_2\langle \bar q_2 q_2\rangle\bigg)
 \nonumber\\
 && \ \ \ -\bigg(m_1\langle \bar q_2 q_2\rangle+m_2\langle \bar q_1 q_1\rangle\bigg)
  \Bigg(\ln\frac{-q^2}{\mu^2}+\frac{1}{3}\Bigg)\Bigg]\nonumber\\
  && +(0+{\cal O}(\alpha_s^2))
  \bigg(\langle \bar q_1 q_1\rangle \langle \bar q_2 q_2\rangle,
  \langle \bar q_1 q_1\rangle^2, \langle \bar q_2
  q_2\rangle^2\bigg), \\
   T^{A,\lambda,\rm OPE}_{^3P_1}(q^2)&=&
 \frac{\alpha_s}{36\pi q^2}
 \Bigg[5\bigg(m_1\langle \bar q_1 q_1\rangle -m_2\langle \bar q_2 q_2\rangle\bigg)
 -4\bigg(m_1\langle \bar q_2 q_2\rangle
  - m_2\langle \bar q_1 q_1\rangle\bigg)
  \Bigg(\ln\frac{-q^2}{\mu^2}-\frac{2}{3}\Bigg)\Bigg]
  \nonumber\\
 && \ \ \
 +0 \cdot \langle \alpha_s G^2\rangle  -\frac{8\pi\alpha_s}{27 q^4}
   \bigg(\langle \bar q_1 q_1\rangle^2 - \langle\bar q_2 q_2\rangle^2\bigg),
  \\
   T^{A,\sigma,\rm OPE}_{^3P_1}(q^2)&=&
   \frac{\alpha_s}{24\pi q^2}
   \Bigg[\bigg(m_1\langle \bar q_1 q_1\rangle -m_2\langle \bar q_2 q_2\rangle\bigg)
  -2\bigg(m_1\langle \bar q_2 q_2\rangle-m_2\langle \bar q_1 q_1\rangle\bigg)
  \Bigg(\ln\frac{-q^2}{\mu^2}-1\Bigg)\Bigg]\nonumber\\
  && +0 \cdot \langle \alpha_s G^2\rangle
   + 0 \cdot \langle \bar q_1 q_1\rangle \langle \bar q_2 q_2\rangle
  +(0+{\cal O}(\alpha_s^2))
 \bigg(\langle \bar q_1 q_1\rangle^2, \langle \bar q_2
  q_2\rangle^2\bigg).
\end{eqnarray}
From the above OPE results, it follows the sum rules:
\begin{eqnarray}
 && e^{-m_{^3\! P_1}^2/M^2} m_{^3\! P_1} f_{^3\! P_1}
 f_{3,^3\! P_1}^A
 + e^{-m_{^1\! P_1}^2/M^2} m_{^1\! P_1} f_{^1\! P_1} a_0^{\parallel,1^1P_1}
 f_{3,^1\! P_1}^A \sigma_{^1P_1}^A\nonumber\\
 &&\ \ \
 = \frac{\alpha_s}{1440\pi^3}\int_0^{s_0^{^3\! P_1}}s\, e^{-s/M^2} ds
  +\frac{1}{72\pi}\, \langle \alpha_s G^2\rangle
  +\frac{\alpha_s}{36\pi}
  \Bigg[-\frac{9}{2}\bigg(m_1\langle \bar q_1 q_1\rangle
   +m_2\langle \bar q_2 q_2\rangle\bigg)
 \nonumber\\
 && \ \ \ \ \ \
 +\bigg(m_1\langle \bar q_2 q_2\rangle+m_2\langle \bar q_1 q_1\rangle\bigg)
  \Bigg( \frac{1}{3} -\gamma_E -\ln\frac{\mu^2}{M^2}
    +{\rm Ei}\Bigg(-\frac{s_0}{M^2}\Bigg) \Bigg) \Bigg]\,,
\label{eq:3P1_SRf3A}
\end{eqnarray}
\begin{eqnarray}
 && e^{-m_{^3\! P_1}^2/M^2} m_{^3\! P_1}
 f_{^3\! P_1}\, f_{3,^3\! P_1}^A  \lambda^A_{^3\! P_1}
 + e^{-m_{^1\! P_1}^2/M^2} m_{^1\! P_1} f_{^1\! P_1} a_0^{\parallel,1^1P_1}
 f_{3,^1\! P_1}^A \nonumber\\
 &&\ \ \
 = -\frac{8\pi\alpha_s}{27 M^2}
   \bigg(\langle \bar q_1 q_1\rangle^2 - \langle\bar q_2 q_2\rangle^2\bigg)
  +\frac{\alpha_s}{36\pi}
  \Bigg[-5\bigg(m_1\langle \bar q_1 q_1\rangle
   - m_2\langle \bar q_2 q_2\rangle\bigg)
 \nonumber\\
 && \ \ \ \ \ \
 +4\bigg(m_1\langle \bar q_2 q_2\rangle - m_2\langle \bar q_1 q_1\rangle\bigg)
  \Bigg( -\frac{2}{3} -\gamma_E -\ln\frac{\mu^2}{M^2}
    +{\rm Ei}\Bigg(-\frac{s_0}{M^2}\Bigg) \Bigg) \Bigg]\,,
\label{eq:3P1_SRf3A_lambda}
\end{eqnarray}
and
\begin{eqnarray}
 && e^{-m_{^3\! P_1}^2/M^2} m_{^3\! P_1}
 f_{^3\! P_1}\, f_{3,^3\! P_1}^A  \Bigg( \frac{3}{7} \lambda^A_{^3\! P_1}
 + \frac{3}{28}\sigma^A_{^3\! P_1} \Bigg)
 + e^{-m_{^1\! P_1}^2/M^2} m_{^1\! P_1} f_{^1\! P_1} a_0^{\parallel,1^1P_1}
 f_{3,^1\! P_1}^A
 \Bigg( \frac{3}{7}
 + \frac{3}{28}\omega^A_{^1\! P_1} \Bigg)\nonumber\\
 &&\ \ \
 = \frac{\alpha_s}{24\pi}
  \Bigg[-\bigg(m_1\langle \bar q_1 q_1\rangle
   -m_2\langle \bar q_2 q_2\rangle\bigg)
 \nonumber\\
 && \ \ \ \ \ \
 +2\bigg(m_1\langle \bar q_2 q_2\rangle- m_2\langle \bar q_1 q_1\rangle\bigg)
  \Bigg( -1 -\gamma_E -\ln\frac{\mu^2}{M^2}
    +{\rm Ei}\Bigg(-\frac{s_0}{M^2}\Bigg) \Bigg) \Bigg]\,.
\label{eq:3P1_SRf3A_sigma}
\end{eqnarray}

\subsubsection{$f^{\perp}_{3,^3\! P_1}$,
$\omega_{^3\! P_1}^\perp$ and
 $\sigma_{^3\! P_1}^\perp$}\label{sec:properties-3pdas-1-3}

To evaluate the coupling constants $f^{\perp}_{3,^3\! P_1}$,
$\omega_{^3\! P_1}^\perp$ and $\sigma_{^3\! P_1}^\perp$ of $1^3P_1$
states, we first consider the following non-diagonal correlation
functions,
 \begin{equation}
 i\int d^4x\, e^{iqx}\,\langle 0 |
 T\Big\{J^{3,\perp}_{1}(x)\,,
  \bar q_1(0) \gamma_\mu\gamma_5 q_2(0) \Big\}|0\rangle
  = T^\perp_{^3P_1}(q^2)\, (q\bar{z}) \bar{z}_\mu +\cdots\,,
   \label{eq:f3perp-3p1}
 \end{equation}
\begin{eqnarray}
 i\int d^4x\, e^{iqx}\,\langle 0 |
  T\Big\{J^{3,\perp}_{2}(x)\,,
  \bar q_1(0) \gamma_\mu\gamma_5 q_2(0) \Big\}|0\rangle
  = T^{\perp,\alpha_g}_{^3P_1}(q^2)\,(q\bar{z})^2 \bar{z}_\mu +\cdots\,,
\end{eqnarray}
\begin{eqnarray}
 i\int d^4x\, e^{iqx}\,\langle 0 |
 T\Big\{J^{3,\perp}_{3}(x)\,,
  \bar q_1(0) \gamma_\mu\gamma_5 q_2(0) \Big\}|0\rangle
  = T^{\perp,\sigma}_{^3P_1}(q^2)\,(q\bar{z})^2 \bar{z}_\mu +\cdots\,,
\end{eqnarray}
where  \begin{eqnarray}
 & & J^{3,\perp}_{1}(0)
 = \bar q_2(0) \bar{z}^\beta
 \bar{z}_\lambda \sigma_{\alpha\beta}\gamma_5 g_s
 G^{\lambda\alpha}(0) q_1(0)\,, \label{eq:c3-1}
 \\
 & & J^{3,\perp}_{2}(0)
  = \bar q_2(0)
\bar{z}^\beta \bar{z}_\lambda \sigma_{\alpha\beta}\gamma_5 g_s
[iD\bar{z} G^{\lambda\alpha}(0)]  q_1(0)\,, \label{eq:c3-2}
\\
 & & J^{3,\perp}_{3}(0)
  = \bar q_2(0)
\bar{z}^\beta \bar{z}_\lambda \sigma_{\alpha\beta}\gamma_5 g_s
i\bar{z}^\delta[G^{\lambda\alpha}(0)\derright_\delta-
\derleft_\delta G^{\lambda\alpha}(0)] q_1(0)\,.\label{eq:c3-3}
 \end{eqnarray}
To including the light quark masses consistently to the ${\cal
O}(\alpha_s$), the currents needs to be replaced by the renormalized
ones as
 \begin{eqnarray}
 J^{3,\perp}_{1} \longrightarrow\,
 \bar J^{3,\perp}_{1} \,
 &=&J^{3,\perp}_{1} + \frac{\alpha_s}{4\pi}\frac{4}{9 \hat\varepsilon}
 (m_1+m_2)\bar z^\alpha iD\cdot \bar z (\bar q_2 \not\!\bar{z} \gamma_5 q_1) \,,
 \label{eq:s1-1}\\
 J^{3,\perp}_{2} \longrightarrow\,
 \bar J^{3,\perp}_{2} \,
 &=& J^{3,\perp}_{2}
    + \frac{\alpha_s}{4\pi}\frac{1}{5 \hat\varepsilon}
 (m_1+m_2)\bar z^\alpha (i D\cdot \bar z)^2 (\bar q_2 \not\!\bar{z} \gamma_5 q_1) \,,
 \label{eq:s1-2}\\
 J^{3,\perp}_{3}  \longrightarrow\,
 \bar J^{3,\perp}_{3} \,
 &=&J^{3,\perp}_{3} - \frac{\alpha_s}{4\pi}\frac{1}{9 \hat\varepsilon}
 (m_1-m_2)\bar z^\alpha (i D\cdot \bar z)^2 (\bar q_2 \not\!\bar{z} \gamma_5 q_1)
 \,, \label{eq:s1-3}
\end{eqnarray}
where
$$\frac{1}{\hat \varepsilon}= \frac{1}{\varepsilon} + \gamma_E - \ln4\pi\,,$$
and $d\,({\rm dimension})=4+2\varepsilon$. As adopted in all the
calculations of this paper, the modified minimal substraction scheme
(${\rm \overline{MS}}$) is used to regularize the divergent
integrals. From the technical point of view, the reason that we have
to take into account the mixings of $J^{3,\perp}_{i}$ and twist-2
operators is because we need to remove the nonphysical
$\ln(-q^2/\mu^2)/\hat\varepsilon$ terms in the calculation.
Physically speaking, $f^{\perp}_{3,^3\! P_1}$ and $f^{\perp}_{3,^3\!
P_1}\omega_{^3\! P_1}^\perp$ mix with $f_{^3P_1}(m_{q_1} +
m_{q_2})$, while $f^{\perp}_{3,^3\! P_1} \sigma_{^3\! P_1}^\perp$
mixes with $f_{^3P_1}(m_{q_1} - m_{q_2})$. For a massive quark, the
above mixings were studied in Ref. \cite{Ball:2006wn} by using the
light-ray-operator technique \cite{Balitsky:1987bk}. According to
their results, (i) $f_{3,^3P_1}^\perp$ can mix with $(m_{q_1}
+m_{q_2})\, f_{^3P_1}$ and $(m_{q_1} +m_{q_2})\, f_{^3P_1}
a_1^{\parallel, ^3P_1}$, (ii) $f_{3,^3P_1}^\perp
\omega_{^3P_1}^\perp$ with $(m_{q_1} +m_{q_2})\, f_{^3P_1}$ and
$(m_{q_1} +m_{q_2})\,  f_{^3P_1} a_1^{\parallel, ^3P_1}$, and
$(m_{q_1} +m_{q_2})\, f_{^3P_1} a_2^{\parallel,^3P_1}$, and (iii)
$f_{3,^3P_1}^\perp \sigma_{^3P_1}^\perp$ with $(m_{q_1} -m_{q_2})\,
f_{^3P_1}$ and $(m_{q_1} -m_{q_2})\, f_{^3P_1} a_1^{\parallel,
^3P_1}$, and $(m_{q_1} -m_{q_2})\, f_{^3P_1} a_2^{\parallel,^3P_1}$.
Neglecting the corrections arising from $a_1^{\parallel,^3P_1}$ and
$a_2^{\parallel,^3P_1}$, our results basically agree with those
given in Ref.~\cite{Ball:2006wn} except a sign difference for
$J^{3,\perp}_{3}$. To LO approximation, the complete RG evolutions
of the relevant parameters are collected in Appendix
\ref{appsec:inputs}.

$T^\perp_{^3P_1}(q^2), T^{\perp,\alpha_g}_{^3P_1}$, and
$T^{\perp,\sigma}_{^3P_1}$ are relevant for the present
consideration and the OPE results are given by
\begin{eqnarray}\label{eq:pi-3da-3p1-3-1}
  T^{\perp,\rm OPE}_{^3P_1}(q^2)&=&
   (m_1 +m_2) \frac{\alpha_s}{144\pi^3}q^2 \ln\frac{-q^2}{\mu^2}
  \bigg[5 -2 \ln\frac{-q^2}{\mu^2} \bigg]
  +(0+{\cal O}(\alpha_s^2))
  \bigg(\langle \bar q_1 q_1\rangle + \langle \bar q_2 q_2\rangle\bigg)
  \nonumber\\
  && - \frac{1}{108 q^2}
  \bigg(\ln\frac{-q^2}{\mu^2} -\frac{547}{96}\bigg)\frac{\alpha_s}{\pi}
  \bigg(\langle \bar q_1 g_s\sigma G q_1\rangle
 +\langle \bar q_2 g_s\sigma G q_2\rangle\bigg)\nonumber\\
 && -\frac{\pi}{9 q^4} \langle \alpha_s G^2\rangle
  (\langle \bar q_1 q_1\rangle +\langle \bar q_2 q_2\rangle)\,,
\end{eqnarray}
\begin{eqnarray}\label{eq:pi-3da-3p1-3-2}
  T^{\perp,\alpha_g,\rm OPE}_{^3P_1}(q^2)
  &=&
   (m_1 +m_2) \frac{\alpha_s}{320\pi^3}q^2 \ln\frac{-q^2}{\mu^2}
  \bigg[\frac{83}{10} - 2 \ln\frac{-q^2}{\mu^2} \bigg]
   +(0+{\cal O}(\alpha_s^2))
  \bigg(\langle \bar q_1 q_1\rangle + \langle \bar q_2 q_2\rangle\bigg)
  \nonumber\\
  && -\frac{1}{216q^2}
  \bigg(\ln\frac{-q^2}{\mu^2}-\frac{823}{96}
  \bigg)\frac{\alpha_s}{\pi}
  \bigg(\langle \bar q_1 g_s\sigma G q_1\rangle
 +\langle \bar q_2 g_s\sigma G q_2\rangle\bigg)\nonumber\\
 & & + (0+{\cal O}(\alpha_s))\langle \alpha_s G^2\rangle
   (\langle \bar q_1 q_1\rangle +\langle \bar q_2 q_2\rangle)\,,
\end{eqnarray}
\begin{eqnarray}\label{eq:pi-3da-3p1-3-3}
  T^{\perp,\sigma,\rm OPE}_{^3P_1}(q^2)
  &=&
   (m_1 - m_2) \frac{\alpha_s}{576\pi^3}q^2 \ln\frac{-q^2}{\mu^2}
  \bigg[\frac{47}{5}-2 \ln\frac{-q^2}{\mu^2} \bigg]
  \nonumber\\
  &&   -\frac{\alpha_s}{6 \pi} \bigg(\ln \frac{-q^2}{\mu^2} + const.\bigg)
 (\langle \bar q_1 q_1\rangle - \langle \bar q_2 q_2\rangle)
 \nonumber\\
  && + \frac{1}{216 q^2}
  \bigg(\ln\frac{-q^2}{\mu^2} -\frac{847}{96}\bigg)\frac{\alpha_s}{\pi}
  \bigg(\langle \bar q_1 g_s\sigma G q_1\rangle
 - \langle \bar q_2 g_s\sigma G q_2\rangle\bigg)\nonumber\\
 &&
 +\frac{\pi}{9 q^4} \langle \alpha_s G^2\rangle
  (\langle \bar q_1 q_1\rangle - \langle \bar q_2 q_2\rangle
 )\,.
\end{eqnarray}
For the coefficient of the quark-gluon condensate in the above
equations, the logarithmic factor arises from the contribution of
the diagrams that contains a loop connecting a quark propagator,
where the quark propagator (which does not belong to the part of the
loop) emits a soft gluon into the condensate. Note that there is no
infrared pole (IR) in the calculations of the present work due to
the fact that the off-shell external momentum $-q^2<0$ regularize
the IR singularity in QCD sum rule approach. (It is interesting to
note that the infrared sensitive terms $\sim \ln(-q^2/m_q^2)$ may
appear as considering the order up to $m_q^2$. However, they can be
absorbed into the condensate \cite{Ball:2006fz}.)

  The sum rules for $f^{\perp}_{3,^3\! P_1}$, $\omega_{^3\!
P_1}^\perp$ and $\sigma_{^3\! P_1}^\perp$ therefore read
\begin{eqnarray}
&& e^{-m_{^3\! P_1}^2/M^2} m_{^3\! P_1}^2 f_{^3\! P_1}
 f^\perp_{3, ^3\! P_1}
 + e^{-m_{^1\! P_1}^2/M^2} m_{^1\! P_1}^2 f_{^1\! P_1}
   a_0^{\parallel,1^1P_1} f^\perp_{3, ^1\! P_1}
   \lambda_{^1P_1}^\perp
 \nonumber\\
 &&= (m_1+m_2)
 \frac{\alpha_s}{144\pi^3} \int_0^{s_0^{^3\! P_1}}s
 \bigg(-5+4 \ln\frac{s}{\mu^2}\bigg) \ e^{-s/M^2} ds
  \nonumber\\
 && -
 \frac{\alpha_s}{108\pi} \Bigg[\frac{547}{96} + \Bigg(\gamma_E +
 \ln\frac{\mu^2}{M^2} -{\rm Ei}\Bigg(-\frac{s_0}{M^2} \Bigg) \Bigg]
 \bigg(\langle \bar q_1 g_s\sigma G q_1\rangle
 +\langle \bar q_2 g_s\sigma G q_2\rangle\bigg)\nonumber\\
  && \ -
  \frac{\pi}{9 M^2}\, \langle \alpha_s G^2\rangle
  (\langle \bar q_1 q_1\rangle +\langle \bar q_2 q_2\rangle)\,,
 \label{eq:3P1-SR-f3T}
\end{eqnarray}
\begin{eqnarray}
&&
 e^{-m_{^3\! P_1}^2/M^2} m_{^3\! P_1}^2 f_{^3\! P_1}
 f^\perp_{3,^3\! P_1}
 \Bigg(\frac{3}{7}+\frac{3}{28}\omega_{^3\! P_1}^\perp\Bigg)
 + e^{-m_{^1\! P_1}^2/M^2} m_{^1\! P_1}^2 f_{^1\! P_1}
   a_0^{\parallel,1^1P_1} f^\perp_{3, ^1\! P_1}
   \Bigg(\frac{3}{7} \lambda_{^1P_1}^\perp
   +\frac{3}{28}\sigma_{^1\! P_1}^\perp\Bigg)
   \nonumber\\
 && \ \ \
  =(m_1+m_2) \frac{\alpha_s}{320\pi^3} \int_0^{s_0^{^3\! P_1}}s
   \bigg(-\frac{83}{10} + 4 \ln\frac{s}{\mu^2}\bigg)\ e^{-s/M^2} ds
  \nonumber\\
  && \ \ \ \ \ \
  - \frac{\alpha_s}{216\pi} \Bigg[\frac{823}{96}+ \Bigg(\gamma_E +
 \ln\frac{\mu^2}{M^2}-{\rm Ei}\Bigg(-\frac{s_0}{M^2} \Bigg)\Bigg) \Bigg]
 \bigg(\langle \bar q_1 g_s\sigma G q_1\rangle
 +\langle \bar q_2 g_s\sigma G q_2\rangle\bigg)\,,\ \ \ \ \ \
\label{eq:3P1-SR-f3T-omega}
\end{eqnarray}
and
\begin{eqnarray}
&& e^{-m_{^3\! P_1}^2/M^2} m_{^3\! P_1}^2 f_{^3\! P_1}
 f^\perp_{3, ^3\! P_1}\, \sigma_{^3\! P_1}^\perp
 + e^{-m_{^1\! P_1}^2/M^2} m_{^1\! P_1}^2 f_{^1\! P_1}
   a_0^{\parallel,1^1P_1} f^\perp_{3, ^1\! P_1}
 \nonumber\\
 && \ \ \ =
 (m_1-m_2)
 \frac{\alpha_s}{576\pi^3} \int_0^{s_0^{^3\! P_1}}s
 \bigg(-\frac{47}{5} +4 \ln\frac{s}{\mu^2}\bigg)
 \ e^{-s/M^2} ds
  \nonumber\\
  &&~~~~ +\frac{\alpha_s}{6 \pi}
  (\langle \bar q_1 q_1\rangle - \langle \bar q_2 q_2\rangle)
  \int_0^{s_0^{^3\! P_1}}\ e^{-s/M^2} ds
 \nonumber\\
 &&~~~~  +
 \frac{\alpha_s}{216\pi} \Bigg[\frac{847}{96} + \Bigg(\gamma_E +
 \ln\frac{\mu^2}{M^2} -{\rm Ei}\Bigg(-\frac{s_0}{M^2} \Bigg) \Bigg]
 \bigg(\langle \bar q_1 g_s\sigma G q_1\rangle
 -\langle \bar q_2 g_s\sigma G q_2\rangle\bigg)\nonumber\\
  &&~~~~ +
  \frac{\pi}{9 M^2}\, \langle \alpha_s G^2\rangle
  (\langle \bar q_1 q_1\rangle -\langle \bar q_2 q_2\rangle)\,.
 \label{eq:3P1-SR-f3T-sigma}
\end{eqnarray}

Unfortunately, one can read from Eq. (\ref{eq:pi-3da-3p1-3-1}) or
Eq. (\ref{eq:3P1-SR-f3T}) that for the $f^{\perp}_{3,^3\! P_1}$ sum
rule the term of dimension-3 in OPE series vanishes in ${\cal
O}(\alpha_s)$, while the terms of dimensions-5 and -7 are comparable
in magnitude but with the opposite signs. As a result, the OPE
series does not show any convergent behavior.

Therefore, to evaluate $f^{\perp}_{3,^3\! P_1}$, we further consider
the following diagonal correlation function,
 \begin{equation}
 i\int d^4x\, e^{iqx}\,\langle 0 |
 T\Big\{J^{3,^3\! P_1}_{\perp}(x)\,,
  J^{3,^3\! P_1}_{\perp}(0) \Big\}|0\rangle
  =-\widetilde{T}^\perp(q^2)\, (q\bar{z})^4\,.
   \label{eq:f3perp-3p1-2}
 \end{equation}
However, at the hadronic level, the lowest-lying resonance for the
above correlation function is the pseudoscalar meson (the $0^-$
state). To subtract the contribution arising from the lowest
pseudoscalar meson, we study the following non-diagonal correlation
function:
 \begin{eqnarray}
 i\int d^4x\, e^{iqx}\,\langle 0 |
  T\Big\{J^{3,^3\! P_1}_{\omega_\perp}(x)\,,
   q_1(0)\gamma_5 q_2(0) \Big\}|0\rangle
  = -\widetilde{T'}^{\perp}(q^2)\,(q\bar{z})^5\,,
 \end{eqnarray}
where
\begin{equation}
 z^\beta z^\mu
 \langle 0|
 q_2 \sigma_{\alpha \beta}\gamma_5 g_s G_{\mu}^{\ \alpha} q_1
 |PS(P)\rangle
 =2i f_{3PS}(Pz)^2
\end{equation}
with $``PS\equiv$ the lowest-lying pseudoscalar meson". Only the
pseudoscalar mesons contribute to the above non-diagonal correlation
function. $\widetilde{T}^\perp(q^2)$ and
$\widetilde{T'}^{\perp}(q^2)$ have been calculated in Ref.
\cite{Ball:2006wn} for studying $f_{3K}$.

Using the results given in Ref.~\cite{Ball:2006wn}, the
corresponding sum rules are
\begin{eqnarray}
 && 4\, e^{-m_{PS}^2/M^2} f_{3PS}^2
 + e^{-m_{^3\! P_1}^2/M^2} (f^\perp_{3, ^3\! P_1})^2
 =
 \frac{\alpha_s}{360\pi^3} \int_0^{s_0^{^3\! P_1}}s \
 e^{-s/M^2} ds
 +\frac{89}{5184}\frac{\alpha_s}{\pi^2}\langle \alpha_s G^2\rangle
 \nonumber\\
& & ~~~~~~~~~~~~~~
 +\frac{\alpha_s}{18\pi} (m_{q_1}\langle\bar q_1 q_1\rangle
    + m_{q_2}\langle\bar q_2 q_2\rangle)
    -\frac{\alpha_s}{108\pi}\frac{1}{M^2}
  \bigg(m_{q_1}\langle \bar q_1 g_s\sigma G q_1\rangle
 + m_{q_2}\langle \bar q_2 g_s\sigma G q_2\rangle\bigg)
 \nonumber\\
&  & ~~~~~~~~~~~~~
 +\frac{71}{729}\frac{\alpha_s^2}{M^2} (\langle\bar q_1 q_1\rangle^2
    + \langle\bar q_2 q_2\rangle^2)
    +\frac{32}{81}\frac{\alpha_s^2}{M^2}
    \langle\bar q_1 q_1\rangle \langle\bar q_2 q_2\rangle
 \label{eq:3P1-SR-f3T-2}\,,
\end{eqnarray}
and
\begin{eqnarray}
 && 2\, e^{-m_{PS}^2/M^2} f_{3PS}
 \frac{f_{PS} m_{PS}^2}{m_{q_1}+ m_{q_2}}
 =
 \frac{\alpha_s}{72\pi^3} \int_0^{s_0^{PS}}s \
 e^{-s/M^2} ds
 +\frac{1}{12\pi}\langle \alpha_s G^2\rangle
 \nonumber\\
& & ~~~~~~~~~~~~~~
 -\frac{\alpha_s}{9\pi}
  (m_{q_1}\langle\bar q_1 q_1\rangle + m_{q_2}\langle\bar q_2 q_2\rangle)
  \nonumber\\
& & ~~~~~~~~~~~~~~
 -\frac{2\alpha_s}{9\pi} (m_{q_1}\langle\bar q_2 q_2\rangle
    + m_{q_2}\langle\bar q_1 q_1\rangle)
    \Bigg( \frac{8}{3} +\gamma_E +\ln\frac{\mu^2}{M^2}
    -{\rm Ei}\Bigg(-\frac{s_0}{M^2}\Bigg) \Bigg) \nonumber\\
 &&  ~~~~~~~~~~~~~~
   -\frac{1}{6M^2} \bigg(m_{q_1}\langle \bar q_2 g_s\sigma G q_2\rangle
 + m_{q_2}\langle \bar q_1 g_s\sigma G q_1\rangle\bigg)
 \nonumber\\
&  & ~~~~~~~~~~~~~~
 +\frac{16}{27}\frac{\pi\alpha_s}{M^2} (\langle\bar q_1 q_1\rangle^2
    + \langle\bar q_2 q_2\rangle^2)
    +\frac{16}{9}\frac{\pi\alpha_s}{M^2}
    \langle\bar q_1 q_1\rangle \langle\bar q_2 q_2\rangle\,.
 \label{eq:3P1-SR-f3T-3}
\end{eqnarray}
 In calculating the $f^\perp_{3, ^3\! P_1}$ sum rule,
we substitute $f_{3PS}$ by using the expression given in
Eq.~(\ref{eq:3P1-SR-f3T-3}). Although the sign of $f^\perp_{3, ^3\!
P_1}$ compared with $f_{^3\! P_1}$ cannot be determined from
Eq.~(\ref{eq:3P1-SR-f3T-2}), there are two indications that the sign
of $f^\perp_{3, ^3\! P_1}$ should be negative. One is that at the
large $M^2$ limit the result of Eq.~(\ref{eq:3P1-SR-f3T}) implies
the negative $f^\perp_{3, ^3\! P_1}$ although the sum rule cannot
offer its reliable magnitude. The other one is that only the
negative $f^\perp_{3, ^3\! P_1}$ can result in a physical value for
$\langle \alpha_g^\perp\rangle$, the average gluon momentum fraction
in a $1^3P_1$ meson,  which should satisfy
 \begin{equation}
 0 \leq\langle \alpha_g^\perp\rangle = \frac{3}{7}
 +\frac{3}{28}\omega_{^3\! P_1}^\perp \leq 1 \,.
 \end{equation}
Note that, as for $\omega_{^3\! P_1}^\perp$ sum rule
(Eq.~(\ref{eq:3P1-SR-f3T-omega})) which is dominated by the
quark-gluon condensate, because the contribution ${\cal
O}(\alpha_s)$ of the dimension-7 term, $\sim\langle \alpha_s
G^2\rangle (\langle \bar q_1 q_1\rangle +\langle \bar q_2
q_2\rangle)$, is absent, we thus do not further study a different
correlation function.

\subsection{Axial-vector mesons with quantum number $1^1P_1$}\label{sec:properties-3pdas-2}

\subsubsection{$f^V_{3, ^1P_1}$, $\lambda^V_{^1P_1}$,
and $\sigma^V_{^1P_1}$}\label{sec:properties-3pdas-2-1}

The coupling constants $f^V_{3, ^1\! P_1}$, $\lambda^V_{^1\! P_1}$,
and $\sigma^V_{^1\! P_1}$ can be obtained through the following
matrix element
\begin{eqnarray}
 \langle 0 | J^{3,V}_{3,\, \mu}(0)| 1^1P_1(P, \lambda)\rangle
 &=& -i f_{3,^1\! P_1}^V \, (P\bar{z})^3
 \epsilon^{(\lambda)}_{\perp,\mu} +{\cal O}(\bar{z}_\mu)\,,\\
 \langle 0 | J^{3,V}_{1,\, \mu}(0)| 1^1P_1(P, \lambda)\rangle
 &=& -i f_{3,^1\! P_1}^V \, \lambda^V_{^1\! P_1} (P\bar{z})^2
 \epsilon^{(\lambda)}_{\perp,\mu} +{\cal O}(\bar{z}_\mu)\,,\\
 \langle 0 | J^{3,V}_{2,\, \mu}(0)| 1^1P_1(P, \lambda)\rangle
 &=& -i f_{3,^1\! P_1}^V \, \Bigg( \frac{3}{7} \lambda^V_{^1\! P_1}
 + \frac{3}{28}\sigma^V_{^1\! P_1} \Bigg) (P\bar{z})^3
 \epsilon^{(\lambda)}_{\perp,\mu} +{\cal O}(\bar{z}_\mu)\,,
\end{eqnarray}
where the currents have been defined in Eqs. (\ref{eq:c1-1}),
(\ref{eq:c1-2}), and (\ref{eq:c1-3}).
 To evaluate $f^V_{3, ^1\! P_1}$, $\lambda^V_{^1\!
P_1}$, and $\sigma^V_{^1\! P_1}$, we consider the following
``non-diagonal'' correlation functions,
\begin{eqnarray}
  \int d^4x\, e^{iqx}
 \langle 0 | T\Big\{J^{3,V}_{3,\, \mu}(x)\,,
 \bar q_1(0) \bar{z}^\lambda \sigma_{\nu\lambda}\gamma_5 q_2(0) \Big\}|0\rangle
  &=& T^{V}_{^1P_1}(q^2)\, (q\bar{z})^4 g_{\mu\nu}^\perp
 +\cdots\,,\\
 \int d^4x\, e^{iqx}
 \langle 0 | T\Big\{J^{3,V}_{1,\, \mu}(x)\,,
 \bar q_1(0) \bar{z}^\lambda \sigma_{\nu\lambda}\gamma_5 q_2(0) \Big\}|0\rangle
  &=& T^{V,\lambda}_{^1P_1}(q^2)\, (q\bar{z})^3 g_{\mu\nu}^\perp
 +\cdots\,,\\
 \int d^4x\, e^{iqx}
 \langle 0 | T\Big\{J^{3,V}_{2,\, \mu}(x)\,,
 \bar q_1(0) \bar{z}^\lambda \sigma_{\nu\lambda}\gamma_5 q_2(0) \Big\}|0\rangle
  &=& T^{V,\sigma}_{^1P_1}(q^2)\, (q\bar{z})^4 g_{\mu\nu}^\perp
 +\cdots\,.
\end{eqnarray}
Concerning the light quark masse corrections to the correlation
functions, we have to replace the currents by the renormalized ones:
 \begin{eqnarray}
 J^{3,V}_{3,\, \mu}  \longrightarrow\,
 \bar J^{3,V}_{3,\, \mu} \,
 &=&J^{3,V}_{3,\, \mu}
  + \frac{\alpha_s}{4\pi}\frac{1}{18 \hat\varepsilon}
 (m_1+m_2)\bar z^\alpha i(i D\cdot \bar z)^2 (\bar q_2
 \sigma_{\alpha\mu}\gamma_5 q_1)\,, \label{eq:s2-1}\\
 J^{3,V}_{1,\, \mu}  \longrightarrow\,
 \bar J^{3,V}_{1,\, \mu} \,
 &=&J^{3,V}_{1,\, \mu}
 - \frac{\alpha_s}{4\pi}\frac{2}{9 \hat\varepsilon}
 (m_1-m_2)\bar z^\alpha i (iD\cdot \bar z)
   (\bar q_2 \sigma_{\alpha\mu}\gamma_5 q_1) \,, \label{eq:s2-2}\\
 J^{3,V}_{2,\, \mu}  \longrightarrow\,
 \bar J^{3,V}_{2,\, \mu} \,
 &=&J^{3,V}_{2,\, \mu}
 - \frac{\alpha_s}{4\pi}\frac{1}{10 \hat\varepsilon}
 (m_1-m_2)\bar z^\alpha i(i D\cdot \bar z)^2
   (\bar q_2 \sigma_{\alpha\mu} \gamma_5 q_1) \,. \label{eq:s2-3}
\end{eqnarray}
The above mixings lead to that $f^{V}_{3,^1\! P_1}$ mixes with
$f_{^1P_1}^\perp (m_{q_1} + m_{q_2})$, while $f^{V}_{3,^1\! P_1}
\lambda_{^1\! P_1}^\perp$ and $f^{V}_{3,^1\! P_1}\sigma_{^1\!
P_1}^\perp $ mix with $f_{^1P_1}^\perp(m_{q_1} - m_{q_2})$. We did
not find any explicit result in the literature that can be used to
compare with the present calculations. However, in analogy to the
discussion after Eq.~(\ref{eq:s1-3}), the relevant parameters can
mix in addition with $f_{^1P_1}^\perp(m_{q_1} \pm
m_{q_2})a_1^{\perp,^1P_1}$ and $f_{^1P_1}^\perp(m_{q_1} \pm
m_{q_2})a_2^{\perp,^1P_1}$, where the upper sign corresponds to the
G-parity conserving parameter $f^{V}_{3,^1\! P_1}$ and the lower
sign to the G-parity violating parameters for which $f^{V}_{3,^1\!
P_1} \lambda_{^1\! P_1}^\perp$ does not mix with
$f_{^1P_1}^\perp(m_{q_1} - m_{q_2})a_2^{\perp,^1P_1}$. We have
neglected the RG-corrections due to $(m_{q_1} \pm
m_{q_2})a_1^{\perp,^1P_1}$ and $(m_{q_1} \pm
m_{q_2})a_2^{\perp,^1P_1}$ in the present calculations. Considering
the mass corrections in the RG equations, the scale dependence of
the parameters relevant to the twist-three three-parton LCDAs is
summarized in Appendix~\ref{appsec:inputs}.

The OPE results of $T^{V}_{^1P_1}(q^2)$,
$T^{V,\lambda}_{^1P_1}(q^2)$ and $T^{V,\alpha}_{^1P_1}(q^2)$ are
\begin{eqnarray}
  T^{V,{\rm OPE}}_{^1P_1}(q^2) &=&
  (m_1 +m_2) \frac{\alpha_s}{1152\pi^3} \ln\frac{-q^2}{\mu^2}
  \bigg[13 -2\ln\frac{-q^2}{\mu^2} \bigg]
  \nonumber\\
  && +\frac{\alpha_s}{18\pi}
    \frac{\langle \bar q_1 q_1\rangle +\langle \bar q_2 q_2\rangle}{q^2}
    +\frac{5}{108}\frac{\alpha_s}{\pi}
    \frac{\langle \bar q_1 g_s\sigma G q_1\rangle
    + \langle \bar q_2 g_s\sigma G q_2\rangle}{q^4}
    \nonumber\\
  &&  +(0+{\cal O}(\alpha_s)) \langle \alpha_s G^2\rangle
  (\langle \bar q_1 q_1\rangle + \langle \bar q_2 q_2\rangle
 )\,,
\end{eqnarray}
\begin{eqnarray}
  T^{V, \lambda,{\rm OPE}}_{^1P_1}(q^2) &=&
 (m_1 -m_2) \frac{\alpha_s}{288\pi^3} \ln\frac{-q^2}{\mu^2}
  \bigg[-11 + 2\ln\frac{-q^2}{\mu^2} \bigg]
  \nonumber\\
  && +\frac{\alpha_s}{9\pi}
     \frac{\langle \bar q_1 q_1\rangle -\langle \bar q_2 q_2\rangle}{q^2}
    +\frac{173}{3456}\frac{\alpha_s}{\pi}
    \frac{\langle \bar q_1 g_s\sigma G q_1\rangle
    - \langle \bar q_2 g_s\sigma G q_2\rangle}{q^4}
   \nonumber\\
  &&  +(0+{\cal O}(\alpha_s)) \langle \alpha_s G^2\rangle
   (\langle \bar q_1 q_1\rangle - \langle \bar q_2 q_2\rangle
 )\,,
\end{eqnarray}
and
\begin{eqnarray}
  T^{V, \sigma,{\rm OPE}}_{^1P_1}(q^2) &=&
 (m_1 -m_2) \frac{\alpha_s}{3200\pi^3} \ln\frac{-q^2}{\mu^2}
  \bigg[-57 + 10 \ln\frac{-q^2}{\mu^2} \bigg]
  \nonumber\\
  && +\frac{\alpha_s}{18\pi}
   \frac{\langle \bar q_1 q_1\rangle -\langle \bar q_2 q_2\rangle}{q^2}
   +\frac{107}{1728}\frac{\alpha_s}{\pi}
   \frac{\langle \bar q_1 g_s\sigma G q_1\rangle
   - \langle \bar q_2 g_s\sigma G q_2\rangle}{q^4}
   \nonumber\\
  &&  +(0+{\cal O}(\alpha_s)) \langle \alpha_s G^2\rangle
   (\langle \bar q_1 q_1\rangle - \langle \bar q_2 q_2\rangle
 )\,,
\end{eqnarray}
respectively. Note that $T^{V}_{^1P_1}$, $T^{V, \lambda}_{^1P_1}$,
and $T^{V, \sigma}_{^1P_1}$ can receive contributions from $^3P_1$
states because $^3P_1$ states have small pseudo-tensor coupling
constants due to the unequal quark masses. Consequently, we obtain
the QCD sum rules
\begin{eqnarray}
 && e^{-m_{^1\! P_1}^2/M^2} f^\perp_{^1\! P_1} f_{3,^1\! P_1}^V
  + e^{-m_{^3\! P_1}^2/M^2} f^\perp_{^3\! P_1} a_0^{\perp,1^3P_1}
  f_{3,^3\! P_1}^V \sigma_{^3P_1}^V
  \nonumber\\
  &&\ \ \ \ \ \ =
 (m_1+m_2)
 \frac{\alpha_s}{1152\pi^3} \int_0^{s_0^{^1P_1}}
 \bigg[-13+ 4\ln\frac{s}{\mu^2}\bigg] \ e^{-s/M^2} ds
  \nonumber\\
  &&\ \ \ \ \ \ \ \ \ -\frac{\alpha_s}{18\pi} \bigg(\langle \bar q_1 q_1\rangle +\langle
\bar q_2 q_2\rangle\bigg)
  +\frac{5}{108}\frac{\alpha_s}{\pi} \frac{\langle \bar q_1 g_s\sigma G q_1\rangle
 + \langle \bar q_2 g_s\sigma G q_2\rangle}{M^2} \,,
 \ \ \ \ \
 \label{eq:1P1_SRf3V}
\end{eqnarray}
\begin{eqnarray}
 && e^{-m_{^1\! P_1}^2/M^2} f^\perp_{^1\! P_1} f_{3,^1\! P_1}^V
 \lambda^V_{^1\! P_1}
 + e^{-m_{^3\! P_1}^2/M^2} f^\perp_{^3\! P_1}  a_0^{\perp,1^3P_1}
  f_{3,^3\! P_1}^V
 \nonumber\\
  &&\ \ \ \ \ \  =
 (m_1-m_2)
 \frac{\alpha_s}{288\pi^3} \int_0^{s_0^{^1P_1}}
 \bigg[11 - 4\ln\frac{s}{\mu^2}\bigg] \ e^{-s/M^2} ds
  \nonumber\\
  &&\ \ \ \ \ \ \ \ \ -\frac{\alpha_s}{9\pi}
  \bigg(\langle \bar q_1 q_1\rangle -\langle \bar q_2 q_2\rangle\bigg)
  +\frac{173}{3456}\frac{\alpha_s}{\pi}
  \frac{\langle \bar q_1 g_s\sigma G q_1\rangle
  - \langle \bar q_2 g_s\sigma G q_2\rangle}{M^2} \,,
 \ \ \ \ \
 \label{eq:1P1_SRf3V_lambda}
 \end{eqnarray}
 and
\begin{eqnarray}
 && e^{-m_{^1\! P_1}^2/M^2} f^\perp_{^1\! P_1}
  f_{3,^1\! P_1}^V
  \Bigg( \frac{3}{7} \lambda^V_{^1\! P_1}
   + \frac{3}{28}\sigma^V_{^1\! P_1} \Bigg)
 + e^{-m_{^3\! P_1}^2/M^2} f^\perp_{^3\! P_1}a_0^{\perp,1^3P_1}
  f_{3,^3\! P_1}^V
 \Bigg( \frac{3}{7} + \frac{3}{28}\omega^V_{^3\! P_1} \Bigg)
 \nonumber\\
 &&~~~~~~~~~~~~~~~~~~ =
 (m_1-m_2)
 \frac{\alpha_s}{3200\pi^3} \int_0^{s_0^{^1P_1}}
 \bigg[57 - 20\ln\frac{s}{\mu^2}\bigg] \ e^{-s/M^2} ds
  \nonumber\\
  &&~~~~~~~~~~~~~~~~~~~~~~ -\frac{\alpha_s}{18\pi}
  \bigg(\langle \bar q_1 q_1\rangle -\langle \bar q_2 q_2\rangle\bigg)
  +\frac{107}{1728}\frac{\alpha_s}{\pi} \frac{\langle \bar q_1 g_s\sigma G q_1\rangle
  - \langle \bar q_2 g_s\sigma G q_2\rangle}{M^2} \,.
 \ \ \ \ \
 \label{eq:1P1_SRf3V_sigma}
\end{eqnarray}
In Eqs.~(\ref{eq:1P1_SRf3V}), the contribution originating from
G-parity breaking parameters $a_0^{\perp,1^3P_1} \sigma_{^3P_1}^V$
relevant to LCDAs of the $1^3\!P_1$ state are relatively suppressed
by ${\cal O}(m_q^2)$ and can thus be neglected. In
Eqs.~(\ref{eq:1P1_SRf3V_lambda}) and (\ref{eq:1P1_SRf3V_sigma}), the
contributions arising from the $1^3\!P_1$ state are of the same
order of magnitude as the G-parity breaking parameters related to
the $1^1\!P_1$ state, and should be taken into account in the
numerical analysis. Analogously, one needs to take into account the
corrections due to the $1^3\!P_1$ state in Eqs.
(\ref{eq:1P1-SR-f3A-sigma}), (\ref{eq:1P1_SRf3t_lambda}) and
(\ref{eq:1P1_SRf3t_sigma}) in the following subsections, whereas
such corrections can be negligible in Eqs. (\ref{eq:1P1-SR-f3A}),
(\ref{eq:1P1-SR-f3A-omega}) and (\ref{eq:1P1_SRf3t}).

\subsubsection{$f^A_{3, ^1P_1}$, $\omega_{^1P_1}^A$
and $\sigma_{^1P_1}^A$}\label{sec:properties-3pdas-2-2}

The parameters $f^A_{3, ^1P_1}$, $\omega_{^1P_1}^A$ and
$\sigma_{^1P_1}^A$ are defined through the matrix elements:
 \begin{eqnarray}
 & & \langle 0| J^{3,A}_{1,\, \mu}(0)|
 1^1P_1(P,\lambda)\rangle =-f^A_{3,^1P_1} (P\bar{z})^2
 \epsilon^{(\lambda)}_{\perp,\mu}
 +{\cal O}(\bar{z}_\mu)\,,\\
& &
 \langle 0| J^{3,A}_{2,\, \mu}(0)| 1^1P_1(P,\lambda)\rangle
 =-f^A_{3,^1P_1} \langle \alpha_g^{A}\rangle (P\bar{z})^3
 \epsilon^{(\lambda)}_{\perp,\mu} +{\cal O}(\bar{z}_\mu)\,,\\
 & &
 \langle 0| J^{3,A}_{3,\, \mu}(0)| 1^1P_1(P,\lambda)\rangle
 =-f^A_{3,^1P_1} \sigma^A_{^1P_1} (P\bar{z})^3
 \epsilon^{(\lambda)}_{\perp,\mu} +{\cal O}(\bar{z}_\mu)\,,
 \end{eqnarray}
where the interpolating currents have been given by Eqs.
(\ref{eq:c2-1}), (\ref{eq:c2-2}), and (\ref{eq:c2-3}). ${\cal
O}(\bar{z}_\mu)$ contains the twist-3 and twist-4 corrections, and
the average gluon momentum fraction $\langle \alpha_g^A\rangle$
satisfies
\begin{equation}
 \langle \alpha_g^A\rangle = \frac{3}{7} +\frac{3}{28}\omega_{^1\!
 P_1}^A \,.
 \end{equation}
$f^A_{3, ^1P_1}$, $\omega_{^1P_1}^A$ and $\sigma_{^1P_1}^A$  can be
therefore evaluated by considering the correlation functions,
\begin{eqnarray}
 i \int d^4x\, e^{iqx}\,
 \langle 0 | T\Big\{J^{3,A}_{1,\, \mu}(x)\,,\,
 \bar q_1(0) \bar{z}^\lambda
\sigma_{\nu\lambda}\gamma_5 q_2(0) \Big\}|0\rangle
 &=&
T^{A}_{^1P_1}(q^2)\, (q\bar{z})^3 g_{\mu\nu}^\perp +\cdots,\\
 i \int d^4x\, e^{iqx}\,
 \langle 0 | T\Big\{J^{3,A}_{2,\, \mu}(x)\,,\,  \bar q_1(0)
\bar{z}^\lambda \sigma_{\nu\lambda}\gamma_5 q_2(0) \Big\}|0\rangle
 &=& T^{A,\alpha_g}_{^1P_1}(q^2)\, (q\bar{z})^4 g_{\mu\nu}^\perp
+\cdots\,, \\
i \int d^4x\, e^{iqx}\,
 \langle 0 | T\Big\{J^{3,A}_{3,\, \mu}(x)\,,\,  \bar q_1(0)
\bar{z}^\lambda \sigma_{\nu\lambda}\gamma_5 q_2(0) \Big\}|0\rangle
 &=& T^{A,\sigma}_{^1P_1}(q^2)\, (q\bar{z})^4 g_{\mu\nu}^\perp
+\cdots\,, \hspace{1cm}
\end{eqnarray}
respectively. To consider the light quark masses consistently to the
${\cal O}(\alpha_s$), the currents needs to be replaced by the
renormalized ones as
 \begin{eqnarray}
 J^{3,A}_{1,\, \mu}  \longrightarrow\,
 \bar J^{3,A}_{1,\, \mu} \,
 &=& J^{3,A}_{1,\, \mu} + \frac{\alpha_s}{4\pi}\frac{2}{9 \hat\varepsilon}
  (m_1+m_2)
  \bar z^\alpha (iD\cdot \bar z) (\bar q_2 \sigma_{\alpha\mu} \gamma_5 q_1)
 \,, \label{eq:s3-1}\\
 J^{3,A}_{2,\, \mu}  \longrightarrow\,
 \bar J^{3,A}_{2,\, \mu} \,
 &=& J^{3,A}_{2,\, \mu} + \frac{\alpha_s}{4\pi}\frac{1}{10 \hat\varepsilon}
  (m_1+m_2)
  \bar z^\alpha (i D\cdot \bar z)^2 (\bar q_2 \sigma_{\alpha\mu} \gamma_5 q_1)\,,
  \label{eq:s3-2}\\
 J^{3,A}_{3,\, \mu}  \longrightarrow\,
 \bar J^{3,A}_{3,\, \mu} \,
 &=& J^{3,A}_{3,\, \mu} - \frac{\alpha_s}{4\pi}\frac{4}{45 \hat\varepsilon}
  (m_1-m_2)
  \bar z^\alpha (i D\cdot \bar z)^2 (\bar q_2 \sigma_{\alpha\mu}\gamma_5 q_1)
  \label{eq:s3-3} \,.
\end{eqnarray}
From the above results we obtain that $f^{A}_{3,^1\! P_1}$ and
$f^{A}_{3,^1\! P_1} \omega_{^1\! P_1}^\perp$ mix with
$f_{^1P_1}^\perp (m_{q_1} + m_{q_2})$, while $f^{A}_{3,^1\!
P_1}\sigma_{^1\! P_1}^\perp$ mixes with $f_{^1P_1}^\perp(m_{q_1} -
m_{q_2})$. As our results in the previous subsection, we did not
find any literature that can be used to compare with the present
calculations. Again, in analogy to the discussions after
Eqs.~(\ref{eq:s1-3}) and (\ref{eq:s2-3}), the relevant parameters
can mix in addition with $f_{^1P_1}^\perp(m_{q_1} +
m_{q_2})a_1^{\perp,^1P_1}$ for $f^{A}_{3,^1\! P_1}$, and with
$f_{^1P_1}^\perp(m_{q_1} \pm m_{q_2})a_1^{\perp,^1P_1}$ and
$f_{^1P_1}^\perp(m_{q_1} \pm m_{q_2})a_2^{\perp,^1P_1}$ for
$f^{A}_{3,^1\! P_1} \omega_{^1\! P_1}^\perp$ corresponding to the
upper sign and $f^{A}_{3,^1\! P_1}\sigma_{^1\! P_1}^\perp $ to the
lower sign. We neglect the RG-corrections due to $(m_{q_1} \pm
m_{q_2})a_1^{\perp,^1P_1}$ and $(m_{q_1} \pm
m_{q_2})a_2^{\perp,^1P_1}$ in the present calculations. The RG
evolutions, containing the quark mass corrections, for the relevant
parameters are summarized in Appendix~\ref{appsec:inputs}.

 We get
\begin{eqnarray}
T^{A, {\rm OPE}}_{^1P_1}(q^2) &=&
 -(m_1 +m_2) \frac{\alpha_s}{1152\pi^3} \ln\frac{-q^2}{\mu^2}
  \bigg[7 + 8\ln\frac{-q^2}{\mu^2} \bigg]
  \nonumber\\
 &&
 -\frac{\alpha_s}{9\pi }\,
  \frac{\langle \bar q_1 q_1\rangle +\langle \bar q_2 q_2\rangle}{q^2}
 +\frac{59}{1728} \frac{\alpha_s}{\pi}
  \frac{\langle \bar q_1 g_s\sigma G q_1\rangle
 +\langle \bar q_2 g_s\sigma G q_2\rangle}{q^4}
  \nonumber\\
  &&  +(0+{\cal O}(\alpha_s)) \langle \alpha_s G^2\rangle
  (\langle \bar q_1 q_1\rangle + \langle \bar q_2 q_2\rangle
 )\,,
\end{eqnarray}

\begin{eqnarray}
 T^{A,\alpha_g, {\rm OPE}}_{^1P_1}(q^2) &=&
 (m_1 +m_2) \frac{\alpha_s}{3200\pi^3} \ln\frac{-q^2}{\mu^2}
  \bigg[57 -10\ln\frac{-q^2}{\mu^2} \bigg]
  \nonumber\\
 && -\frac{\alpha_s}{18\pi}\, \frac{\langle \bar q_1 q_1\rangle
 +\langle \bar q_2 q_2\rangle} {q^2}
 +\frac{29}{864} \frac{\alpha_s}{\pi}
 \frac{\langle \bar q_1 g_s\sigma G q_1\rangle
 +\langle \bar q_2 g_s\sigma G q_2\rangle}{q^4}
  \nonumber\\
  &&  +(0+{\cal O}(\alpha_s)) \langle \alpha_s G^2\rangle
  (\langle \bar q_1 q_1\rangle + \langle \bar q_2 q_2\rangle)\,,
\end{eqnarray}
and
\begin{eqnarray}
 T^{A,\sigma, {\rm OPE}}_{^1P_1}(q^2) &=&
 -(m_1 - m_2) \frac{\alpha_s}{1440\pi^3} \ln\frac{-q^2}{\mu^2}
  \bigg[\frac{185}{24} - \ln\frac{-q^2}{\mu^2} \bigg]
  \nonumber\\
 && -\frac{\alpha_s}{18\pi}\,
  \frac{\langle \bar q_1 q_1\rangle -\langle \bar q_2 q_2\rangle}{q^2}
 + \frac{\alpha_s}{27\pi}
 \frac{\langle \bar q_1 g_s\sigma G q_1\rangle
 - \langle \bar q_2 g_s\sigma G q_2\rangle}{q^4}
  \nonumber\\
  &&  +(0+{\cal O}(\alpha_s)) \langle \alpha_s G^2\rangle
  (\langle \bar q_1 q_1\rangle - \langle \bar q_2 q_2\rangle)\,.
\end{eqnarray}

The resulting QCD sum rules read
\begin{eqnarray}
 e^{-m_{^1P_1}^2/M^2} f^\perp_{^1P_1} f_{3,^1P_1}^A
&+& e^{-m_{^3P_1}^2/M^2} f^\perp_{^3P_1} a_0^{\perp,1^3P_1}
  f_{3,^3P_1}^A \lambda_{^3\!P_1}^A
 \nonumber\\
 &=& (m_1+m_2)\frac{\alpha_s}{1152\pi^3}
 \int_0^{s_0^{^1P_1}} \bigg(7+ 16\ln\frac{s}{\mu^2}\bigg)
 \ e^{-s/M^2} ds
  \nonumber\\
 &&+
 \frac{\alpha_s}{9\pi}\, (\langle \bar q_1 q_1\rangle
 +\langle \bar q_2 q_2\rangle)
 +\frac{59}{1728} \frac{\alpha_s}{\pi}
 \frac{\langle \bar q_1 g_s\sigma G q_1\rangle
 +\langle \bar q_2 g_s\sigma G q_2\rangle}{M^2}\,,\nonumber\\
 \label{eq:1P1-SR-f3A}
\end{eqnarray}
\begin{eqnarray}
 & & e^{-m_{^1P_1}^2/M^2} f^\perp_{^1P_1} f_{3,^1P_1}^A
 \Bigg(\frac{3}{7} +\frac{3}{28}\omega_{^1\! P_1}^A\Bigg)
 + e^{-m_{^3P_1}^2/M^2} f^\perp_{^3P_1} a_0^{\perp,1^3P_1}
  f_{3,^3P_1}^A
 \Bigg(\frac{3}{7} \lambda_{^3\! P_1}^A +\frac{3}{28}\sigma_{^3\! P_1}^A\Bigg)
 \nonumber\\
 &&\ \ \ \ \ \
 = (m_1+m_2)\frac{\alpha_s}{3200\pi^3}
 \int_0^{s_0^{^1P_1}} \bigg(-57+ 20\ln\frac{s}{\mu^2}\bigg)
 \ e^{-s/M^2} ds
  \nonumber\\
 && \ \ \ \ \ \ \ \ \ \
 +
 \frac{\alpha_s}{18\pi}\, (\langle \bar q_1 q_1\rangle
  +\langle \bar q_2 q_2\rangle ) +\frac{29}{864} \frac{\alpha_s}{\pi}
 \frac{\langle \bar q_1 g_s\sigma G q_1\rangle
  +\langle \bar q_2 g_s\sigma G q_2\rangle}{M^2}\,,
 \label{eq:1P1-SR-f3A-omega}
\end{eqnarray}
and
\begin{eqnarray}
 && e^{-m_{^1P_1}^2/M^2} f^\perp_{^1P_1} f_{3,^1P_1}^A
 \sigma^A_{^1P_1}+ e^{-m_{^3P_1}^2/M^2} f^\perp_{^3P_1} a_0^{\perp,1^3P_1}
  f_{3,^3P_1}^A
 \nonumber\\
 &&\ \ \ \ \ \
 = (m_1-m_2)\frac{\alpha_s}{1440\pi^3}
 \int_0^{s_0^{^1P_1}} \bigg(\frac{185}{24} -2 \ln\frac{s}{\mu^2}\bigg)
 \ e^{-s/M^2} ds
  \nonumber\\
 && \ \ \ \ \ \ \ \ \ \
 +
 \frac{\alpha_s}{18\pi}\, (\langle \bar q_1 q_1\rangle
  -\langle \bar q_2 q_2\rangle ) +  \frac{\alpha_s}{27\pi}
 \frac{\langle \bar q_1 g_s\sigma G q_1\rangle
  -\langle \bar q_2 g_s\sigma G q_2\rangle}{M^2}\,.
 \label{eq:1P1-SR-f3A-sigma}
\end{eqnarray}

\subsubsection{$f^\perp_{3, ^1\! P_1}$, $\lambda^\perp_{^1\! P_1}$,
and $\sigma^\perp_{^1\! P_1}$}\label{sec:properties-3pdas-2-3}

The coupling constants $f^\perp_{3, ^1\! P_1}$, $\lambda^\perp_{^1\!
P_1}$, and $\sigma^\perp_{^1\! P_1}$ for $1^1P_1$ mesons are defined
as the following matrix elements
\begin{eqnarray}
 \langle 0 | J^{3,\perp}_3(0)| 1^1P_1(P, \lambda)\rangle
 &=&  i f_{3,^1\! P_1}^\perp\, (P\bar{z})^2
 \big(\epsilon^{(\lambda)} \bar{z} \big)\,, \\
 \langle 0 | J^{3,\perp}_{1}(0)| 1^1P_1(P, \lambda)\rangle
 &=&  i f_{3,^1\! P_1}^\perp\, \lambda^\perp_{^1\! P_1} (P\bar{z})
 \big(\epsilon^{(\lambda)} \bar{z} \big)\,, \\
 \langle 0 | J^{3,\perp}_{2}(0)| 1^1P_1(P, \lambda)\rangle
 &=&  i f_{3,^1\! P_1}^\perp\, \Bigg( \frac{3}{7} \lambda^\perp_{^1\! P_1}
 + \frac{3}{28}\sigma^\perp_{^1\! P_1} \Bigg) (P\bar{z})^2
 \big(\epsilon^{(\lambda)} \bar{z} \big)\,,
\end{eqnarray}
where the interpolating currents have been given by Eqs.
(\ref{eq:c3-1}), (\ref{eq:c3-2}), and (\ref{eq:c3-3}). To calculate
these three parameters, we consider the following correlation
functions,
\begin{eqnarray}
 && \int d^4x\, e^{iqx}
 \langle 0 | T\Big\{J^{3,\perp}_3(x), \,\,
 \bar q_1(0) \bar{z}^\nu \sigma_{\mu\nu}\gamma_5 q_2(0) \Big\}|0\rangle
 = - T^{\perp}_{^1P_1}(q^2)\, (q\bar{z})^3 \bar{z}_\mu
+\cdots\,, \label{eq:correlator-1p1-t3-3.1} \\
&& \int d^4x\, e^{iqx}
 \langle 0 | T\Big\{J^{3\perp}_{1}(x), \,\,
 \bar q_1(0) \bar{z}^\nu \sigma_{\mu\nu}\gamma_5 q_2(0) \Big\}|0\rangle
 = - T^{\perp,\lambda}_{^1P_1}(q^2)\, (q\bar{z})^2 \bar{z}_\mu
+\cdots\,,\\
&& \int d^4x\, e^{iqx}
 \langle 0 | T\Big\{J^{3,\perp}_{2}(x), \,\,
 \bar q_1(0) \bar{z}^\nu \sigma_{\mu\nu}\gamma_5 q_2(0) \Big\}|0\rangle
 = - T^{\perp,\sigma}_{^1P_1}(q^2)\, (q\bar{z})^3 \bar{z}_\mu
+\cdots\,.
\end{eqnarray}
 It is interesting to note that the above correlation
functions receive no contributions from $1^-$ states. The OPE
results of $ T^{\perp}_{^1P_1}(q^2)$ have the following forms
\begin{eqnarray}\label{eq:pi-1p1-t3-3.1}
 && T^{\perp,{\rm OPE}}_{^1P_1}(q^2)
 \nonumber\\
 &&\ \ \ =
 -\frac{\alpha_s}{720\pi^3}\, q^2\ln \frac{-q^2}{\mu^2}
 -\frac{\langle \alpha_s G^2\rangle}{36\pi q^2}
 +\frac{16\pi\alpha_s}{27 q^4}
 (\langle \bar q_1 q_1\rangle^2 +\langle \bar q_2 q_2\rangle^2 )
 +\frac{8\pi\alpha_s}{9 q^4} \langle \bar q_1 q_1\rangle \langle \bar q_2 q_2\rangle)
 \nonumber\\
 &&\ \ \ \ \ \ +\frac{\alpha_s}{18\pi q^2}
 \Bigg[\frac{9}{2}\bigg(m_1\langle \bar q_1 q_1\rangle
  +m_2\langle \bar q_2 q_2\rangle\bigg)
  -\bigg(m_1\langle \bar q_2 q_2\rangle+m_2\langle \bar q_1 q_1\rangle\bigg)
  \Bigg(\ln\frac{-q^2}{\mu^2}+\frac{1}{3}\Bigg)\Bigg]\,,\ \ \ \ \ \ \
  \ \
\end{eqnarray}
\begin{eqnarray}
 && T^{\perp,\lambda,{\rm OPE}}_{^1P_1}(q^2)
  =0 \cdot \langle \alpha_s G^2\rangle
 -\frac{16\pi\alpha_s}{27 q^4}
 (\langle \bar q_1 q_1\rangle^2 -\langle \bar q_2 q_2\rangle^2 )
 \nonumber\\
 &&~~~~~ -\frac{\alpha_s}{18\pi q^2}
 \Bigg[7\bigg(m_1\langle \bar q_1 q_1\rangle
  -m_2\langle \bar q_2 q_2\rangle\bigg)
  +4\bigg(m_1\langle \bar q_2 q_2\rangle - m_2\langle \bar q_1 q_1\rangle\bigg)
  \Bigg(\ln\frac{-q^2}{\mu^2}-\frac{2}{3}\Bigg)\Bigg].\ \ \ \ \ \ \
\end{eqnarray}
\begin{eqnarray}
 && T^{\perp,\sigma,{\rm OPE}}_{^1P_1}(q^2)\nonumber\\
 &&\ \  =
  -\frac{\alpha_s}{36\pi q^2}
 \Bigg[5\bigg(m_1\langle \bar q_1 q_1\rangle
  -m_2\langle \bar q_2 q_2\rangle\bigg)
  +4\bigg(m_1\langle \bar q_2 q_2\rangle - m_2\langle \bar q_1 q_1\rangle\bigg)
  \Bigg(\ln\frac{-q^2}{\mu^2}-\frac{7}{12}\Bigg)\Bigg]\nonumber\\
  &&\ \ \ \ \ +0 \cdot \langle \alpha_s G^2\rangle
   + 0 \cdot \langle \bar q_1 q_1\rangle \langle \bar q_2 q_2\rangle
  +(0+{\cal O}(\alpha_s^2))
 \bigg(\langle \bar q_1 q_1\rangle^2, \langle \bar q_2
  q_2\rangle^2\bigg).\ \ \ \ \ \ \
\end{eqnarray}
Consequently, we obtain the sum rules
\begin{eqnarray}
&& e^{-m_{^1P_1}^2/M^2} m_{^1P_1} f^\perp_{^1P_1} f_{3,^1P_1}^\perp
 +  e^{-m_{^3P_1}^2/M^2} m_{^3P_1} f^\perp_{^3P_1} a_0^{\perp,1^3P_1}
f_{3,^3P_1}^\perp \sigma_{^3P_1}^\perp\nonumber\\
 &&\ \ \ \  =
\frac{\alpha_s}{720\pi^3}\,\int_0^{s_0^{^1P_1}}\!\! ds\,s e^{-s/M^2}
+ \frac{\langle \alpha_s G^2\rangle}{36\pi}
 +\frac{16\pi\alpha_s}{27M^2} (\langle \bar q_1 q_1\rangle^2
  +\langle \bar q_2 q_2\rangle^2 )
  +\frac{8\pi\alpha_s}{9M^2}
  \langle \bar q_1 q_1\rangle\langle \bar q_2 q_2\rangle\nonumber\\
 &&\ \ \ \ \ \ \ +\frac{\alpha_s}{18\pi}
  \Bigg[-\frac{9}{2}\bigg(m_1\langle \bar q_1 q_1\rangle
   +m_2\langle \bar q_2 q_2\rangle\bigg)
 \nonumber\\
 && \ \ \ \ \ \ \ \ \ \ \ \ \ \ \ \ \
 +\bigg(m_1\langle \bar q_2 q_2\rangle+m_2\langle \bar q_1 q_1\rangle\bigg)
  \Bigg( \frac{1}{3} -\gamma_E -\ln\frac{\mu^2}{M^2}
    +{\rm Ei}\Bigg(-\frac{s_0^{^1P_1}}{M^2}\Bigg) \Bigg) \Bigg]\,,
    \label{eq:1P1_SRf3t}
\end{eqnarray}
\begin{eqnarray}
&& e^{-m_{^1P_1}^2/M^2} m_{^1P_1} f^\perp_{^1P_1} f_{3,^1P_1}^\perp
\lambda^\perp_{^1\! P_1}
 +  e^{-m_{^3P_1}^2/M^2} m_{^3P_1} f^\perp_{^3P_1} a_0^{\perp,1^3P_1}
f_{3,^3P_1}^\perp
 \nonumber\\
 & & \ \  \ \ \ \ \ \
  =
 -\frac{16\pi\alpha_s}{27M^2} (\langle \bar q_1 q_1\rangle^2
  -\langle \bar q_2 q_2\rangle^2 ) +\frac{\alpha_s}{18\pi}
  \Bigg[7\bigg(m_1\langle \bar q_1 q_1\rangle
   - m_2\langle \bar q_2 q_2\rangle\bigg)
 \nonumber\\
 && \ \ \ \ \ \ \ \ \ \ \
  + 4 \bigg(m_1\langle \bar q_2 q_2\rangle- m_2\langle \bar q_1
q_1\rangle\bigg)
  \Bigg( -\frac{2}{3} -\gamma_E -\ln\frac{\mu^2}{M^2}
    +{\rm Ei}\Bigg(-\frac{s_0^{^1P_1}}{M^2}\Bigg) \Bigg) \Bigg]\,,
    \label{eq:1P1_SRf3t_lambda}
\end{eqnarray}
and
\begin{eqnarray}
&& e^{-m_{^1P_1}^2/M^2} m_{^1P_1} f^\perp_{^1P_1}
f_{3,^1P_1}^\perp\Bigg( \frac{3}{7} \lambda^\perp_{^1\! P_1}
 + \frac{3}{28}\sigma^\perp_{^1\! P_1} \Bigg)
+  e^{-m_{^3P_1}^2/M^2} m_{^3P_1} f^\perp_{^3P_1} a_0^{\perp,1^3P_1}
f_{3,^3P_1}^\perp\Bigg( \frac{3}{7}
 + \frac{3}{28}\omega^\perp_{^3\! P_1} \Bigg)
 \nonumber\\
&&\ \ \ \  =
  \frac{\alpha_s}{36\pi}\Bigg[5\bigg(m_1\langle \bar q_1 q_1\rangle
   - m_2\langle \bar q_2 q_2\rangle\bigg)
 \nonumber\\
 && \ \ \ \ \ \ \ \ \ \ \ \ \ \ \ \ \
 + 4 \bigg(m_1\langle \bar q_2 q_2\rangle- m_2\langle \bar q_1 q_1\rangle\bigg)
  \Bigg( -\frac{7}{12} -\gamma_E -\ln\frac{\mu^2}{M^2}
    +{\rm Ei}\Bigg(-\frac{s_0^{^1P_1}}{M^2}\Bigg) \Bigg) \Bigg]\,.
    \label{eq:1P1_SRf3t_sigma}
\end{eqnarray}
Note that the calculation of Eq.~(\ref{eq:correlator-1p1-t3-3.1}) is
actually analogous to that of $f_{3\rho}^T$ for the $\rho$
\cite{Ball:1998sk}, where there is no $\gamma_5$ for the $\rho$.
Neglecting the mass corrections in Eq.~(\ref{eq:1P1_SRf3t}), our
perturbative and gluon-condensate contributions agree with
Eq.~(C.15) in Ref.~\cite{Ball:1998sk}, but the term of dimension-6
is different from theirs, where the sign difference due to
$\gamma_5$ has been considered.

\subsection{Results}\label{sec:3pdas-3}

In the numerical analysis, we use $f_{^3P_1}$, $f_{^1P_1}^\perp$,
$s_0^{^3P_1}$,  $s_0^{^1P_1}$, and the masses for axial-vector
mesons, which have been obtained in the previous section, as inputs.
We also adopt the parameters given in Appendix~\ref{appsec:inputs}.
The Borel window can thus be determined by means of that the
contributions arising from the higher resonances, including the
continuum, and from the term of the highest dimension in the OPE
series are well under control. We find that the suitable Borel
window is 1.5~GeV$^2< M^2 < 2.5$~GeV$^2$ except that the Borel
window is 2.5~GeV$^2< M^2 < 3.5$~GeV$^2$ for the $\omega_{^3P_1}^V$
sum rule. The reason that we have to choose a higher Borel window
for the $\omega_{^3P_1}^V$ sum rule is because a lower Borel mass
will lead to a slowly convergence at the quark-gluon level; for
instance, at $M^2= 2.0$~GeV$^2$, the highest dimension (dimension=6)
term still gives a large correction, $\sim 33\%$, at the quark-gluon
level. Instead, during 2.5~GeV$^2< M^2 < 3.5$~GeV$^2$, the
contribution arising from the highest dimension term is about 14\%
$\sim$ 5\%. On the other hand, because the average gluon momentum
fraction in an axial-vector meson is not less than zero, therefore
we should have $\omega_{^3P_1}^V\geq -4$, so that the region for
$M^2<1.6$ GeV$^2$ is strongly disfavored by the $\omega_{^3P_1}^V$
sum rule.

 As we evaluate G-parity invariant
parameters for $1^3P_1$ states, the corrections receiving from
$1^1P_1$ states are negligible, and vice versa. Nevertheless, if
calculating G-parity violating parameters for the $K_{1A}$
($K_{1B}$) state, the corrections originating from the $K_{1B}$
($K_{1A}$) state cannot be ignored. In the numerical study we use
$a_0^{\perp, K_{1A}}$ and $a_0^{\parallel, K_{1B}}$, given in
Table~\ref{tab:Gegenbauer}, as inputs. To obtain the relevant
parameter in a sum rule, we replace the other twist-3 parameters
using the corresponding sum rules. For instance, in evaluating
$\sigma_{^3P_1}^V$, we substitute $f_{3,^3P_1}^V$ and
$f_{3,^1P_1}^V$ by the expressions given in
Eqs.~(\ref{eq:3P1-SR-f3V}) and (\ref{eq:1P1_SRf3V}) into
Eq.~(\ref{eq:3P1-SR-f3V-sigma}). Note that for $\omega_{^3P_1}^V$,
because the working Borel window is different from the others, we
thus adopt directly the values of $f_{3,^3P_1}^V$ given in
Table~\ref{tab:para-3p-t3} in the study.

\begin{figure}[ht!]
\epsfxsize=11.1cm \centerline{\epsffile{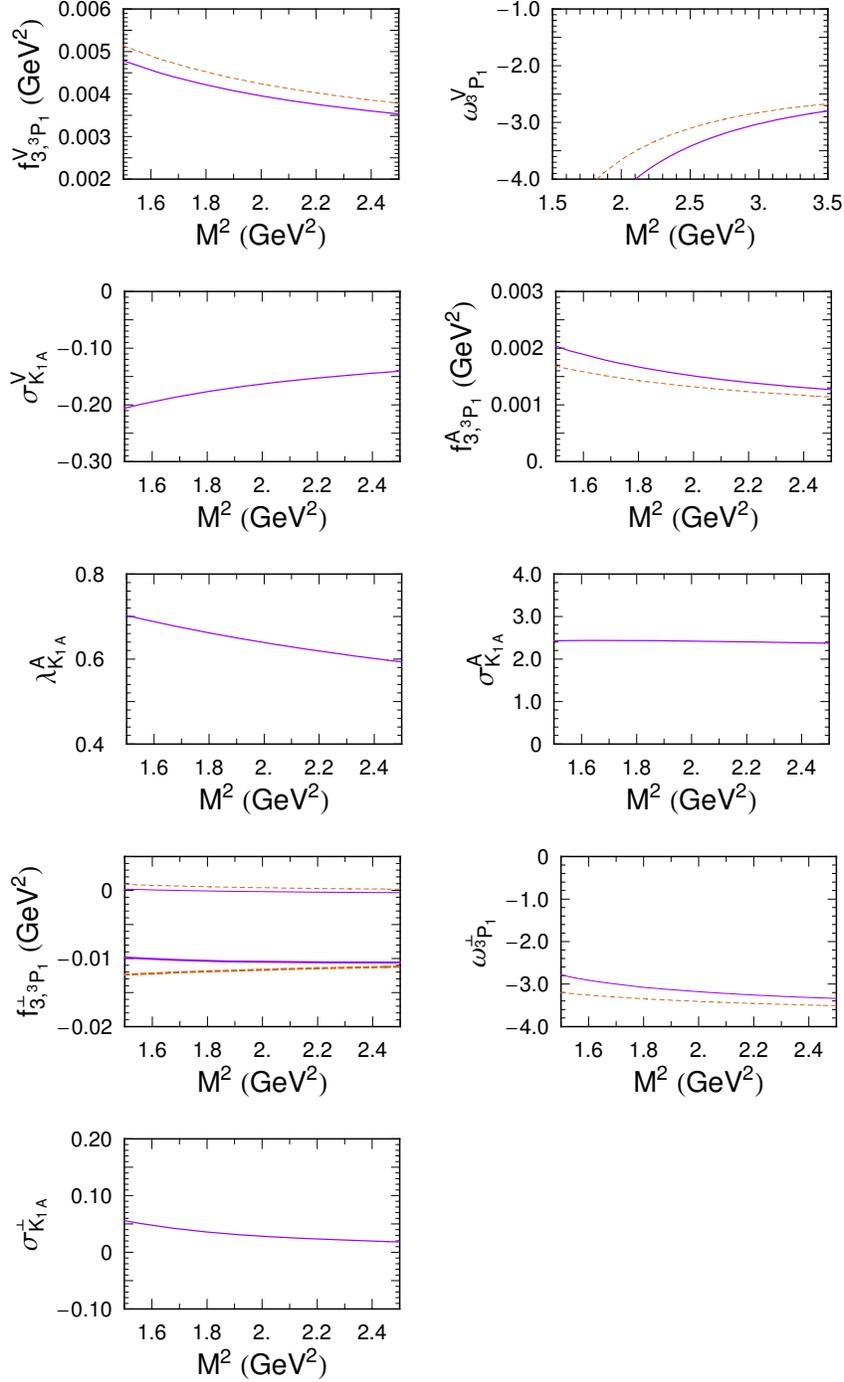}}
\centerline{\parbox{14cm}{\caption{\label{fig:t3-3P1} Some relevant
parameters in determinations of three-parton distribution amplitudes
of twist-3 for the $K_{1A}$ (solid curve) and $a_1 (1260)$ (dashed
curve) as functions of the Borel mass squared, where the central
values of input parameters have been used. The renormalization scale
is set at $\mu\simeq$ 1.4 GeV except that $\omega_{^3P_1}^V$ is at
$\mu\simeq 1.7$~GeV. For $f_{3,^3P_1}^\perp$, the lower two curves
are derived from Eqs.~(\ref{eq:3P1-SR-f3T-2}) and
(\ref{eq:3P1-SR-f3T-3}), and the upper two curves from
Eq.~(\ref{eq:3P1-SR-f3T}).}}}
\end{figure}

\begin{figure}[ht!]
\epsfxsize=11.7cm \centerline{\epsffile{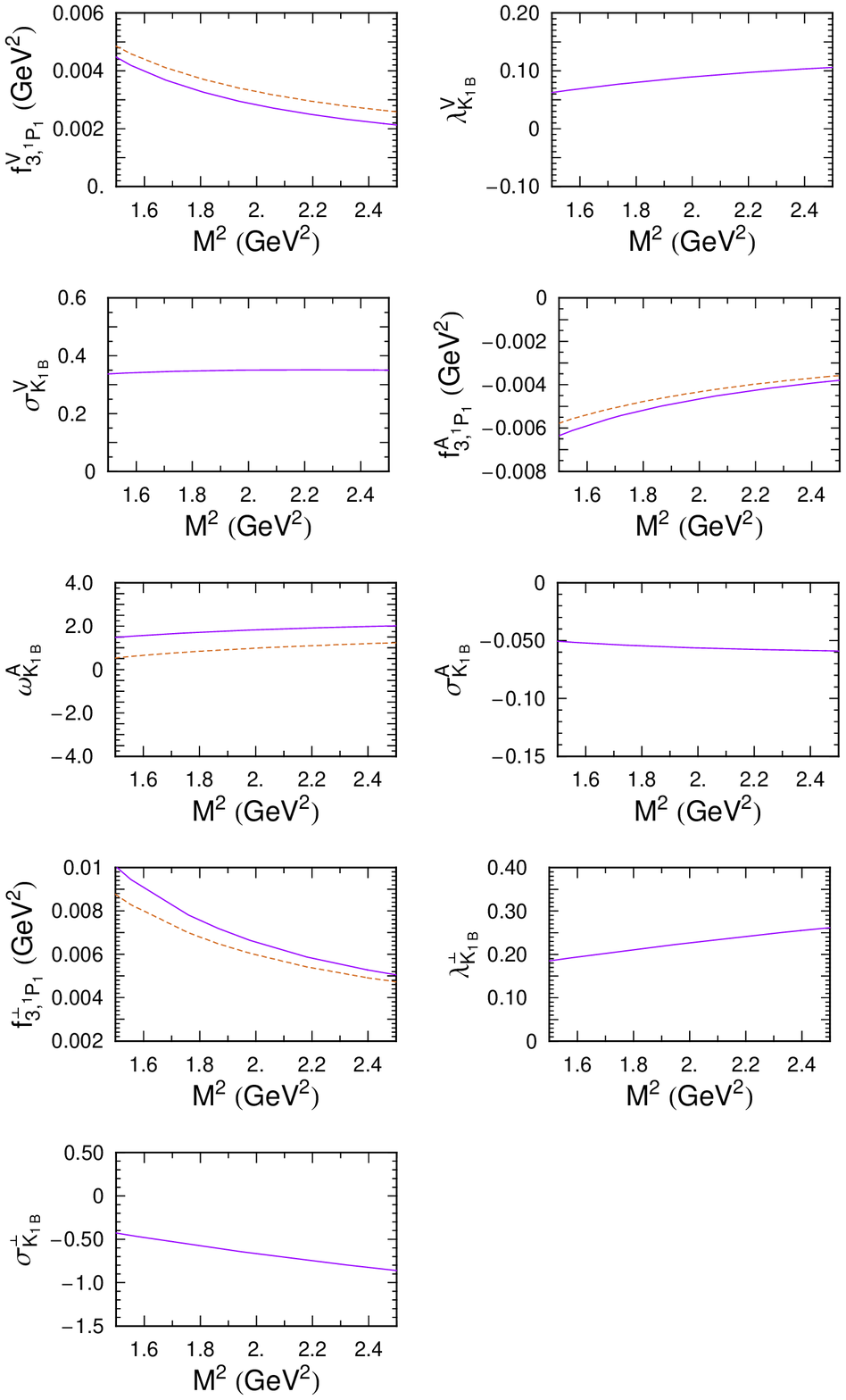}}
\centerline{\parbox{14cm}{\caption{\label{fig:t3-1P1} Some relevant
parameters in determinations of three-parton distribution amplitudes
of twist-3 for the $K_{1B}$ (solid curve) and $b_1 (1235)$ (dashed
curve) as functions of the Borel mass squared, where the central
values of input parameters have been used. The renormalization scale
is set at $\mu\simeq$ 1.4 GeV.}}}
\end{figure}

We summarize the numerical results in Tables~\ref{tab:para-3p-t3}
and \ref{tab:para-3p-t1}, where the theoretical errors are due to
variation of all inputs and the predictive uncertainty within the
Borel window. To illustrate the sum rule results, we plot the
parameters for the $a_1(1260), K_{1A}$ as functions of the Borel
mass squared in Figs.~\ref{fig:t3-3P1} and $b_1(1235), K_{1B}$ in
Fig.~\ref{fig:t3-1P1}, where the central values of input parameters
given in Tables \ref{tab:mass-decay-constant-3p1},
\ref{tab:mass-decay-constant-1p1}, \ref{tab:Gegenbauer}, and in
Appendix~\ref{appsec:inputs} are used. For simplicity, we do not
plot the results for $f_1, f_8, h_1$ and $h_8$, which are analogous
to the above ones. The $f_{3, ^3P_1}^\perp$ sum rule results
obtained from Eq.~(\ref{eq:3P1-SR-f3T-2}) and from
Eq.~(\ref{eq:3P1-SR-f3T}) are depicted. Because the sum rule given
in Eq.~(\ref{eq:3P1-SR-f3T}) exhibits badly convergent behavior (see
the discussion after Eq.~(\ref{eq:3P1-SR-f3T-sigma})), we therefore
choose to use the diagonal sum rule given in
Eq.~(\ref{eq:3P1-SR-f3T-2}), where the lowest lying pseudoscalar
contribution is substituted by using Eq.~(\ref{eq:3P1-SR-f3T-3}).
Although the sum rule in Eq.~(\ref{eq:3P1-SR-f3T-2}) cannot
determine the sign of $f_{3, ^3P_1}^\perp$, however there are two
reasons that its sign should be negative. First,
Eq.~(\ref{eq:3P1-SR-f3T}) yields a negative $f_{3, ^3P_1}^\perp$ at
the large $M^2$. Second, only a negative $f_{3, ^3P_1}^\perp$ can
result in a physical value of $\omega_{^3P_1}^\perp$ that should
satisfy $0 \leq \omega_{^3P_1}^\perp\leq 16/3$.

Finally, it should be noted that the main contributions of some sum
rules for parameters are due to the quark and quark-gluon
condensates. These sum rules may suffer from the contributions of
higher resonances at hadronic level and radiative corrections at
quark-gluon level. The two effects may partially cancel each other
and give about $\lesssim 10\%$ corrections to the numerical results
for parameters. See also the discussions in
subsection~\ref{subsec:gegenbauer-3p1-result}. Here we do not
include these effects in Tables~\ref{tab:para-3p-t3} and
\ref{tab:para-3p-t1}.

\begin{table}[t]
\caption{Parameters for twist-3 distribution amplitudes of $1^3P_1$
mesons at the scales $\mu=1$ GeV and 2.2 GeV (shown in parentheses),
where $f^V_{3,^3P_1}, f^A_{3,^3P_1}$, and $f^\perp_{3,^3P_1}$ are in
units of GeV$^2$. } \label{tab:para-3p-t3} \vskip0.3cm
\begin{tabular}{|c|cccc|} \hline\hline
  & $a_1 (1260)$ & $f_1$ & $f_8$ & $K_{1A}$  \\ \hline
 $f^V_{3,^3P_1}$    & $0.0055\pm 0.0027$ & $0.0055\pm 0.0027$ & $0.0054\pm 0.0027$ & $0.0052\pm 0.0027$\\
                    &$(0.0036\pm 0.0018)$&$(0.0036\pm 0.0018)$&$(0.0035\pm 0.0018)$&$(0.0034\pm 0.0018)$\\ \hline
 $\omega_{^3P_1}^V$ & $-2.9  \pm 0.9$    &  $-2.8  \pm 0.9$   &  $-3.0  \pm 1.1$   & $-3.1  \pm 1.1$   \\
                    & $(-2.9  \pm 0.9)$  &  $(-2.8  \pm 0.9)$ &  $(-3.0  \pm 1.0)$ & $(-3.1  \pm 1.1)$ \\ \hline
 $\sigma_{^3P_1}^V$ & ---                &  ---               &   ---              &  $-0.13\pm0.16$ \\
                    & ---                &  ---               &   ---              & $(-0.13\pm0.16)$\\ \hline\hline
 $f^A_{3,^3P_1}$    & $0.0022\pm 0.0009$ & $0.0022\pm 0.0009$ & $0.0028\pm 0.0009$ & $0.0026\pm 0.0013$\\
                    &$(0.0012\pm 0.0005)$&$(0.0012\pm 0.0005)$&$(0.0015\pm 0.0005)$&$(0.0014\pm 0.0007)$\\ \hline
 $\lambda_{^3P_1}^A$& ---                &  ---               &   ---              & $0.57\pm0.39$  \\
                    & ---                &  ---               &   ---              & $(0.70\pm0.46)$ \\ \hline
 $\sigma_{^3P_1}^A$ & ---                &  ---               &   ---              & $2.4\pm 2.0$  \\
                    & ---                &  ---               &   ---              & $(2.4\pm 2.0)$  \\ \hline\hline
 $f^\perp_{3,^3P_1}$& $-0.013\pm 0.002$  & $-0.012\pm 0.002$   & $-0.012\pm 0.002$ & $-0.012\pm 0.002$\\
                    & $(-0.009\pm 0.002)$& $(-0.009\pm 0.002)$&$(-0.009\pm 0.002)$ &$(-0.009\pm 0.002)$\\ \hline
 $\omega_{^3P_1}^\perp$& $-3.7\pm0.4$    &     $-3.4\pm0.4$   &    $-3.2\pm0.6$    &  $-3.4\pm 0.6$ \\
                    &  $(-2.9\pm0.3)$    &    $(-2.6\pm0.3)$   &   $(-2.4\pm0.4)$   & $(-2.6\pm0.4)$\\ \hline
 $\sigma_{^3P_1}^\perp$& ---             &  ---               &   ---              &  $0.07\pm 0.21$ \\
                    &    ---             &  ---               &   ---              &  $(0.05\pm0.15)$ \\
  \hline\hline
\end{tabular}
\end{table}
\begin{table}[t]
\caption{The same as Table~\ref{tab:para-3p-t3} except for $1^1P_1$
mesons, where $f^V_{3,^1P_1}, f^A_{3,^1P_1}$, and
$f^\perp_{3,^1P_1}$ are in units of GeV$^2$.} \label{tab:para-3p-t1}
\vskip0.3cm
\begin{tabular}{|c|cccc|}
\hline\hline
  & $b_1 (1235)$ & $h_1$ & $h_8$ & $K_{1B}$  \\ \hline
 $f^V_{3,^1P_1}$    & $0.0052\pm 0.0018$ & $0.0046\pm 0.0021$ & $0.0045\pm 0.0020$ & $0.0049\pm 0.0021$\\
                    &$(0.0030\pm 0.0011)$&$(0.0027\pm 0.0012)$&$(0.0027\pm 0.0012)$&$(0.0029\pm 0.0012)$\\ \hline
 $\lambda_{^1P_1}^V$& ---                &  ---               &   ---              &   $0.07\pm0.19$ \\
                    & ---                &  ---               &   ---              &  $(0.09\pm0.24)$\\ \hline
 $\sigma_{^1P_1}^V$ & ---                &  ---               &   ---              &   $0.35\pm0.73$ \\
                    & ---                &  ---               &   ---              &  $(0.31\pm0.68)$\\ \hline
 $f^A_{3,^1P_1}$    &$-0.0058\pm 0.0023$ & $-0.0053\pm 0.0023$& $-0.0055\pm 0.0023$& $-0.0065\pm 0.0029$\\
                   &$(-0.0036\pm 0.0014)$&$(-0.0033\pm 0.0014)$&$(-0.0035\pm 0.0014)$&$(-0.0041\pm 0.0018)$\\ \hline
 $\omega_{^1P_1}^A$ & $-1.5 \pm 0.4$     &  $-1.9  \pm 0.6$   &  $-3.5  \pm 1.0$   &  $-1.9  \pm 0.6$   \\
                    & $(-1.4  \pm 0.3)$  &  $(-1.7  \pm 0.4)$ &  $(-2.9  \pm 0.8)$ &  $(-1.7 \pm 0.4)$ \\ \hline
 $\sigma_{^1P_1}^A$ & ---                &  ---               &   ---              &    $-0.06\pm0.05$ \\
                    & ---                &  ---               &   ---              &   $(-0.05\pm0.04)$\\ \hline\hline
 $f^\perp_{3,^1P_1}$& $0.011\pm 0.006$   & $0.012\pm 0.006$   & $0.012\pm 0.005$   &   $0.012\pm 0.005$\\
                    & $(0.006\pm 0.003)$ & $(0.006\pm 0.003)$ & $(0.006\pm 0.003)$ &  $(0.006\pm 0.003)$\\ \hline
 $\lambda_{^1P_1}^\perp$& ---            &  ---               &   ---              &    $0.17\pm0.17$ \\
                    & ---                &  ---               &   ---              &   $(0.25\pm0.25)$\\ \hline
 $\sigma_{^1P_1}^\perp$& ---             &  ---               &   ---              &    $-0.71\pm0.53$ \\
                    & ---                &  ---               &   ---              &   $(-0.76\pm0.56)$\\
  \hline\hline
\end{tabular}
\end{table}


\section{Models for LCDAs}\label{sec:lcda}

\subsection{Two-parton LCDAs of twist-two}\label{subsec:2plcda-t2}

We have calculated the first few Gegenbauer moments of leading-twist
light-cone distribution amplitudes of $1^3P_1$  and $1^1P_1$ mesons
in Sec.~\ref{subsec:moments-t2} using the QCD sum rule technique.
The Gegenbauer moments of higher conformal spins may not be
predictive in the QCD sum rule approach owing to the divergence of
the OPE series for relevant correlation functions. Therefore the
models for light-cone distribution amplitudes depend on the
truncated conformal expansions with the reliable Gegenbauer moments.

Here we take into account the approximate forms of twist-2 distributions
for $1^3P_1$ mesons as follows:
\begin{eqnarray}
\Phi_\parallel(u) & = & 6 u \bar u \left[ 1 + 3 a_1^\parallel\, \xi +
a_2^\parallel\, \frac{3}{2} ( 5\xi^2  - 1 )
 \right], \label{eq:lcda-3p1-t2-1}\\
 \Phi_\perp(u) & = & 6 u \bar u \left[ a_0^\perp + 3 a_1^\perp\, \xi +
a_2^\perp\, \frac{3}{2} ( 5\xi^2  - 1 ) \right], \label{eq:lcda-3p1-t2-2}
\end{eqnarray}
where $\xi=2u-1$ and $a_1^\parallel, a_0^\perp, a_2^\perp$ are
non-zero only for strange mesons. The above LCDAs of {\it pure}
$1^3P_1$ states are normalized as the normalization conditions
\begin{eqnarray}
 \int_0^1 du \Phi_\parallel(u) &=& 1,
 \label{eq:nom-3P1-t2-1}\\
 \int_0^1 du \Phi_\perp (u) &=& a^\perp_0 \,.
 \label{eq:nom-3P1-t2-2}
\end{eqnarray}
In our convention, $u$ is the momentum fraction carried by the $q_1$
quark in an axial-vector meson (and is therefore equivalent to the
momentum fraction carried by the $s$ quark in a strange meson). Note
that in this paper the strange mesons that we discuss should contain
an $s$ quark, while for the LCDAs of strange mesons involving an
$\bar s$, the replacement $u\leftrightarrow 1-u$ has to be made,
namely $\xi \to -\xi$. As for $1^1P_1$ mesons, we take the following
approximation (see also the discussions given in
Ref.~\cite{Yang:2005gk}):
\begin{eqnarray}
 \Phi_\parallel(u) & = & 6 u \bar u \left[ a_0^\parallel + 3
a_1^\parallel\, \xi +
a_2^\parallel\, \frac{3}{2} ( 5\xi^2  - 1 ) \right], \label{eq:lcda-1p1-t2-1}\\
\Phi_\perp(u) & = & 6 u \bar u \left[ 1 + 3 a_1^\perp\, \xi +
a_2^\perp\, \frac{3}{2} ( 5\xi^2  - 1 ) \right],
\label{eq:lcda-1p1-t2-2}
\end{eqnarray}
where $a_0^\parallel, a_2^\parallel$, and $a_1^\perp$ are non-zero
only for strange mesons, so that the LCDAs of {\it pure} $1^1P_1$
states are normalized as the normalization conditions
\begin{eqnarray}
 \int_0^1 du \Phi_\parallel(u) &=& a^\parallel_0,
 \label{eq:nom-1P1-t2-1}\\
 \int_0^1 du \Phi_\perp (u) &=& 1 \,.
 \label{eq:nom-1P1-t2-2}
\end{eqnarray}
Due to mixtures, we define the LCDAs of physical $h_1(1170),
h_1(1380)$, $f_1(1285), f_1(1420)$ mesons in Appendix
\ref{appsec:def-f1-h1}, and, on other hand, the LCDAs for
$K_1(1270)$ and $K_1(1400)$ are consequently given by
 \begin{eqnarray}
 \Phi_\parallel^{K_1(1270)}(u)
 &=&
 \frac{f_{K_{1A}} m_{K_{1A}}}{f_{K_1(1270)} m_{K_1(1270)}}
  \Phi_\parallel^{K_{1A}}(u)\sin{\theta_K}
    + \frac{f_{K_{1B}} m_{K_{1B}}}{f_{K_{1}(1270)} m_{K_1(1270)}}
   \Phi_\parallel^{K_{1B}}(u)\cos{\theta_K}, ~~~~\\
  \Phi_\parallel^{K_1(1400)}(u)
 &=&
 \frac{f_{K_{1A}} m_{K_{1A}}}{f_{K_1(1400)} m_{K_1(1400)}}
  \Phi_\parallel^{K_{1A}}(u)\cos{\theta_K}
    - \frac{f_{K_{1B}} m_{K_{1B}}}{f_{K_1(1400)} m_{K_1(1400)}}
   \Phi_\parallel^{K_{1B}}(u)\sin{\theta_K},
 \end{eqnarray}

 \begin{eqnarray}
 \Phi_\perp^{K_1(1270)}(u)
 &=&
 \frac{f_{K_{1A}}^\perp}{f_{K_1(1270)}^\perp}
  \Phi_\perp^{K_{1A}}(u)\sin{\theta_K}
    + \frac{f_{K_{1B}}^\perp }{f_{K_{1}(1270)}^\perp}
   \Phi_\perp^{K_{1B}}(u)\cos{\theta_K}, ~~~~\\
  \Phi_\perp^{K_1(1400)}(u)
 &=&
 \frac{f_{K_{1A}}^\perp }{f_{K_1(1400)}^\perp }
  \Phi_\perp^{K_{1A}}(u)\cos{\theta_K}
    - \frac{f_{K_{1B}}^\perp}{f_{K_1(1400)}^\perp }
   \Phi_\perp^{K_{1B}}(u)\sin{\theta_K}.
 \end{eqnarray}

\begin{figure}[t!]
\epsfxsize=14cm \centerline{\epsffile{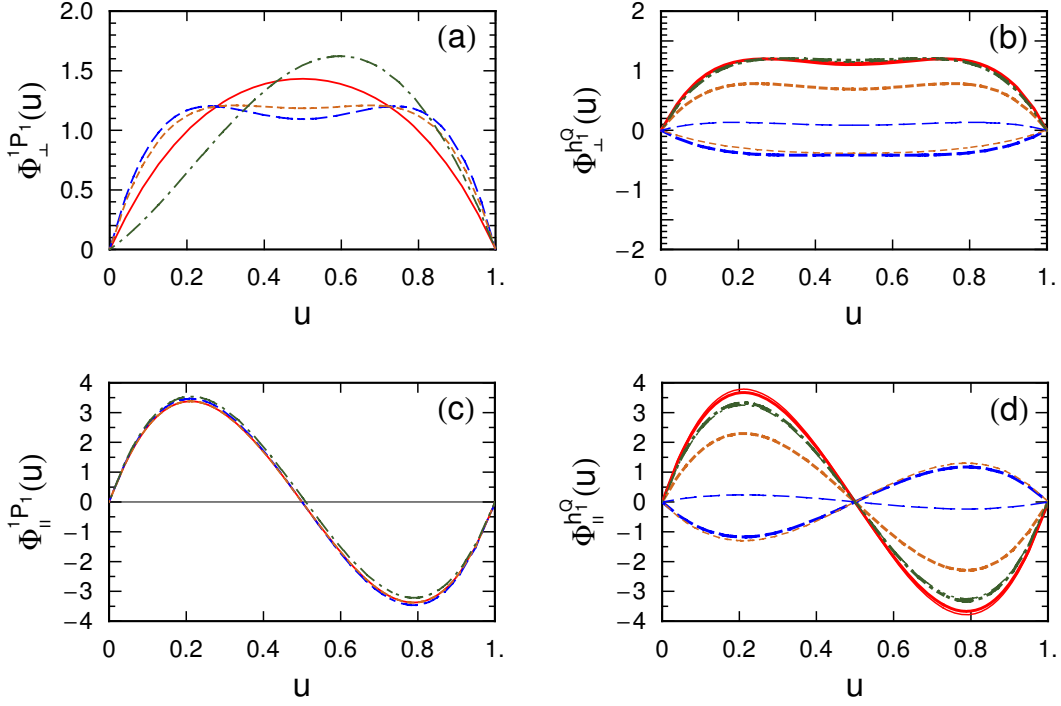}}
\centerline{\parbox{14cm}{\caption{\label{fig:lcda-t2-1p1}
Leading-twist light-cone distribution amplitudes, normalized at the
scale $\mu=$1~GeV, for $1^1P_1$ states, where the central values of
Gegenbauer moments given in Table \ref{tab:Gegenbauer} are used. $u$
($\bar u\equiv 1-u$) is the meson momentum fraction carried by the
quark (antiquark). In (a) and (c), the solid, long-dashed,
short-dashed and dot-dashed curves correspond to $b_1(1235),
h_1(singlet), h_8(octet)$ and $K_{1B}$, respectively. In (b) and
(d), the solid [short-dashed] and long-dashed [dot-dashed] curves
correspond to the $\bar uu\, [\bar ss]$ contents of $h_1(1170)$ and
$h_1(1380)$, respectively, where $\theta_{^1P_1}=10^\circ
(45^\circ)$ have been used for heavier (lighter) curves. The
definitions for the LCDAs of $h_1(1170)$ and $h_1(1380)$ have been
given in Appendix \ref{appsec:def-f1-h1}.}}}
\end{figure}

\begin{figure}[t!]
\epsfxsize=14cm \centerline{\epsffile{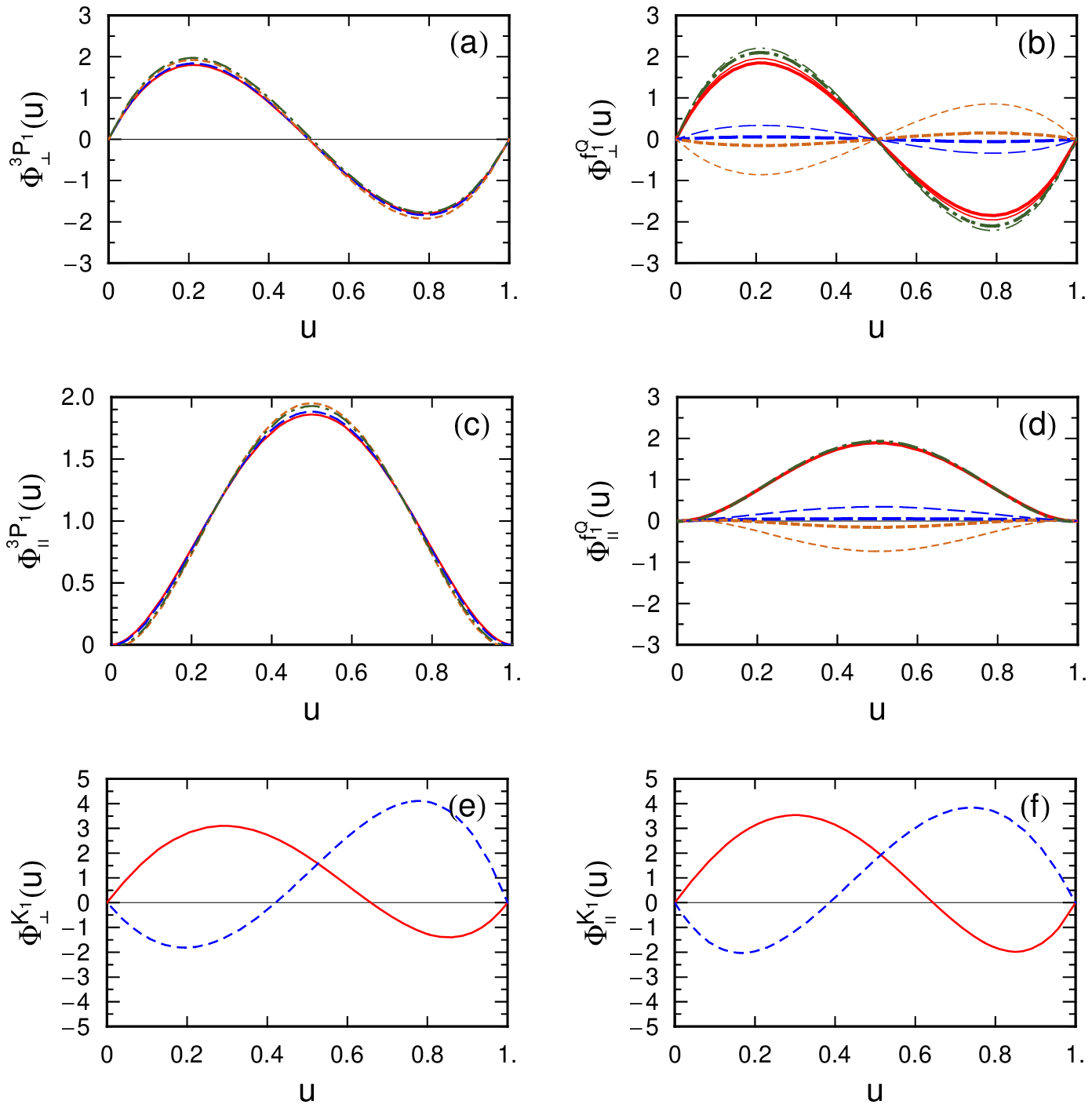}}
\centerline{\parbox{14cm}{\caption{\label{fig:lcda-t2-3p1}
Leading-twist light-cone distribution amplitudes, normalized at the
scale $\mu=$1~GeV, for $1^3P_1$ states, and for $K_1(1270)$ and
$K_1(1400)$ mesons, where the central values of Gegenbauer moments
given in Table \ref{tab:Gegenbauer} are used. $u$ ($\bar u\equiv
1-u$) is the meson momentum fraction carried by the quark
(antiquark). In (a) and (c), the solid, long-dashed, short-dashed
and dot-dashed curves correspond to $a_1(1260), f_1(singlet),
f_8(octet)$ and $K_{1A}$, respectively. In (b) and (d), the solid
[short-dashed] and long-dashed [dot-dashed] curves respectively
correspond to the $\bar uu\, [\bar ss]$ contents of $f_1(1285)$ and
$f_1(1420)$, where $\theta=38^\circ (50^\circ)$ have been used for
heavier (lighter) curves. The definitions for the LCDAs of
$f_1(1285)$ and $f_1(1420)$ have been given in Appendix
\ref{appsec:def-f1-h1}. In (e) and (f), the solid (dashed) and
dashed (solid) curves correspond to $K_1(1270)$ and $K_1(1400)$,
respectively, for $\theta_K=45^\circ$ ($\theta_K=-45^\circ$).}}}
\end{figure}

In Figs.~\ref{fig:lcda-t2-1p1} and \ref{fig:lcda-t2-3p1}, we plot
the twist-2 light-cone distribution amplitudes for $1^1P_1$ and
$1^3P_1$ states, including results for the physical mesons,
$h_1(1170), h_1(1380)$, $f_1(1285), f_1(1420)$, $K_1(1270)$ and
$K_1(1400)$, at the scale $\mu=1$~GeV. $\Phi_\perp^{^1P_1}(u)$ and
$\Phi_\parallel^{^3P_1}(u)$ are symmetric under $u \leftrightarrow
1-u$ if neglecting SU(3) breaking effects, whereas
$\Phi_\parallel^{^1P_1}(u)$ and $\Phi_\perp^{^3P_1}(u)$ are
antisymmetric.

The contents of $h_1(1170)$ and $h_1(1380)$ are dominated by their
$\bar uu$ and $\bar ss$ components, respectively, which are slightly
different for $\theta_{^1P_1}=10^\circ$ and $45^\circ$. However, the
$\bar ss$ component of $h_1(1170)$ and $\bar uu$ component of
$h_1(1380)$ become significant for $\theta_{^1P_1}=10^\circ$.
Analogously, considering the real states $f_1(1285)$ and
$f_1(1420)$, we find that only the $\bar s s$ fraction of
$f_1(1285)$ is a little sensitive to the singlet-octet mixing angle
$\theta_{^3P_1}$ as changing $\theta_{^3P_1}=38^\circ$ to be
$50^\circ$. In particular, the $\Phi_\perp$ and $\Phi_\parallel$ for
$f_1(1285)$ (or for $f_1(1420)$) are predominated by their $\bar uu$
(or $\bar ss$) component. In Fig.~\ref{fig:lcda-t2-3p1}(e), it is
interesting to note that due to the mixture between $K_{1A}$ and
$K_{1B}$, where the axial-vector mesons contain an $s$ quark and a
light anti-quark $\bar q$, we find that the $\bar q$ (or $s$)
carries a larger momentum fraction  for the $K_1(1270)$\, (or
$K_1(1400))$ meson with respect to $\theta=45^\circ$. Nevertheless,
if $\theta=-45^\circ$,  the $s$ (or $\bar q$) instead carries a
larger momentum fraction  for the $K_1(1270)$\, (or $K_1(1400))$
meson.

%
%

\subsection{Two-parton LCDAs of twist-three}\label{subsec:2plcda-t3}

Using the equations of motion allows one to rewrite the two-parton
LCDAs of twist-three in terms of the leading-twist LCDAs and
three-parton  LCDAs of twist-3. Thus, substituting the twist-2 LCDAs
specified by Eqs. (\ref{eq:lcda-3p1-t2-1}),
(\ref{eq:lcda-3p1-t2-2}), (\ref{eq:lcda-1p1-t2-1}), and
(\ref{eq:lcda-1p1-t2-2}), and three-parton LCDAs of twist-3
specified by Eqs.~(\ref{eq:lcda-3p1-t3-1})-(\ref{eq:lcda-1p1-t3-3})
into Eqs.~(\ref{eq:even-sol1}), (\ref{eq:even-sol2}),
(\ref{eq:odd-sol1}), and (\ref{eq:odd-sol2}), we get the approximate
expressions in linear in quark masses (valid up to conformal spin
$9/2$):
\begin{eqnarray}
 g_\perp^{(a)}(u) & = &  \frac{3}{4}(1+\xi^2)
+ \frac{3}{2}\, a_1^\parallel\, \xi^3
 + \left(\frac{3}{7} \,
a_2^\parallel + 5 \zeta_{3,^3P_1}^V \right) \left(3\xi^2-1\right)
 \nonumber\\
& & {}+ \left( \frac{9}{112}\, a_2^\parallel + \frac{105}{16}\,
 \zeta_{3,^3P_1}^A - \frac{15}{64}\, \zeta_{3,^3P_1}^V \omega_{^3P_1}^V
 \right) \left( 35\xi^4 - 30 \xi^2 + 3\right) \nonumber\\
 & &
 + 5\Bigg[ \frac{21}{4}\zeta_{3,^3P_1}^V \sigma_{^3P_1}^V
  + \zeta_{3,^3P_1}^A \bigg(\lambda_{^3P_1}^A -\frac{3}{16}
 \sigma_{^3P_1}^A\Bigg) \Bigg]\xi(5\xi^2-3)
 \nonumber\\
& & {}-\frac{9}{2} \bar{a}_1^\perp
\,\widetilde{\delta}_+\,\left(\frac{3}{2}+\frac{3}{2}\xi^2+\ln u
 +\ln\bar{u}\right) - \frac{9}{2} \bar{a}_1^\perp\,\widetilde{\delta}_-\, (
3\xi + \ln\bar{u} - \ln u), \label{eq:ga-3p1}\\
g_\perp^{(v)}(u) & = & 6 u \bar u \Bigg\{ 1 +
 \Bigg(a_1^\parallel + \frac{20}{3} \zeta_{3,^3P_1}^A
 \lambda_{^3P_1}^A\Bigg) \xi\nonumber\\
 && + \Bigg[\frac{1}{4}a_2^\parallel + \frac{5}{3}\,
 \zeta^V_{3,^3P_1} \left(1-\frac{3}{16}\, \omega^V_{^3P_1}\right)
 +\frac{35}{4} \zeta^A_{3, ^3P_1}\Bigg] (5\xi^2-1) \nonumber\\
 &&+ \frac{35}{4}\Bigg(\zeta_{3,^3P_1}^V
 \sigma_{^3P_1}^V -\frac{1}{28}\zeta_{3,^3P_1}^A
 \sigma_{^3P_1}^A \Bigg) \xi(7\xi^2-3) \Bigg\}\nonumber\\
& & {} -18 \, \bar{a}_1^\perp\widetilde{\delta}_+ \,  (3u \bar{u} +
\bar{u} \ln \bar{u} + u \ln u ) - 18\,
\bar{a}_1^\perp\widetilde{\delta}_- \,  (u \bar{u}\xi + \bar{u} \ln \bar{u} -
u \ln u),
 \label{eq:gv-3p1}\\
h_\parallel^{(t)}(u) &= & 3a_0^\perp\xi^2+ \frac{3}{2}\,a_1^\perp
\,\xi (3 \xi^2-1) + \frac{3}{2} \Bigg[a_2^\perp \xi +
 \zeta^\perp_{3,^3P_1}\Bigg(5
-\frac{\omega_{^3P_1}^{\perp}}{2}\Bigg)\Bigg]\, \xi \,(5\xi^2-3)
\nonumber\\
 && +\frac{35}{4}\zeta^\perp_{3,^3P_1} \sigma^\perp_{^3P_1}
 (35\xi^4-30\xi^2+3)
  + 18 \bar{a}_2^\parallel
  \Bigg[\delta_+ \xi -\frac{5}{8}\delta_- (3\xi^2-1)\Bigg]\nonumber\\
 && -
  \frac{3}{2}\, \Bigg( \delta_+ \, \xi [2 +  \ln (\bar{u}u)]
   +\,\delta_- \, [ 1 + \xi \ln (\bar{u}/u) ]\Bigg)
   (1+ 6 \bar{a}_2^\parallel)
 ,\label{eq:ht-3p1}\\
h_\parallel^{(p)}(u) & = & 6u\bar u \Bigg\{ a_0^\perp +
\Bigg[a_1^\perp +5\zeta^\perp_{3,
^3P_1}\Bigg(1-\frac{1}{40}(7\xi^2-3)
 \omega_{^3P_1}^{\perp} \Bigg)\Bigg] \xi \nonumber\\
 && \ \ \ \ \ \ \ + \Bigg( \frac{1}{4}a_2^\perp
 +
 \frac{35}{6} \zeta^\perp_{3,^3P_1} \sigma^\perp_{^3P_1} \Bigg)
 (5\xi^2-1)
 -5\bar{a}_2^\parallel
  \Bigg[\delta_+ \xi + \frac{3}{2} \delta_- (1-\bar{u} u) \Bigg]\Bigg\}
  \nonumber\\
 & & {}- 3[\, \delta_+\, (\bar{u} \ln \bar{u} - u \ln u)
 + \,\delta_-\,  ( u \bar{u} + \bar{u} \ln \bar{u} + u \ln u)]
 (1+ 6 \bar{a}_2^\parallel),
 \label{eq:hp-3p1}
 \end{eqnarray}
with the normalization conditions
\begin{eqnarray}
 \int_0^1 du g_\perp^{(a)}(u)     &=&\int_0^1 dug_\perp^{(v)}(u)=1\,,
 \label{eq:nomalization-3P1-1}\\
 \int_0^1 du h_\parallel^{(t)}(u) &=& a^\perp_0,
 \label{eq:nomalization-3P1-2}\\
 \int_0^1 du h_\parallel^{(p)}(u) &=& a^\perp_0 + \delta_- \,,
 \label{eq:nomalization-3P1-3}
\end{eqnarray}
for {\it pure} $^3P_1$ states, and
\begin{eqnarray}
 g_\perp^{(a)}(u) & = & \frac{3}{4} a_0^\parallel (1+\xi^2)
+ \frac{3}{2}\, a_1^\parallel\, \xi^3
 + 5\left[\frac{21}{4} \,\zeta_{3,^1P_1}^V
 + \zeta_{3,^1P_1}^A \Bigg(1-\frac{3}{16}\omega_{^1P_1}^A\Bigg)\right]
 \xi\left(5\xi^2-3\right)
 \nonumber\\
& & {}+ \frac{3}{16}\, a_2^\parallel \left(15\xi^4 -6 \xi^2 -1\right)
 + 5\, \zeta^V_{3,^1P_1}\lambda^V_{^1P_1}\left(3\xi^2 -1\right)
 \nonumber\\
& & {}+ \frac{105}{16}\left(\zeta^A_{3,^1P_1}\sigma^A_{^1P_1}
-\frac{1}{28} \zeta^V_{^1P_1}\sigma^V_{^1P_1}\right)
 \left(35\xi^4 -30 \xi^2 +3\right)\nonumber\\
 & & {}-15\bar{a}_2^\perp \bigg[ \widetilde{\delta}_+ \xi^3 +
 \frac{1}{2}\widetilde{\delta}_-(3\xi^2-1) \bigg] \nonumber\\
& & {}
  -\frac{3}{2}\,\bigg[\widetilde{\delta}_+\, ( 2 \xi + \ln\bar{u} -\ln u)
 +\, \widetilde{\delta}_-\,(2+\ln u + \ln\bar{u})\bigg](1+6\bar{a}_2^\perp)
 ,\label{eq:ga-1p1}\\
g_\perp^{(v)}(u) & = & 6 u \bar u \Bigg\{ a_0^\parallel +
a_1^\parallel \xi +
 \Bigg[\frac{1}{4}a_2^\parallel
  +\frac{5}{3} \zeta^V_{3,^1P_1}
  \Bigg(\lambda^V_{^1P_1} -\frac{3}{16} \sigma^V_{^1P_1}\Bigg)
  +\frac{35}{4} \zeta^A_{3,^1P_1}\sigma^A_{^1P_1}\Bigg](5\xi^2-1) \nonumber\\
  & & {}  + \frac{20}{3}\,  \xi
 \left[\zeta^A_{3, ^1P_1}
 + \frac{21}{16}
 \Bigg(\zeta^V_{3,^1P_1}- \frac{1}{28}\, \zeta^A_{3,^1P_1}\omega^A_{^1P_1}
  \Bigg)
 (7\xi^2-3)\right]\nonumber\\
 & & {} -5\, \bar{a}_2^\perp [2\widetilde\delta_+ \xi + \widetilde\delta_- (1+\xi^2)]
 \Bigg\}\nonumber\\
 & & {} - 6 \bigg[\, \widetilde{\delta}_+ \, (\bar{u} \ln\bar{u} -u\ln u )
  +\, \widetilde{\delta}_- \, (2u \bar{u} + \bar{u} \ln \bar{u} + u \ln u)\bigg]
  (1+6\bar{a}_2^\perp) ,
 \label{eq:gv-1p1}\\
h_\parallel^{(t)}(u) &= & 3\xi^2+ \frac{3}{2}\,a_1^\perp \,\xi
(3\xi^2-1) + \Bigg[\frac{3}{2} a_2^\perp\, \xi
 + \frac{15}{2}\zeta^\perp_{3, ^1P_1} \Bigg(\lambda^\perp_{^1\!
 P_1} - \frac{1}{10} \sigma^\perp_{^1\! P_1}\Bigg)
 \Bigg]
   \, \xi(5\xi^2-3) \nonumber\\
 && {} +\frac{35}{4}\zeta^\perp_{3,^1P_1}(35\xi^4-30\xi^2+3)\nonumber\\
 & & {} +\frac{9}{2} \bar{a}_1^\parallel\, \xi
 \Bigg[\delta_+\, (\ln u - \ln \bar{u} -3\xi)
  - \delta_-\,  \Bigg( \ln u + \ln \bar{u}
  +\frac{8}{3}\Bigg)\Bigg]\,,
\label{eq:ht-1p1}\\
 h_\parallel^{(p)}(u) & = & 6u\bar u \Bigg\{ 1 + a_1^\perp \xi
 +  \left(\frac{1}{4}a_2^\perp +
 \frac{35}{6}\,\zeta^\perp_{3,^1P_1} \right)(5\xi^2-1)
  \nonumber\\
 & & {} +5\zeta^\perp_{3,^1P_1}
 \Bigg[\lambda^\perp_{^1\! P_1}-\frac{1}{40}(7\xi^3-3)\sigma^\perp_{^1\! P_1}
 \Bigg] \, \xi
 \Bigg\}  \nonumber\\
 & & {} -9\bar{a}_1^\parallel\, \delta_+\, (3 u \bar{u} + \bar{u} \ln
\bar{u} + u \ln u)
 -9\bar{a}_1^\parallel\,\delta_-\,  \Bigg( \frac{2}{3}\xi u\bar{u}
 + \bar{u} \ln \bar{u} - u \ln u \Bigg)\,, \label{eq:hp-1p1}
\end{eqnarray}
with the normalization conditions
\begin{eqnarray}
 \int_0^1 dug_\perp^{(a)}(u) &=& a^\parallel_0 , \label{eq:nomalization-1P1-1}\\
 \int_0^1 du g_\perp^{(v)}(u) &=& a^\parallel_0 + \widetilde\delta_-,
 \label{eq:nomalization-1P1-2}\\
 \int_0^1 du h_\parallel^{(t)}(u) &=& \int_0^1 du h_\parallel^{(p)}(u) = 1\,,
 \label{eq:nomalization-1P1-3}
\end{eqnarray}
for {\it pure} $^1P_1$ states. Note that to include the corrections
consistently in linear in quark masses, in Eqs.
(\ref{eq:ga-3p1})-(\ref{eq:hp-3p1}) and
(\ref{eq:ga-1p1})-(\ref{eq:hp-1p1}) the parameters with the ``{\it
bar}" should be replaced by the corresponding ones in the massless
quark limit. For the physical $h_1(1170), h_1(1380)$, $f_1(1285)$
and $f_1(1420)$ mesons, their two-parton LCDAs of twist-3 are
defined in Appendix \ref{appsec:def-f1-h1}. The LCDAs for
$K_1(1270)$ and $K_1(1400)$ are given by
 \begin{eqnarray}
 g_\perp^{(a,v)K_1(1270)}
 &=&
 \frac{f_{K_{1A}} m_{K_{1A}}}{f_{K_1(1270)} m_{K_1(1270)}}
  g_\perp^{(a,v)K_{1A}} \sin{\theta_K}
    + \frac{f_{K_{1B}} m_{K_{1B}}}{f_{K_{1}(1270)} m_{K_1(1270)}}
   g_\perp^{(a,v)K_{1B}} \cos{\theta_K}, ~~~~\\
 g_\perp^{(a,v) K_1(1400)}
 &=&
 \frac{f_{K_{1A}} m_{K_{1A}}}{f_{K_1(1400)} m_{K_1(1400)}}
  g_\perp^{(a,v) K_{1A}} \cos{\theta_K}
    - \frac{f_{K_{1B}} m_{K_{1B}}}{f_{K_1(1400)} m_{K_1(1400)}}
  g_\perp^{(a,v) K_{1B}} \sin{\theta_K},
 \end{eqnarray}

 \begin{eqnarray}
 h_\parallel^{(t,p)K_1(1270)}
 &=&
 \frac{f_{K_{1A}}^\perp m_{K_{1A}}^2}{f_{K_1(1270)}^\perp m_{K_1(1270)}^2}
  h_\parallel^{(t,p) K_{1A}} \sin{\theta_K}
    + \frac{f_{K_{1B}}^\perp m_{K_{1B}}^2}{f_{K_{1}(1270)}^\perp m_{K_1(1270)}^2}
   h_\parallel^{(t,p) K_{1B}}\cos{\theta_K}, ~~~~\\
 h_\parallel^{(t,p) K_1(1400)}
 &=&
 \frac{f_{K_{1A}}^\perp m_{K_{1A}}^2}{f_{K_1(1400)}^\perp m_{K_1(1400)}^2}
  h_\parallel^{(t,p) K_{1A}} \cos{\theta_K}
    - \frac{f_{K_{1B}}^\perp m_{K_{1B}}^2}{f_{K_1(1400)}^\perp m_{K_1(1400)}^2}
  h_\parallel^{(t,p) K_{1B}} \sin{\theta_K}.
 \end{eqnarray}

\begin{figure}[t!]
\begin{center}
 \centerline{
{\epsfxsize 5.8in \epsffile{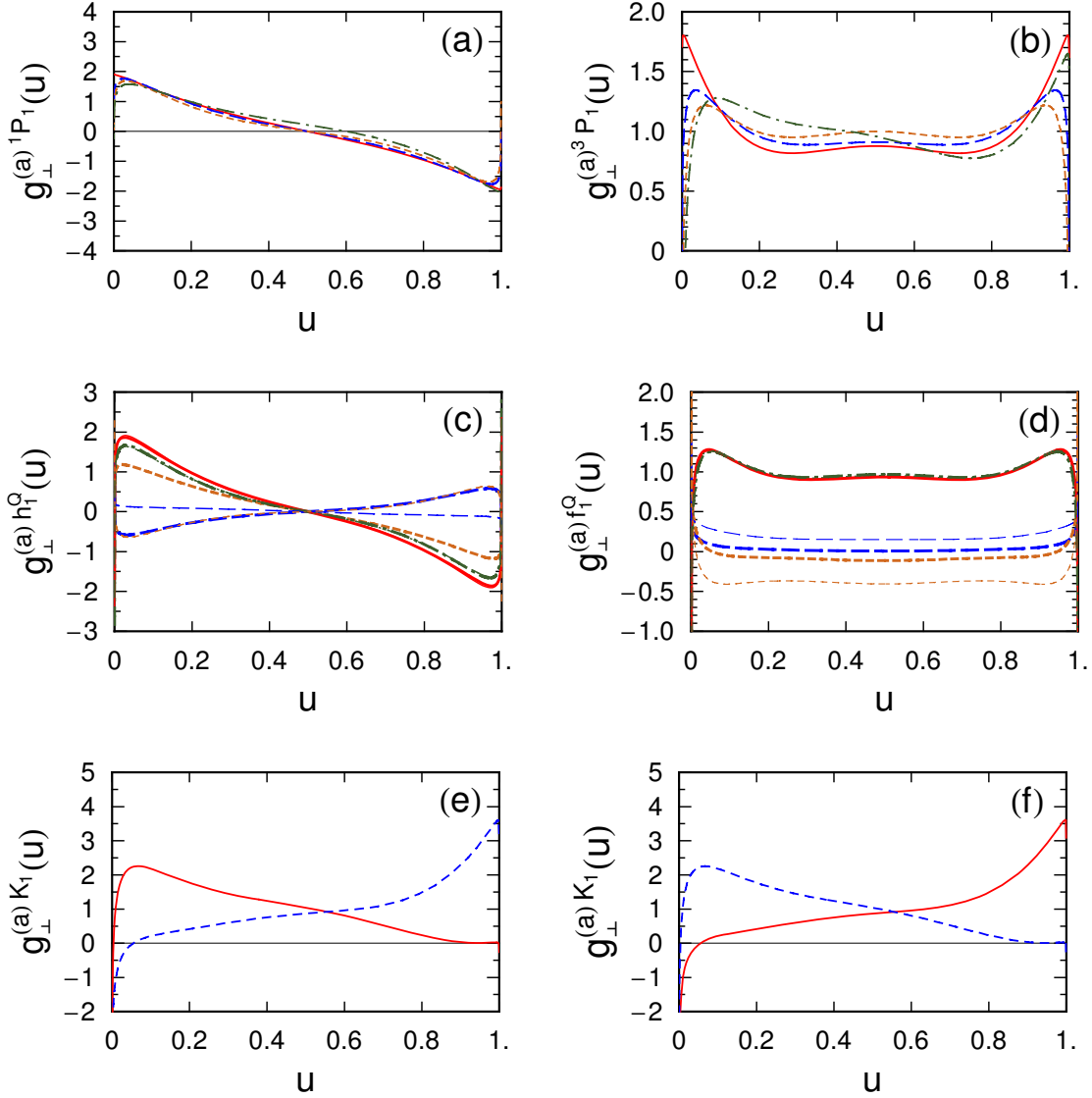}} }
\centerline{\parbox{14.1cm}{\caption{\label{fig:t3-ga} Twist-3
light-cone distribution amplitudes $g_\perp^{(a)}$ at the scale
$\mu=$1~GeV, where the central values of parameters given in Tables
\ref{tab:Gegenbauer}, \ref{tab:para-3p-t3} and \ref{tab:para-3p-t1}
are used. In (e), where $\theta_K=45^\circ$, and in (f), where
$\theta_K=-45^\circ$, the solid and dashed curves correspond to
$K_1(1270)$ and $K_1(1400)$, respectively. Others are the same as
Figs.~\ref{fig:lcda-t2-1p1} and \ref{fig:lcda-t2-3p1}. }}}
\end{center}
\end{figure}

\begin{figure}[t!]
\begin{center}
 \centerline{
{\epsfxsize 5.8in \epsffile{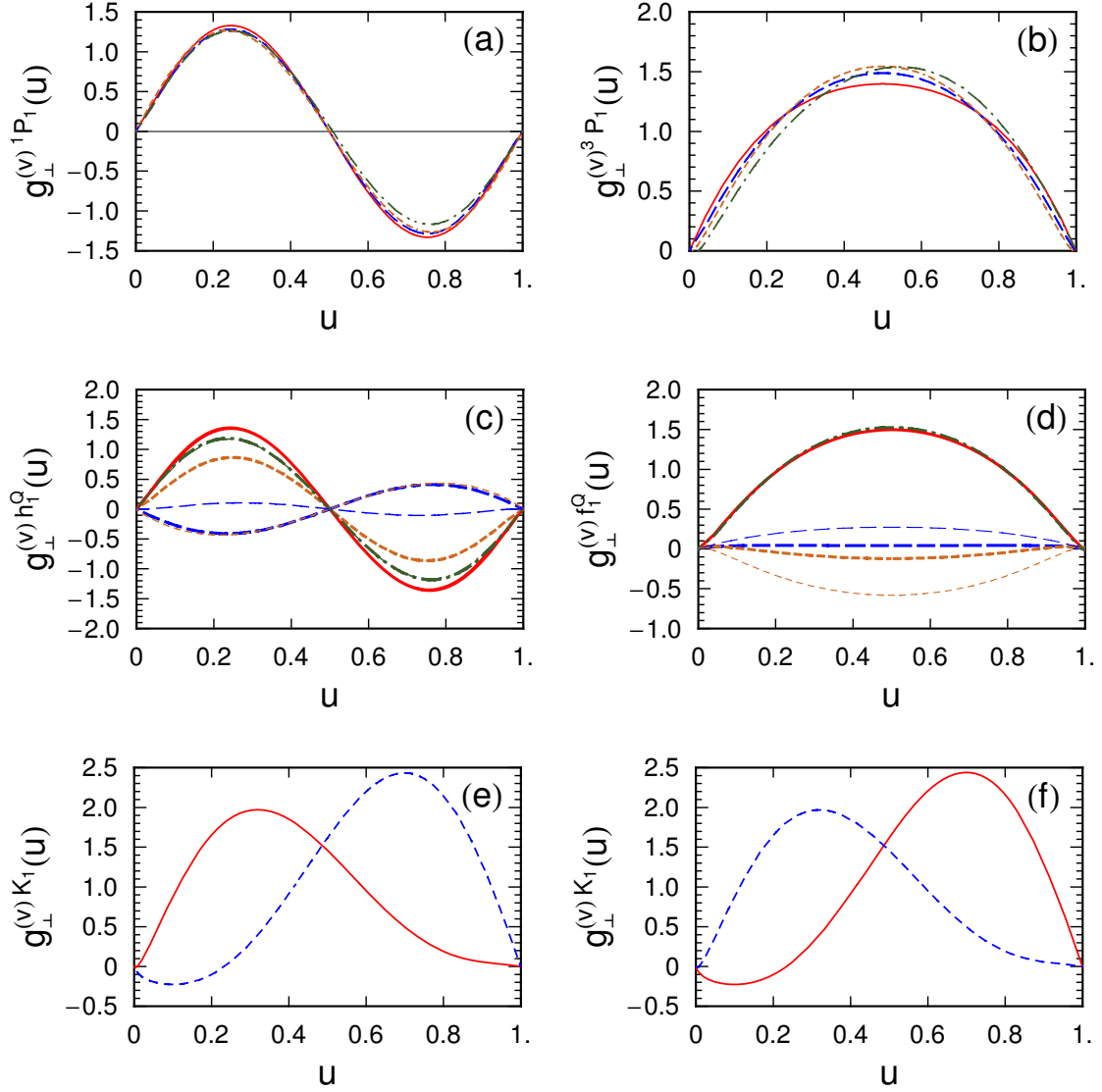}}           }
\centerline{\parbox{14.1cm}{\caption{\label{fig:t3-gv} Twist-3
light-cone distribution amplitudes $g_\perp^{(v)}$ at the scale
$\mu=$1~GeV. Others are the same as Fig.~\ref{fig:t3-ga}.}}}
\end{center}
\end{figure}

\begin{figure}[t!]
\begin{center}
 \centerline{
{\epsfxsize 5.8in \epsffile{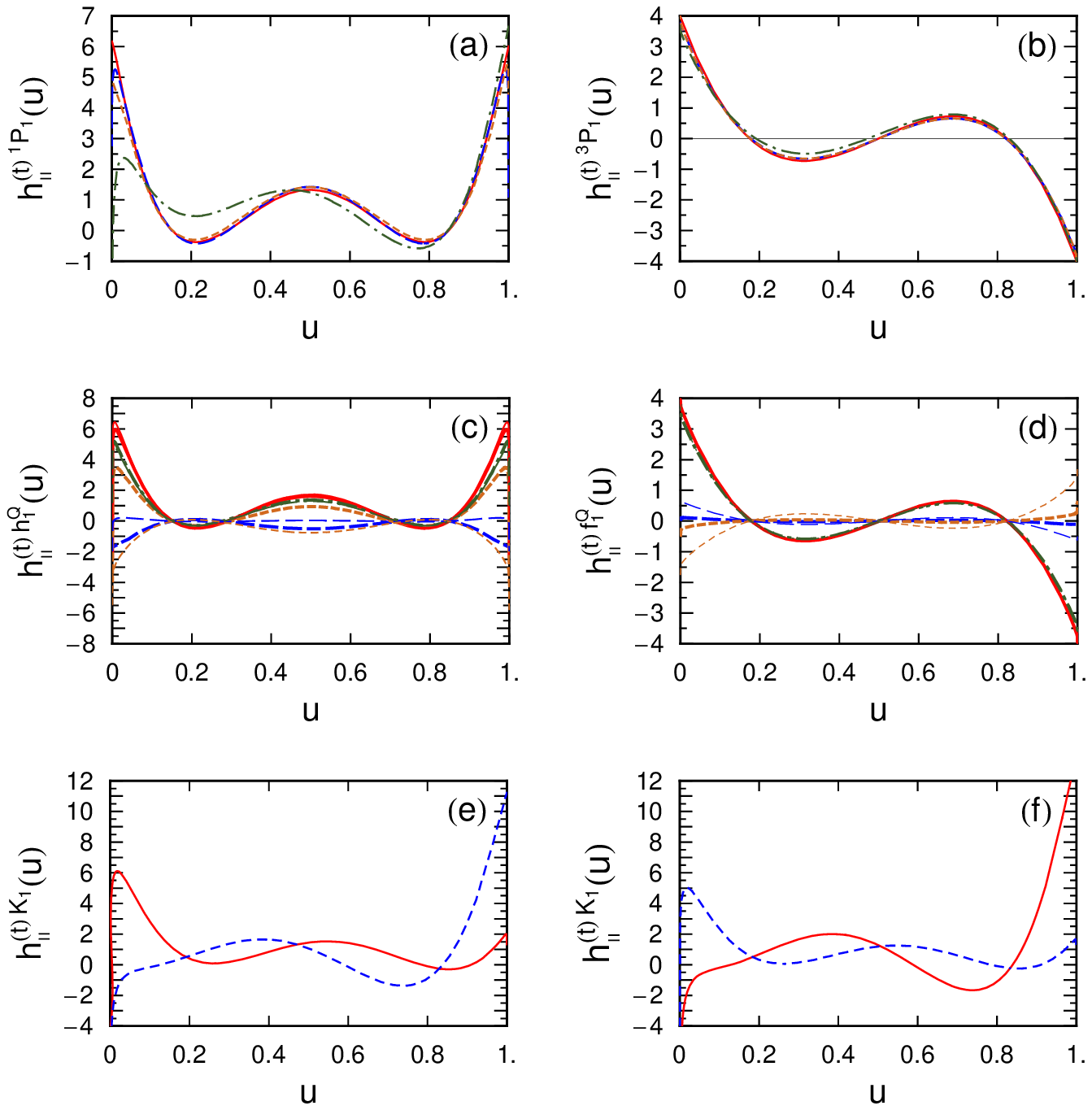}}           }
\centerline{\parbox{14cm}{\caption{\label{fig:t3-ht} Twist-3
light-cone distribution amplitudes $h_\parallel^{(t)}$ at the scale
$\mu=$1~GeV. Others are the same as Fig.~\ref{fig:t3-ga}. }}}
\end{center}
\end{figure}

\begin{figure}[t!]
\begin{center}
 \centerline{
{\epsfxsize 5.8in \epsffile{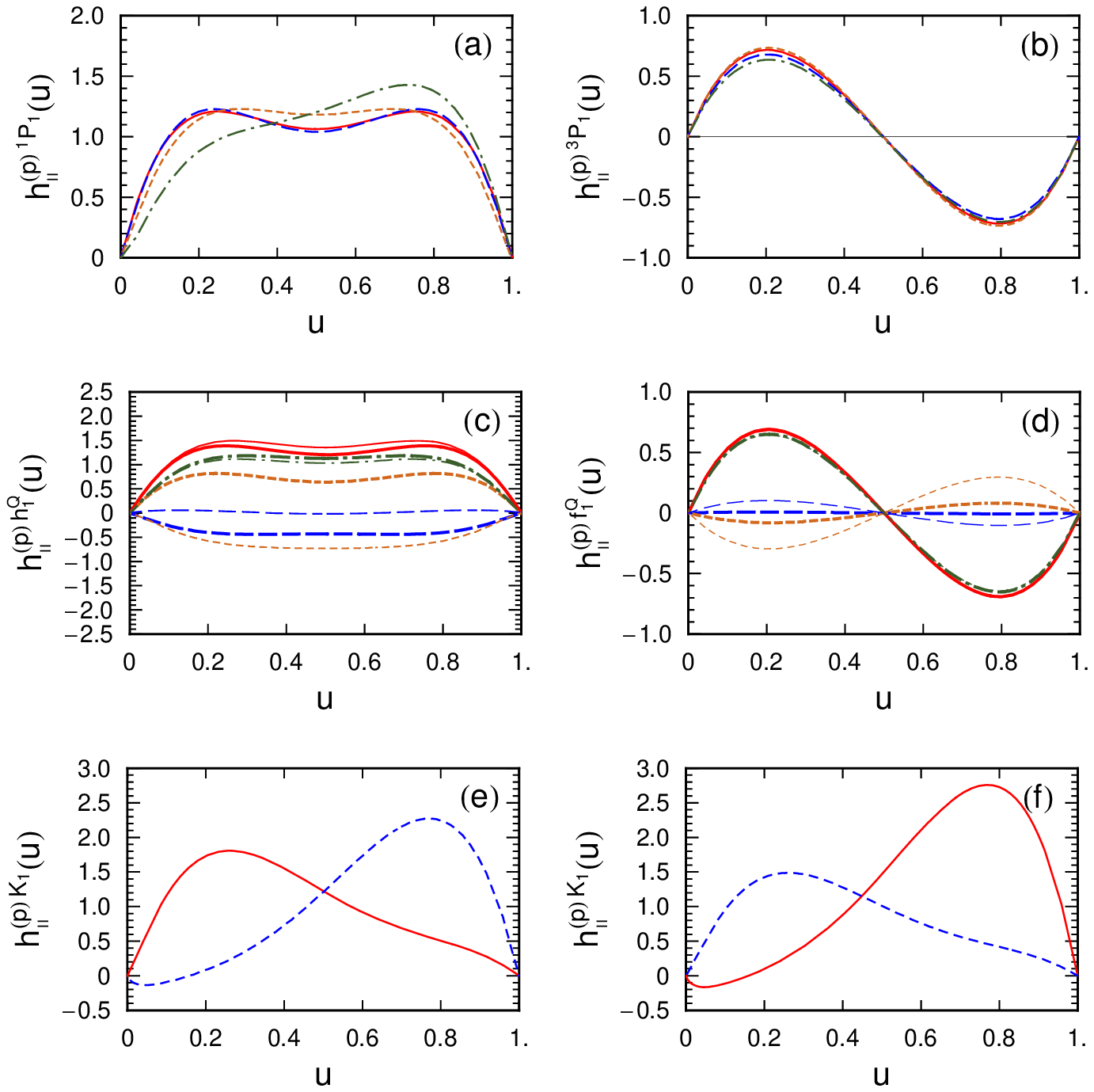}}           }
\centerline{\parbox{14cm}{\caption{\label{fig:t3-hp} Twist-3
light-cone distribution amplitudes $h_\parallel^{(p)}$ at the scale
$\mu=$1~GeV. Others are the same as Fig.~\ref{fig:t3-ga}.}}}
\end{center}
\end{figure}

Substituting the central values of parameters given in
Tables~\ref{tab:Gegenbauer}, \ref{tab:para-3p-t3}, and
\ref{tab:para-3p-t1} into the above equations, in
Figs.~\ref{fig:t3-ga}-\ref{fig:t3-hp} we plot two-parton LCDAs of
twist-3 at the scale $\mu=1$~GeV. The properties of twist-3
two-parton LCDAs are analogous to the cases of the leading-twist
LCDAs that we have given previously. It is interesting to note again
that the LCDAs, $g_\perp^{(a)},g_\perp^{(v)}, h_\parallel^{(t)}$,
and $h_\parallel^{(p)}$, for $f_1(1285)$ (or $f_1(1420)$) are
dominated by the $\bar uu$ (or $\bar ss$) content which is
insensitive to singlet-octet mixing angle $\theta_{^3P_1}$ in the
range $38^\circ < \theta_{^3P_1} <50^\circ$. Moreover,
$g_\perp^{(a)},g_\perp^{(v)}, h_\parallel^{(t)}$, and
$h_\parallel^{(p)}$ for $h_1(1170)$ (or $h_1(1380)$) are dominated
by the $\bar uu$ (or $\bar ss$) content for
$\theta_{^1P_1}=45^\circ$ but the $\bar ss$ (or $\bar uu$) content
becomes significant for $\theta_{^1P_1}=10^\circ$.

\subsection{Three-parton LCDAs of twist-three}\label{subsec:3plcda-t3}

\begin{figure}[t!]
\begin{center}
 \centerline{
{\epsfxsize 5.in \epsffile{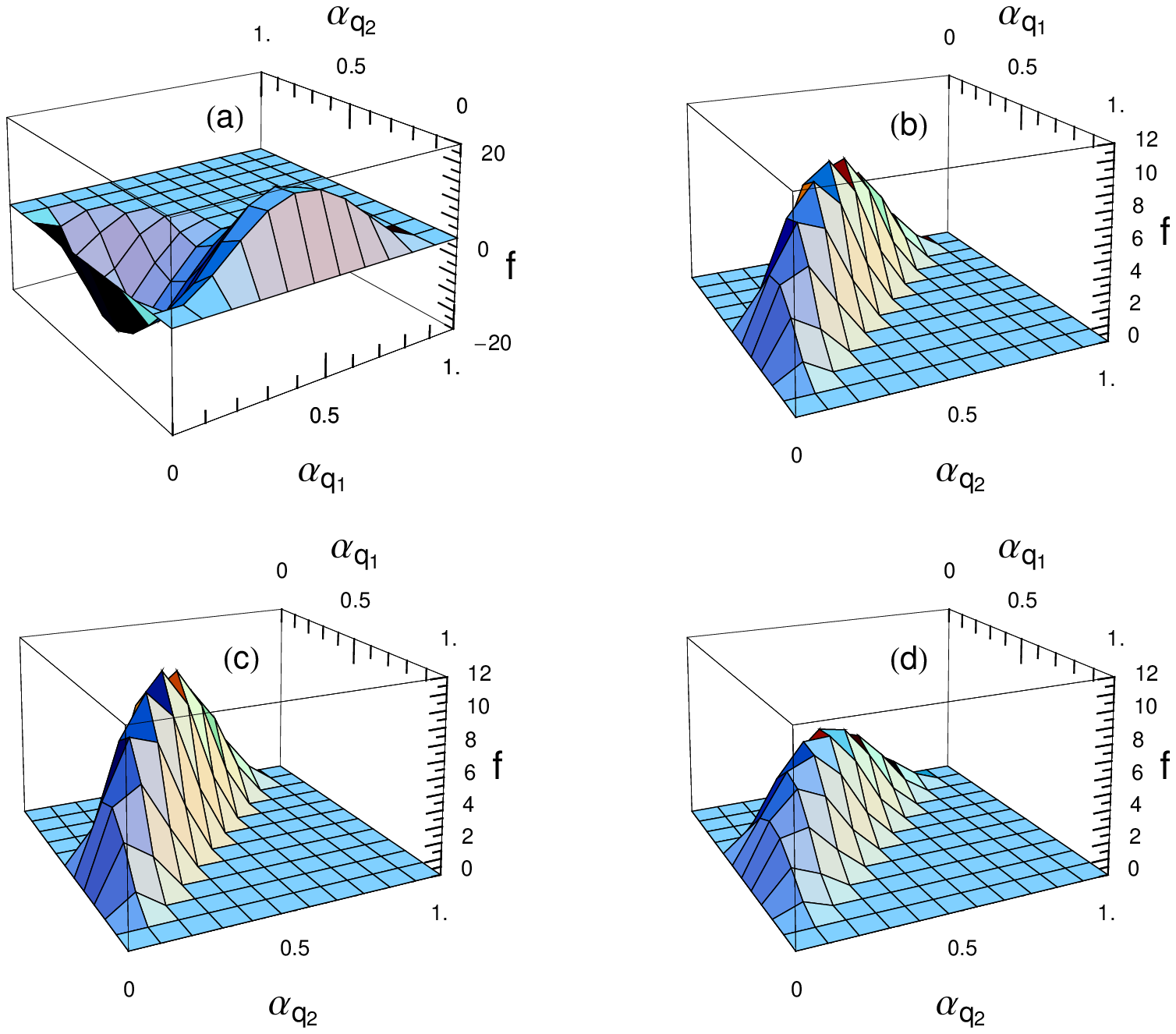}}           }
\centerline{\parbox{14cm}{\caption{\label{fig:3p-lcda} Twist-3
three-parton light-cone distribution amplitudes for the $a_1(1260)$
and $b_1(1235)$ mesons at the scale $\mu=$1~GeV: (a) $f\equiv {\cal
A}$ for $a_1(1260)$ or $f\equiv {\cal V, T}$ for $b_1(1235)$, (b)
$f\equiv {\cal V}$ for $a_1(1260)$, (c) $f\equiv {\cal T}$ for
$a_1(1260)$, and (d) $f\equiv {\cal A}$ for $b_1(1235)$, where we
have taken the central values of $\omega_{a_1}^\perp,
\omega_{a_1}^V$, and $\omega_{b_1}^A$, given in Tables
\ref{tab:para-3p-t3} and \ref{tab:para-3p-t1}, as inputs. Here
$\alpha_{q_1}$ and $\alpha_{q_2}$ are momentum fractions carried by
the quark and anti-quark in an axial-vector meson, respectively. The
gluon momentum fraction is substituted as
$\alpha_g=1-\alpha_{q_1}-\alpha_{q_2}$.}}}
\end{center}
\end{figure}

The approximate three-parton LCDAs of twist-3 are given in
Eqs.~(\ref{eq:lcda-3p1-t3-1})-(\ref{eq:lcda-1p1-t3-3}) and the
relevant parameters are summarized in Tables \ref{tab:para-3p-t3}
and \ref{tab:para-3p-t1}. For completeness and simplicity, we plot
the LCDAs for the $a_1(1260)$ and $b_1(1235)$ mesons in
Fig.~\ref{fig:3p-lcda} to illustrate their behaviors.

\section{Summary}\label{sec:summary}

The light-cone distribution amplitudes specified by the collinear
twist can be expanded in terms of the series of the so-called
conformal partial waves. Each partial wave is characterized by a
specific conformal spin. For each conformal spin, the dependence of
the distribution amplitudes on the transverse coordinates is
governed by the renormalization group equation, while the dependence
on the longitudinal coordinates is involved in "spherical harmonics"
of the $SL(2,\mathbb{R})$ group.

We have presented a detailed study of twist-2 and twist-3 light-cone
distribution amplitudes of axial-vector mesons, based on the QCD
conformal partial wave expansion. The equations of motion allow us
to obtain the relations among the twist-two and twist-three
light-cone distribution amplitudes \cite{Ball:1998sk}, so that we
can use a minimal number of independent nonperturbative parameters
to describe the distribution amplitudes. The conformal partial wave
related by equations of motion should correspond to the same
conformal spin since equations of motion in the QCD perturbative
theory respect all symmetries given in the classical level.

Our main results are as follows:

\begin{itemize}

\item
In subsections~\ref{subsec:decay-constant-1} and
\ref{subsec:tensor_1p1} we have shown the sum rule results for the
axial-vector (tensor) decay constants of $1^3P_1$ ($1^1P_1$)
axial-vector mesons, where we have updated the values for $1^1P_1$
states obtained in Ref.~\cite{Yang:2005tv}.

\item Using Gell-Mann-Okubo mass formula, we have obtained the
mixing angles for the $f_8$ (octet) and $f_1$ (singlet) of $1^3P_1$
states to be $\theta_{^3P_1}\sim 38^\circ$, and for $h_8$ (octet)
and $h_1$ (singlet) of $1^1P_1$ states to be $\theta_{^1P_1}\sim
10^\circ$. Thus the decay constants and light-cone distribution
amplitudes for these states are determined.

\item The sum rules for the first few
Gegenbauer moments of the leading-twist light-cone distribution
amplitudes together with their numerical results have been given in
subsection~\ref{subsec:moments-t2}, where the SU(3) breaking effects
relevant to the $K_{1A}$ and $K_{1B}$ states are included. The
results for $1^3P_1$ states, $h_1, h_8$ and for G-parity violating
Gegenbauer moments are new, while the results of G-parity invariant
Gegenbauer moments for $b_1$ and $K_{1B}$ (which are $1^1P_1$
states) are updated. In the sum rules, because the G-parity
violating Gegenbauer moments of $1^3P_1$ and $1^1P_1$ are always
mixed togther, we thus add a reasonable constraint
\begin{eqnarray}
 \frac{a_0^{\perp,K_{1A}}}
  { a_0^{\parallel,K_{1B}}
 \frac{m_{K_{1B}} f_{K_{1B}} f_{K_{1B}}^\perp}
  { m_{K_{1A}} f_{K_{1A}} f_{K_{1A}}^\perp}}
 =1.0 \pm 0.3\,, \nonumber
\end{eqnarray}
which is a good approximation for G-invariant Gegenbauer moments, to
obtain qualitative estimates. See the detailed discussions in
subsection~\ref{subsec:gegenbauer-3p1-result}.

\item In Sec.~\ref{sec:properties-3pdas}, using the QCD sum rules, the relevant
G-parity invariant and violating parameters for expanding the
three-parton distribution amplitudes of twist-3 in terms of
conformal partial waves with conformal spin up to $9/2$ have been
evaluated, where the SU(3) corrections have been contained. To
determined not only the magnitudes but also the relative signs for
the parameters, one of the interpolating currents in the two-point
correlation functions is chosen to be the local axial-vector (or
pseudo-tensor) current in calculating the parameters for the $1
^3P_1$ (or $1 ^1P_1$) state. All the results are new. We have
checked all the calculations very carefully since nobody did these
before. Only the sum rule calculation for $f_{3,^1P_1}^\perp$ is
very similar to that for $f_{3\rho}^T$ in SU(3) limit (See the
discussions after Eq.~(\ref{eq:1P1_SRf3t_sigma})). On the other
hand, the resulting $f_{3,^3P1}^\perp$ sum rule is not reliable
since the calculated OPE series is not well convergent. We further
resort to the diagonal sum rule for $f_{^3P1}^\perp$ and this
calculation can be found in Ref.~\cite{Ball:2006wn}, where the sum
rule is used to study the coupling of the kaon. The diagonal sum
rule is stable and the sign for $f_{^3P1}^\perp$ can be determined
in an indirect way (See Sec.~\ref{sec:properties-3pdas-1-3} for more
discussions). It should be noted that as G-parity violating
parameters for $1^3P_1$ ($1^1P_1$) states are computed, the
corrections receiving from $1^1P_1$ ($1^3P_1$) states have to be
considered.

\item  Adapting the EOM formulas derived in
Ref.~\cite{Ball:1998sk} for vector mesons to the present case, the
two-parton LCDAs of twist-three can be written in terms of the
leading-twist LCDAs and three-parton  LCDAs of twist-3. The detailed
results for twist-three three-parton LCDAs are shown in
Sec.~\ref{subsec:2plcda-t3}. In the SU(3) limit, the detailed
symmetric properties of LCDAs of axial-vector mesons, as compared
with that of the vector mesons, are summarized in
Table~\ref{tab:symmetric}.

\item Using the conformal partial expansion, we presented the models
for light-cone distribution amplitudes, containing contributions
with conformal spin up to 9/2, in Sec.~\ref{sec:lcda}.

\item We have considered the strange quark mass
corrections to distribution amplitudes for the strange axial-vector
mesons, $K_1(1270)$ and $K_1(1400)$. It is interesting to note that
$\Phi_\perp^{^3P_1}(u), \Phi_\parallel^{^1P_1}(u),
g_\perp^{(a)^1P_1}, g_\perp^{(v)^1P_1}, h_\parallel^{(t)^3P_1},
h_\parallel^{(p)^3P_1}$ have significant antisymmetric behaviors,
which should be phenomenologically attractive.

\end{itemize}
It should be noted that because corrections due the higher resonance
and radiative correction in OPE may partially cancel each other for
non-diagonal sum rules, it is estimated that the resultant errors
for parameters of LCDAs may be $\sim$ 10\%. On the other hand,
because we do not calculate the radiative corrections to the
perturbative term in diagonal sum rules, these corrections may also
lead to $\sim$ 10\% errors for results. However, since we do not do
the qualitative calculations about these effects, we thus do not
include this possible error in Tables \ref{tab:Gegenbauer},
\ref{tab:para-3p-t3}, and \ref{tab:para-3p-t1}.

Recently, Belle has measured $B^- \to K_1^- (1270) \gamma$ and given
an upper bound on $B^- \to K_1^-(1400) \gamma$~\cite{Yang:2004as}.
Interestingly, the recent calculations~\cite{Kwon:2004ri,Lee:2004ju}
of adopting LCSR (light-cone sum rule) form factors
\cite{Safir:2001cd} gave too small predictions for ${\cal B}(B^- \to
K_1^- (1270) \gamma)$ as compared with the data. Since the physical
states $K_1(1270)$ and $K_1(1400)$ are the mixture of $K_{1A}$ and
$K_{1B}$ which are respectively the pure $1^3P_1$ and $1^1P_1$
states, the light-cone distribution amplitudes of $K_{1A}$ and
$K_{1B}$ are relevant to the results of $B\to K_1(1270)$ and
$K_1(1400)$ transition form factors. It is known that for $K_{1B}$,
$\Phi_\parallel$ is antisymmetric, while $\Phi_\perp$ is symmetric
in the SU(3) limit due to G-parity. Nevertheless, for $K_{1A}$,
$\Phi_\parallel$ becomes symmetric, while $\Phi_\perp$ is
antisymmetric. The above properties were not correctly studied in
the literature. Some related researches will be published
elsewhere~\cite{kcymoments}.

\subsubsection*{Acknowledgments}

This work was supported in
part by the National Science Council of R.O.C. under Grant No:
NSC94-2112-M-033-001.

\vfil
\newpage

\appendix
\renewcommand{\theequation}{\Alph{section}.\arabic{equation}}

\section*{Appendices}

\section{Spin Projections and {\it Collinear} Twist}\label{appsec:spin}
\setcounter{equation}{0}

Considering infinitesimal rotation $x_\mu\to x'_\mu =
x_\mu+\epsilon_{\mu\nu}x^\nu$ in the four dimension,
the general field $\Phi(x)$ transforms as $\Phi'(x') = [1
-\epsilon_{\mu\nu}(x^\mu\partial^\nu-x^\nu\partial^\mu
- \Sigma^{\mu\nu})] \Phi(x)
= [1 +i \epsilon_{\mu\nu}M^{\mu\nu}] \Phi(x)$, where  $\Sigma^{\mu\nu}$ is called
the generator of spin rotations of the
field $\Phi$.
For scalar, quark and gluon fields, we have
\begin{eqnarray}
 \Sigma^{\mu\nu}\phi(x) =0, \qquad \Sigma^{\mu\nu}\psi =
\frac{i}{2}\sigma^{\mu\nu}\psi, \qquad
 \Sigma^{\mu\nu}A^{\alpha} =
g^{\nu\alpha}A^{\mu}- g^{\mu\alpha}A^{\nu}\,,
\end{eqnarray}
respectively, where $A_\mu\equiv T^a A^{a}_\mu$.
We can apply the following spin projections
\begin{eqnarray}\label{app-projection-quark-1}
P_+ = \frac12 \gamma_-\gamma_+\,, \qquad
P_- = \frac12 \gamma_+\gamma_-\,, \qquad
P_+ + P_- = 1
\end{eqnarray}
on a quark field $\psi$
to project its typical spin component, so that a quark field with
a given spin component $s$ on the moving direction can be measured to be
\begin{eqnarray}\label{app-projection-quark-2}
\Sigma_{+-} P_{\pm}\psi =s_{\pm} P_{\pm}\psi=\pm \frac{1}{2} P_\pm\psi \,.
\end{eqnarray}
Recalling that the canonical dimension\footnote{Here we do not
distinguish the canonical dimension $\ell^{\rm can}$ and scaling
dimension $\ell$.} of a quark field is $\ell=3/2$, we can therefore
decompose an arbitrary light-ray quark-antiquark current into
different ({\it collinear}) {\it twists} (= dimension $-$ spin
projection on the moving direction) components:
\begin{eqnarray}
 \bar\psi_2(x_{2-}) \Gamma \psi_1(x_{1-})
 &=&
 \underbrace{\bar\psi_2(x_{2-}) P_- \Gamma P_+\psi_1(x_{1-})}_{\mbox{twist-2}}\nonumber\\
 &&{}+ \underbrace{\bar\psi_2(x_{2-}) P_+ \Gamma P_+\psi_1(x_{1-}) +
   \bar\psi_2(x_{2-}) P_- \Gamma P_-\psi_1(x_{1-})}_{\mbox{twist-3}}\nonumber\\
 &&{}+ \underbrace{\bar\psi_2(x_{2-}) P_+ \Gamma P_-\psi_1(x_{1-})}_{\mbox{twist-4}}\,,
\end{eqnarray}
where $\Gamma$ stands for a generic Dirac matrix structure. Note
that, taking $\bar\psi_2(x_{2-}) P_- \Gamma P_+\psi_1(x_{1-})$ as an
example, its ({\it collinear}) twist exactly equals to
$t=\ell_1+\ell_2-s_1-s_2=2$, while its conformal spin is
$j=(\ell_1+\ell_2+s_1+s_2+n)/2=j_1+j_2 +n$ with $n=0,1,2,\cdots$; in
other words, $j$ is not a fixed value for a LCDA defined by a
non-local composite field as seen in
Eq.~(\ref{eq:conformal-expansion-2p}). On the other hand, for a
gluon field with the canonical dimension $\ell=2$, one can find that
\begin{eqnarray}\label{app-projection-gluon}
\Sigma_{+-} G_{+\perp} &=& 1 \cdot G_{+\perp}\,, \nonumber\\
\Sigma_{+-} G_{\perp\perp} &=& 0 \cdot G_{\perp\perp}\,, \qquad
\Sigma_{+-} G_{+-} =0 \cdot G_{+-}\,, \nonumber\\
\Sigma_{+-} G_{-\perp} &=& -1 \cdot G_{-\perp}\,,
\end{eqnarray}
where, for instance, $G_{+\perp}\equiv G_{\mu\nu}n^\mu g^{\nu\alpha}_\perp$
and $g^{\nu\alpha}_\perp=g^{\nu\alpha} - n^\nu \bar n^\alpha -n^\alpha\bar n^\nu$.
In analogy to the previous discussion, we can decompose an arbitrary light-ray
quark-gluon-antiquark current into currents with different twists:
\begin{eqnarray}
 && \bar\psi_2(x_{2-}) \Gamma g_s G_{\mu\nu}(x_{3-})\psi_1(x_{1-})
 =
 \underbrace{\bar\psi_2(x_{2-}) P_- \Gamma  g_s G_{\alpha\beta}(x_{3-})
 P_+ \psi_1(x_{1-})(g^{\alpha\mu}_\perp n^\beta\bar n^\nu
  + n^\alpha\bar n^\mu g^{\beta\nu}_\perp) }_{\mbox{twist-3}}\nonumber\\
 &&{}+ \underbrace{ \bar\psi_2(x_{2-}) P_+ \Gamma g_s G_{\alpha\beta}(x_{3-}) P_+\psi_1(x_{1-})
(g^{\alpha\mu}_\perp n^\beta\bar n^\nu + n^\alpha\bar n^\mu g^{\beta\nu}_\perp)}_{\mbox{twist-4}}
   \nonumber\\
 &&{}+ \underbrace{
   \bar\psi_2(x_{2-}) P_- \Gamma g_s G_{\alpha\beta}(x_{3-})
   P_-\psi_1(x_{1-}) (g^{\alpha\mu}_\perp n^\beta\bar n^\nu
  + n^\alpha\bar n^\mu g^{\beta\nu}_\perp)}_{\mbox{twist-4}}\nonumber\\
 &&{}+ \underbrace{\bar\psi_2(x_{2-}) P_- \Gamma  g_s G_{\alpha\beta}(x_{3-})
 P_+ \psi_1(x_{1-})(g^{\alpha\mu}_\perp g^{\beta\nu}_\perp
  + n^\alpha\bar n^\mu \bar n^{\beta} n^\nu
  + \bar n^\alpha n^\mu  n^{\beta} \bar n^\nu) }_{\mbox{twist-4}}\nonumber\\
  &&{} +{\cal O}({\mbox{twist-5,6,7}})\,.
\end{eqnarray}

\section{Operator Identities}\label{appsec:EOM}
 \setcounter{equation}{0}

The operator identities, which are used to obtain the integral equations,
Eqs.~(\ref{eq:eom-even1}), (\ref{eq:eom-even2}), (\ref{eq:eom-odd1}), and
(\ref{eq:eom-odd2}), are as follows:
\begin{eqnarray}
\bar q_2 (x)\gamma_\mu  q_1 (-x) &=&\int_0^1\, dt\,
{ \partial \over \partial x^\mu}\,\bar q_2 (tx)\not\! x
q_1 (-tx)\nonumber\\
&&{}+ \int_0^1\,dt\,t\,\int_{-t}^t\,dv\, \bar q_2(tx) g_s \tilde{G}_{\mu\nu}(vx)
x^\nu \not\! x \gamma_5 q_1 (-tx)\nonumber\\
&&{}+i
\int_0^1\,dt\,\int_{-t}^t\,dv\,v\, \bar q_2 (tx) g_s G_{\mu\nu}(vx)x^\nu
\not x q_1 (-tx)\nonumber\\
&&{}- i\epsilon_{\mu\nu\alpha\beta}\int_0^1\,dt\,t\,x^\nu\partial^\alpha
\left[ \bar q_2 (tx)\gamma^\beta\gamma_5 q_1 (-tx)\right]\nonumber\\
&&{}+ (m_{q_2} -m_{q_1}) x^\nu \int_0^1\,dt\,t\,\bar q_2 (tx)\sigma_{\nu\mu}
q_1 (-tx)\,, \label{app:vector-id}
\end{eqnarray}
\begin{eqnarray}
\bar q_2 (x)\gamma_\mu\gamma_5 q_1(-x) &=&\int_0^1\, dt\,
{ \partial \over \partial x^\mu}\,\bar q_2(tx)\not\! x \gamma_5
q_1(-tx)\nonumber\\
& &{}+ \int_0^1\,dt\,t\,\int_{-t}^t\,dv\,\bar q_2(tx) g_s\tilde{G}_{\mu\nu}(vx)
x^\nu \not\! x q_1(-tx)\nonumber\\
&&{}+i
\int_0^1\,dt\,\int_{-t}^t\,dv\,v\, \bar q_2 (tx) g_s G_{\mu\nu}(vx)x^\nu
\not\! x\gamma_5 q_1(-tx)\nonumber\\
&&{}- i\epsilon_{\mu\nu\alpha\beta}\int_0^1\,dt\,t\,x^\nu\partial^\alpha
\left[ \bar q_2(tx)\gamma^\beta q_1(-tx)\right]\nonumber\\
&&{}+ (m_{q_2} + m_{q_1}) x^\nu \int_0^1\,dt\,t\,\bar q_2
(tx)\sigma_{\nu\mu}\gamma_5 q_1 (-tx),
\label{app:axial-id}
\end{eqnarray}

\begin{eqnarray}
\frac{\partial}{\partial x_\mu} \,\bar q_2(x)
\sigma_{\mu\nu}\gamma_5 q_1(-x) & = & -i\partial_\nu \bar q_2(x)
\gamma_5 q_1(-x) -
\int_{-1}^1 dv\, \bar q_2(x) x^\alpha g_s G_{\alpha\nu}(vx)\gamma_5 q_1(-x)\nonumber\\
& & {} + i\int_{-1}^1 dv\, v \bar q_2 (x) x_\rho g_s G^{\rho\mu}(vx)
\sigma_{\mu\nu} \gamma_5 q_1(-x)\nonumber\\
&  & {} -(m_{q_2}+m_{q_1})\bar q_2(x) \gamma_\nu \gamma_5 q_1(-x)\,, \label{app:tensor-id}
\end{eqnarray}
\begin{eqnarray}
\bar q_2(x) \gamma_5 q_1(-x) - \bar q_2(0) \gamma_5 q_1(0) &=&
 -\int_{0}^{1} dt \int_{-t}^{t}dv\, \bar q_2(tx)
x^{\alpha} \sigma_{\alpha \beta} x_{\mu} g_s G^{\mu \beta} (vx)
\gamma_5 q_1(-tx) \nonumber\\
&&  + i \int_{0}^{1}\!dt\, \partial^{\alpha}\left\{ \bar q_2(tx)
\sigma_{\alpha \beta}x^{\beta} \gamma_5 q_1(-tx) \right\} \nonumber\\
&& + i (m_{q_2} - m_{q_1}) \int_{0}^{1} \!dt \,\bar q_2(tx) \!\not\!
x \gamma_5 \: q_1(-tx), \label{app:pseudo-id}
\end{eqnarray}
where we have adopted the following notation to stand for the total derivative:
\begin{equation}
\partial_\mu \left\{ \bar q_2(x)\Gamma q_1(-x)\right\} \equiv
\left.\frac{\partial}{\partial y_\mu}\,\left\{ \bar q_2(x+y)
[x+y,-x+y]
    \Gamma q_1(-x+y)\right\}\right|_{y\to 0}.
\end{equation}
Eqs.~(\ref{app:vector-id}), (\ref{app:axial-id}), and
(\ref{app:tensor-id}) have been obtained in
Refs.~\cite{Balitsky:1987bk,Ball:1998sk}, whereas
Eq.~(\ref{app:pseudo-id}) is new.

\section{Input parameters}\label{appsec:inputs}
The theoretical input parameters, used in our analysis, together
with their respective ranges of uncertainty are summarized here. We
take into account $\alpha_s(1~{\rm GeV})=0.497\pm 0.005$,
corresponding to the world average $\alpha_s(m_Z)=0.1176 \pm 0.0020$
\cite{PDG}, and the following relevant parameters at the scale
$\mu=1$~GeV~\cite{Yang:1993bp,Ball:2005vx,Ball:2006wn,Ball:2004rg}:
\begin{eqnarray}
\begin{array}{lcl}
  \langle \alpha_s G_{\mu\nu}^a G^{a\mu\nu} \rangle
     =(0.474\pm 0.120)\ {\rm GeV}^4/(4\pi)\,, &  &   \\
  \langle \bar uu \rangle \cong \langle \bar dd \rangle =-(0.24\pm 0.010)^3~ {\rm
  GeV}^3 \,,
  &   & \langle \bar ss \rangle = (0.8\pm 0.1) \langle \bar uu \rangle \,, \\
  (m_u+m_d)/2=(5\pm 2)\ {\rm MeV}\,, &  & m_s=(140\pm 20)~ {\rm MeV}\,,\\
  \langle g_s \bar u\sigma Gu \rangle \cong \langle g_s\bar
d\sigma Gd \rangle =-(0.8\pm 0.1)\langle \bar uu \rangle, & &\langle
g_s \bar s\sigma Gs \rangle = (0.8\pm 0.1) \langle g_s\bar u\sigma
Gu \rangle,\\
 f_{3PS}=  (0.0045\pm 0.0015)~{\rm GeV}^2\,, &  &   \\
 s_0^\pi = 0.8 ~{\rm GeV}^2\,, & & s_0^{f_1}\simeq s_0^{f_8} \simeq s_0^K=1.1~{\rm
 GeV}^2\,,\\
 a_1^{\perp, K^*}=0.04\pm 0.03\,, & & f_{K^*}^\perp =(185\pm
 10)~{\rm MeV}\,,\\
 a_2^{\perp, \rho}=0.2\pm 0.1\,,& & a_2^{\perp, \phi}=0.0\pm 0.1\,,\\
 a_2^{\perp, K^*}=0.13\pm 0.08\,. & &
\end{array}\label{eq:parameters}
 \end{eqnarray}
where the scale-dependence of operators is given
by~\cite{Yang:1993bp}:
 \begin{eqnarray}\label{app:anamolous}
 && m_{q} (Q)= m_{q} (\mu)
  \left(\frac{\alpha_s(Q)}{\alpha_s(\mu)}\right)^{4/b},
  \nonumber\\
 &&\langle \bar q q\rangle(Q) = \langle \bar q q\rangle (\mu)
 \left(\frac{\alpha_s(Q)}{\alpha_s(\mu)}\right)^{-4/b},\nonumber\\
 && \langle g_s \bar q\sigma\cdot G q\rangle(Q)=
 \langle g_s \bar q\sigma\cdot G q\rangle (\mu)
 \left(\frac{\alpha_s(Q)}{\alpha_s(\mu)}\right)^{2/(3b)},\nonumber\\
 && \langle \alpha_s G^2\rangle(Q) = \langle \alpha_s
 G^2\rangle(\mu),
 \end{eqnarray}
with $b=(11 N_c -2n_f)/3$. As described in
Eq.~(\ref{eq:factorization}), we adopt the vacuum saturation
approximation for describing the four-quark condensates, i.e.,
\begin{eqnarray}
\langle 0|\bar q \Gamma_i \lambda^a q \bar q \Gamma_i \lambda^a
q|0\rangle =-\frac{1}{16N_c^2}{\rm Tr}(\Gamma_i\Gamma_i) {\rm
Tr}(\lambda^a \lambda^a) \langle \bar qq\rangle^2 \,.
\end{eqnarray}
Performing the analysis in analogy to that given in
Ref.~\cite{Ball:1998sk} and using the results in
Refs.~\cite{Koike:1994st,Koike:1996bs,Koike:1998ry} and in Eqs.
(\ref{eq:s1-1}), (\ref{eq:s1-2}), (\ref{eq:s1-3}), (\ref{eq:s2-1}),
(\ref{eq:s2-2}), (\ref{eq:s2-3}), (\ref{eq:s3-1}), (\ref{eq:s3-2}),
(\ref{eq:s3-3}), we obtain the LO scale-dependence of the twist-3
parameters with the light quark mass corrections as below

\noindent [G-parity invariant components]:
\begin{eqnarray}
f_{3, ^3P_1}^{V}(Q) = L^{\Gamma^+_2/b}f_{3, ^3P_1}^{V}(\mu),
 \qquad
 \Gamma_2^+ = -{1\over 3} C_F + 3C_G,  \label{app:scale1-twist3-1}
\end{eqnarray}

\begin{eqnarray}
 & &
 \left(
\begin{array}{c}
f_{3, ^3P_1}^{V}\omega^{V}_{^3P_1} + {28\over 3} f_{3, ^3P_1}^{A}\\
f_{3, ^3P_1}^{V}\omega^{V}_{^3P_1} - {28\over 3} f_{3, ^3P_1}^{A}
\end{array}
\right)_{Q} = L^{\Gamma^{-}_{3}/b} \left(
\begin{array}{c}
f_{3, ^3P_1}^{V}\omega^{V}_{^3P_1} + {28\over 3} f_{3, ^3P_1}^{A}\\
f_{3, ^3P_1}^{V}\omega^{V}_{^3P_1} - {28\over 3} f_{3, ^3P_1}^{A}
\end{array}
\right)_{\mu},\nonumber\\
& &\qquad\qquad\Gamma_{3}^{-} = \left(
\begin{array}{cc}
 \frac{1}{6}C_{F} + {4}C_{G}\ \ &
 \frac{5}{3}C_{F}-\frac{4}{3}C_{G} \\
 \frac{2}{3}C_{F} - \frac{2}{3}C_{G}\ \ &
 \frac{8}{3}C_{F}+\frac{7}{3}C_{G}
\end{array}
\right), \label{app:scale1-twist3-2} \end{eqnarray}
\begin{eqnarray}
 f_{3, ^3P_1}^{\perp} (Q) &=& L^{\Gamma^{T^+_2}/b}f_{3, ^3P_1}^{\perp}(\mu)
 -\frac{4}{19} \bigg(L^{4/b} -
 L^{\Gamma^{T^+_2}/b}\bigg)(m_{q_1}+m_{q_2})(\mu) f_{^3P_1},\nonumber\\
 && \qquad\qquad
 \Gamma^{T^+_2} = {7\over 3} C_F + C_G ,  \label{app:scale1-twist3-3}
\end{eqnarray}
\begin{eqnarray}
f_{3, ^3P_1}^{\perp}\omega^{\perp}_{^3P_1}(Q)
 &=&
 L^{\Gamma^{T^-_3}/b}
 f_{3, ^3P_1}^{\perp}\omega^{\perp}_{^3P_1}(\mu)-\frac{1}{85} \bigg(L^{4/b} -
 L^{\Gamma^{T^-_3}/b}\bigg)(m_{q_1}+m_{q_2})(\mu) f_{^3P_1},\nonumber\\
 && \qquad\qquad
  \Gamma^{T^-_3} = {7\over 6} C_F + \frac{10}{3}C_G ,
  \label{app:scale1-twist3-4}
\end{eqnarray}
[G-parity violating components]:
\begin{eqnarray}
f_{3, ^3P_1}^{A} \lambda_{^3P_1}^{A}(Q) =
 L^{\Gamma^-_2/b}f_{3, ^3P_1}^{A} \lambda_{^3P_1}^{A}(\mu),
 \qquad
 \Gamma_2^- = -{1\over 3} C_F + 3C_G,  \label{app:scale1-twist3-gv-1}
\end{eqnarray}
\begin{eqnarray}
 & & \left(
\begin{array}{c}
  {28\over 3} f_{3, ^3P_1}^{V}\sigma^{V}_{^3P_1}
 - f_{3, ^3P_1}^{A}\sigma^{A}_{^3P_1} \\
  {28\over 3} f_{3, ^3P_1}^{V}\sigma^{V}_{^3P_1}
 + f_{3,^3P_1}^{A}\sigma^{A}_{^3P_1}
\end{array}
\right)_{Q} = L^{\Gamma^{+}_{3}/b} \left(
\begin{array}{c}
 {28\over 3} f_{3, ^3P_1}^{V}\sigma^{V}_{^3P_1}
  - f_{3, ^3P_1}^{A}\sigma^{A}_{^3P_1} \\
 {28\over 3} f_{3, ^3P_1}^{V}\sigma^{V}_{^3P_1}
  + f_{3, ^3P_1}^{A}\sigma^{A}_{^3P_1}
\end{array}
\right)_{\mu},\nonumber\\
& &\qquad\qquad\Gamma_{3}^{+} = \left(
\begin{array}{cc}
\frac{8}{3}C_{F} + {7\over 3}C_{G}\ \ &
\frac{2}{3}C_{F}-\frac{2}{3}C_{G} \\
\frac{5}{3}C_{F} - \frac{4}{3}C_{G}\ \ & \frac{1}{6}C_{F} + 4C_{G}
\end{array}
\right), \label{app:scale1-twist3-gv-2}
\end{eqnarray}
\begin{eqnarray}
 f_{3, ^3P_1}^{\perp}\sigma^{\perp}_{^3P_1}(Q)
 &=& L^{\Gamma^{T^+_3}/b}
 f_{3, ^3P_1}^{\perp}\sigma^{\perp}_{^3P_1}(\mu)+\frac{1}{37} \bigg(L^{4/b} -
 L^{\Gamma^{T^+_3}/b}\bigg)(m_{q_1}-m_{q_2})(\mu) f_{^3P_1},\nonumber\\
 && \qquad\qquad
 \Gamma^{T^+_3} = {23\over 6} C_F + C_G ,\label{app:scale1-twist3-pv-3}
\end{eqnarray}
for $1^3P_1$ states, and\\
\noindent [G-parity invariant components]:
\begin{eqnarray} f_{3, ^1P_1}^{A}(Q) =
L^{\Gamma^-_2/b}f_{3, ^1P_1}^{A}(\mu)-\frac{2}{29} \bigg(L^{16/(3b)}
 - L^{\Gamma^-_2/b}\bigg)(m_{q_1}+m_{q_2})(\mu) f_{^1P_1}^\perp (\mu) ,
   \label{app:scale2-twist3-1}
\end{eqnarray}
\begin{eqnarray}
 \left(
\begin{array}{c}
{28\over 3} f_{3, ^1P_1}^{V}- f_{3, ^1P_1}^{A}\omega^{A}_{^1P_1} \\
{28\over 3} f_{3, ^1P_1}^{V}+ f_{3, ^1P_1}^{A}\omega^{A}_{^1P_1}
\end{array}
\right)_{Q} &=&
 L^{\Gamma^{+}_{3}/b} \left(
\begin{array}{c}
{28\over 3} f_{3, ^1P_1}^{V}- f_{3, ^3P_1}^{A}\omega^{A}_{^1P_1} \\
{28\over 3} f_{3, ^1P_1}^{V}+ f_{3, ^3P_1}^{A}\omega^{A}_{^1P_1}
\end{array}
\right)_{\mu} \nonumber\\
 && -\frac{8}{795} \left(
\begin{array}{c}
 L^{16/(3b)} - L^{560/(9b)}\\
-L^{16/(3b)} + L^{-560/(9b)}
\end{array}
\right) (m_{q_1}+m_{q_2})(\mu) f_{^1P_1}^\perp(\mu) ,\qquad\quad
 \label{app:scale2-twist3-2}
\end{eqnarray}
\begin{eqnarray}
 f_{3, ^1P_1}^{\perp}(Q)
 = L^{\Gamma^{T^+_3}/b}
 f_{3, ^1P_1}^{\perp}(\mu),
  \label{app:scale2-twist3-3}
\end{eqnarray}
[G-parity violating components]:
\begin{eqnarray}
f_{3, ^1P_1}^{V}\lambda^{V}_{^1P_1}(Q) = L^{\Gamma^+_2/b}
 f_{3,^1P_1}^{V}\lambda^{V}_{^1P_1}(\mu)+\frac{2}{29} \bigg(L^{16/(3b)} -
 L^{\Gamma^+_2/b}\bigg)(m_{q_1}-m_{q_2}) (\mu) f_{^1P_1}^\perp (\mu) ,
  \label{app:scale2-twist3-gv-1}
\end{eqnarray}
\begin{eqnarray}
   \left(
\begin{array}{c}
f_{3, ^1P_1}^{V}\sigma^{V}_{^1P_1}
 + {28\over 3} f_{3, ^1P_1}^{A}\sigma^{A}_{^1P_1}\\
f_{3, ^1P_1}^{V}\sigma^{V}_{^1P_1}
 - {28\over 3} f_{3, ^1P_1}^{A}\sigma^{A}_{^1P_1}
\end{array}
\right)_{Q} &=& L^{\Gamma^{-}_{3}/b} \left(
\begin{array}{c}
f_{3, ^1P_1}^{V}\sigma^{V}_{^1P_1}
 + {28\over 3} f_{3, ^1P_1}^{A}\sigma^{A}_{^1P_1}\\
f_{3, ^1P_1}^{V}\sigma^{V}_{^1P_1}
 - {28\over 3} f_{3, ^1P_1}^{A}\sigma^{A}_{^1P_1}
\end{array}
\right)_{\mu}\nonumber\\
 && -\frac{59}{7080} \left(
\begin{array}{c}
 L^{16/(3b)} - L^{-896/(9b)}\\
L^{16/(3b)} - L^{896/(9b)}
\end{array}
\right) (m_{q_1}-m_{q_2})(\mu) f_{^1P_1}^\perp(\mu),
\nonumber\\
 \label{app:scale2-twist3-gv-2}
\end{eqnarray}
\begin{eqnarray}
 f_{3, ^1P_1}^{\perp}\lambda^{\perp}_{^1P_1} (Q) =
L^{\Gamma^{T^+_2}/b}
 f_{3,^1P_1}^{\perp}\lambda^{\perp}_{^1P_1}(\mu),
  \label{app:scale2-twist3-gv-3}
\end{eqnarray}
\begin{eqnarray}
 f_{3, ^1P_1}^{\perp}\sigma^{\perp}_{^1P_1}(Q)=
 L^{\Gamma^{T^-_3}/b}
 f_{3, ^1P_1}^{\perp}\sigma^{\perp}_{^1P_1}(\mu),
 \label{app:scale2-twist3-gv-4}
\end{eqnarray}
for $1^1P_1$ states, where $C_F = (N_c^2-1)/(2N_c), C_G=N_c$ and $L
\equiv \alpha_s(Q)/\alpha_s(\mu)$.

\section{The definitions of LCDAs of physical $f_1$ and $h_1$
mesons}\label{appsec:def-f1-h1}

For $f_1(1285), f_1(1420), h_1(1170)$ and $h_1(1380)$ mesons,
denoted by $B$ in the following, we define the chiral-even LCDAs to
be (with $Q\equiv u,d,$ or $s$)
\begin{eqnarray}
  \langle B(P,\lambda)|\bar Q(y) \gamma_\mu \gamma_5 Q(x)|0\rangle
  && = if_B^D m_B \, \int_0^1 du \,  e^{i (u \, p y + \bar u p x)}
   \left\{p_\mu \,
    \frac{\epsilon^{*(\lambda)} z}{p z} \, \Phi_\parallel^{B^Q}(u)
         +\epsilon_{\perp\mu}^{*(\lambda)} \, g_\perp^{(a)B^Q}(u) \right. \nonumber\\
    &&\ \ \left. - \frac{1}{2}z_{\mu}
\frac{\epsilon^{*(\lambda)} z}{(p  z)^{2}} m_{B}^{2} g_{3}^{B^Q}(u) \right\},
\label{app:evendef1} \\
  \langle B(P,\lambda)|\bar Q(y) \gamma_\mu Q(x)|0\rangle
  && = - i f_B^D m_B \,\epsilon_{\mu\nu\rho\sigma} \,
\epsilon^{*\nu}_{(\lambda)} p^{\rho} z^\sigma \, \int_0^1 du \, e^{i
(u \, p y + \bar u\, p  x)} \, \frac{g_\perp^{(v) B^Q}(u)}{4},
\label{app:evendef2}
\end{eqnarray}
and the chiral-odd LCDAs to be
\begin{eqnarray}
 &&\langle B(P,\lambda)|\bar Q(y) \sigma_{\mu\nu}\gamma_5 Q(x)|0\rangle
  =  f_B^{\perp,D} \,\int_0^1 du \, e^{i (u \, p y +
    \bar u\, p x)} \,
\Bigg\{(\epsilon^{*(\lambda)}_{\perp\mu} p_{\nu} -
  \epsilon_{\perp\nu}^{*(\lambda)}  p_{\mu})
  \Phi_\perp^{B^Q}(u)\nonumber\\
&& \hspace*{+5cm}
  + \,\frac{m_B^2\,\epsilon^{*(\lambda)} z}{(p z)^2} \,
  (p_\mu z_\nu - p_\nu  z_\mu) \, h_\parallel^{(t) {B^Q}}(u)\nonumber\\
 && \hspace*{+5cm} + \frac{1}{2}
(\epsilon^{*(\lambda)}_{\perp \mu} z_\nu
-\epsilon^{*(\lambda)}_{\perp \nu} z_\mu) \frac{m_{B}^{2}}{p\cdot z}
 h_{3}^{B^Q}(u)
\Bigg\},\label{app:odddef1}\\
&&\langle B(P,\lambda)|\bar Q(y) \gamma_5 Q(x) |0\rangle
  =  f_B^{\perp,D}
 m_{B}^2 (\epsilon^{*(\lambda)} z)\,\int_0^1 du \, e^{i (u \, p y +
    \bar u\, p x)}  \, \frac{h_\parallel^{(p) {B^Q}}(u)}{2}\,, \label{app:odddef2}
\end{eqnarray}
where the distribution amplitudes are subject to the choices of the
normalization constants $f_B^{\perp,D}$  with (i) $D\equiv u$ for
$f_1(1285), h_1(1170)$, and (ii) $D\equiv s$ for $f_1(1420),
h_1(1380)$. Here we use the conventions for the decay constants that
$f_{f_1(1285)}^{\perp, u}=f_{f_1(1285)}^u$, $f_{f_1(1420)}^{\perp,
s}=f_{f_1(1285)}^s$, $f_{h_1(1170)}^u = f_{h_1(1170)}^{\perp,
u}(1~{\rm GeV})$, and $f_{h_1(1380)}^s = f_{h_1(1380)}^{\perp,
s}(1~{\rm GeV})$. The relevant decay constants can be found in Eqs.
(\ref{eq:decay-f1-1})$-$(\ref{eq:decay-f1-4}) and
(\ref{eq:decay-h1-1})$-$(\ref{eq:decay-h1-4}).

 \end{document}